\documentclass[%
 aip,
 jmp,%
 amsmath,amssymb,
 reprint,%
]{revtex4-1}

\usepackage{graphicx}
\usepackage{dcolumn}
\usepackage{bm}

\begin{document}


\title[Chimera resonance in networks of chaotic maps]{Chimera resonance in networks of chaotic maps}

\author{Elena Rybalova}
\email{rybalovaev@gmail.com}
 \affiliation{%
Institute of Physics, Saratov  State University, 83 Astrakhanskaya Street, Saratov, 410012, Russia}

\author{Vasilii Nechaev}
\email{nechaev.vas2021@bk.ru}
 \affiliation{%
Institute of Physics, Saratov  State University, 83 Astrakhanskaya Street, Saratov, 410012, Russia}

\author{Eckehard Sch\"{o}ll}
\email{schoell@physik.tu-berlin.de}
 \affiliation{%
Institut f\"ur Theoretische Physik, Technische Universit\"at Berlin, 10623 Berlin, Germany}
 \affiliation{%
Bernstein Center for Computational Neuroscience Berlin, 10115 Berlin, Germany}
 \affiliation{%
Potsdam Institute for Climate Impact Research, 14473 Potsdam, Germany}

\author{Galina Strelkova}
\email{strelkovagi@sgu.ru}
 \affiliation{%
Institute of Physics, Saratov  State University, 83 Astrakhanskaya Street, Saratov, 410012, Russia}

\date{\today}

\begin{abstract}
We explore numerically the impact of additive Gaussian noise on the spatio-temporal dynamics of ring networks of nonlocally coupled chaotic maps. The local dynamics of network nodes is described by the logistic map, the Ricker map, and the Henon map. 2D distributions of the probability of observing chimera states are constructed in terms of the coupling strength and the noise intensity and for several choices of the local dynamics parameters. It is shown that the coupling strength range can be the widest at a certain optimum noise level at which chimera states are observed with a high probability for a large number of different realizations of randomly distributed initial conditions and noise sources. This phenomenon demonstrates a constructive role of noise in analogy with the effects of stochastic and coherence resonance and may be referred to as chimera resonance. 
\end{abstract}

\pacs{Valid PACS appear here}
\keywords{network; nonlocal coupling; chimera states; noise; resonance; logistic map; Henon map; Ricker map}
\maketitle

\begin{quotation}

Random perturbations are inevitably and sometimes permanently present in many real-world systems and thus can significantly affect their functioning and characteristics. Investigating the impact of random influences on the system dynamics enables one to obtain more stable regimes of operation and to find efficient ways of their control.   Despite the generally accepted interpretation of noise as a source of destruction, noise impacts can sometimes play a counterintuitive beneficial role in the system behavior by enhancing the degree of order or improving its characteristics. Since real-world systems consist of interacting nodes with different individual dynamics and coupling topology and can demonstrate various complex nonlinear patterns, studying the robustness of spatio-temporal structures, such as, e.g., chimera and solitary states~\cite{Kuramoto:2002uu,Abrams:2004vx,Maistrenko:2014tm,Jaros:2015tv}, towards noise influences has become one of the prominent research directions in different scientific fields~\cite{Horsthemke:1984wd,Arnold:1995tz,Chacron:2003wm,McDonnell:2011th,Destexhe:2012ve}. 
Recently, it has been shown that introducing noise in complex networks of coupled nonlinear oscillators can induce new structures which cannot exist without noise~\cite{Semenova:2016aa,Khatun:2023un,Zhu:2023ws} and can effect the lifetime of amplitude chimera states~\cite{Loos:2016aa,Zakharova:2020uu,Rybalova:2019un}. 
Multiplexing noise can promote and control synchronization of complex structures in multilalyer networks~\cite{Vadivasova:2020vw,Rybalova:2022wi}. 
It has also been established that  noise sources can increase the probability of observing chimera states in networks of nonlocally coupled chaotic maps within a rather wide range of the noise intensity and the coupling strength~\cite{Rybalova:2022wa}. In this case noise plays a constructive role which deserves to be studied further in more detail. In the present paper we explore numerically the impact of additive white noise on the dynamics and existence of chimeras in networks of nonlocally coupled logistic maps, Henon maps, and Ricker maps. We elucidate how the probability of chimera observation depends on the local dynamics of individual nodes, on the noise intensity and the coupling parameters. Our numerical results show that there is a resonance-like dependence of the high probability of observing chimeras with respect to the noise level and the coupling strength. In this sense, the revealed phenomenon may be referred to as chimera resonance. 

\end{quotation}

\section{Introduction}

The word ``noise'' is ordinarily associated with the term ``obstruction''. It was traditionally considered that the presence of noise can only worsen or even destroy the operation of the system. However, many recent research works have shown that noise can sometimes play a constructive or beneficial role in nonlinear dynamics. Studying the noisy dynamics is highly important for understanding the processes which take place in real-world systems and has a significant practical importance for technological, infrastructural and communication networks, biological, epidemiological, climate and social processes, as well as in neurodynamics and medicine~\cite{Horsthemke:1984wd,Arnold:1995tz,Chacron:2003wm,McDonnell:2011th,Destexhe:2012ve}, etc.

Noise sources in complex dynamical systems are able to induce completely new regimes and patterns that cannot exist without noise~\cite{Horsthemke:1984wd,Gammaitoni:1989wf,Neiman:1994ua,Shulgin:1995uy,Pikovsky:1997vf,Anishchenko:1999ti,Lindner:1999tv,Bashkirtseva:2009ws}. These effects were called noise-induced transitions \cite{Horsthemke:1984wd}. Various studies have convincingly demonstrated that noise can enhance the degree of order or coherence in a system or evokes improvement of its performance~\cite{Gang:1993aa,Pikovsky:1997vf,Lindner:1999tv}. It can lead to the formation of more regular temporal and spatial structures, noise-induced and noise-enhanced synchronization effects of various complex spatio-temporal patterns in networks~\cite{Neiman:1994ua,Kurrer:1995um,Hempel:1999th,Neiman:1999ut,Freund:2000vc,Janson:2004aa}, cause the amplification of weak signals accompanied by growth of their signal-to-noise ratio (stochastic resonance)~\cite{Benzi:1981us,Benzi:1982wv,Gammaitoni:1989wf,Anishchenko:1999ti}. Numerical studies have shown that noise sources can also be used to stabilize and/or to effectively control the operating modes of complex systems and networks~\cite{Horsthemke:1984wd,Neiman:1994ua,Arnold:1995tz,Schimansky-Geier:1993up,Shulgin:1995uy,Arnold:1996uv,Han:1999wi,Bashkirtseva:2009ws}.

Recently, the interest in the study of noise effect on the behavior of complex nonlinear systems  has significantly increased with the discovery of special spatiotemporal structures, such as chimera states~\cite{Kuramoto:2002uu,Abrams:2004vx,Omelchenko:2011uc,Omelchenko:2015uu,Panaggio:2015uu,Bogomolov:2017wq,Scholl:2020uk,Zakharova:2020uu,Scholl:2021wt} and solitary states~\cite{Maistrenko:2014tm,Jaros:2018ve,Berner:2020um,Semenov:2016wt}. A chimera state represents an intriguing example of partial synchronization patterns when spatially localized domains with coherent (synchronized) and incoherent (desynchronized) dynamics coexist in networks of coupled nonlinear oscillators \cite{Kuramoto:2002uu,Abrams:2004vx}. Chimera states appear along a transition from complete synchronization (coherence) to fully desynchronized spatio-temporal dynamics (incoherence)~\cite{Omelchenko:2012tv,Semenova:2017wn,Rybalova:2017tl}. They were studied theoretically and numerically \cite{Kuramoto:2002uu,Abrams:2004vx,Shima:2004aa,Omelchenko:2011uc,Panaggio:2013tr,Zakharova:2014td,Panaggio:2015uu,Semenova:2015tt,Ulonska:2016tx,Scholl:2016vm,Sawicki:2017um,Bogomolov:2017wq} and were also observed experimentally \cite{Hagerstrom:2012vd,Tinsley:2012tc,Larger:2013ub,Martens:2013wq,Wickramasinghe:2013tp,Gambuzza:2014vj,Kapitaniak:2014vc,Rosin:2014wy,Schmidt:2014ui}. The performed investigations  have shown that chimera structures can be related to the processes occurring in the brain and are also associated with various manifestations of the nervous and brain activity of humans and animals \cite{Bansal:2019wy,Majhi:2019ue,Scholl:2021um}.  

Solitary states represent another partial synchronization patterns. In this case an instantaneous spatial profile of the network dynamics consists of a coherent part and a single or a set of isolated spikes corresponding to elements which behave differently from the coherent nodes. It has been shown that nonlocal coupling in a network induces bistable dynamics of individual elements that leads to the appearance of solitary states~\cite{Maistrenko:2014tm,Jaros:2015tv}. These states have been found numerically in a number of networks of oscillators~\cite{Maistrenko:2014tm,Jaros:2015tv,Jaros:2018ve,Berner:2020um,Wu:2018ws}, discrete-time systems~\cite{Semenova:2015tt,Semenova:2017wn,Rybalova:2017tl,Semenova:2018th}, neural models~\cite{Mikhaylenko:2019uq,Rybalova:2019vg,Schulen:2019td,Schulen:2021tn}, models of power grids~\cite{Berner:2021wd,Hellmann:2020wr,Taher:2019wz}, and also experimentally in a network of coupled pendula~\cite{Kapitaniak:2014vc}.

Noise sources can be introduced into complex systems and networks in different ways and can have various characteristics and statistics. They can be added additively to all network elements or multiplicatively to the control parameters of a network.  Noise can influence the network dynamics via noise-modulated intra- or inter-layer coupling. The latter is known as multiplexing noise~\cite{Nikishina2022,Rybalova:2022wi}. The coupling coefficients can be modulated both independent noise sources and a common noise source. It has been found that introducing noise can induce novel spatio-temporal structures, such as a coherence-resonance chimera in neural networks~\cite{Semenova:2016aa} and a solitary state chimera in networks of chaotic oscillators~\cite{Rybalova:2018we}. 
The presence of noise sources can either vanish or increase the lifetime of certain types of chimeras in ring networks of harmonic and chaotic systems~\cite{Loos:2016aa,Zakharova:2020uu,Rybalova:2019un}. It has recently been established that multiplexing noise can induce and control partial and complete inter-layer synchronization of complex structures in multilayer networks independently of the characteristics of the dynamics of an individual element and on the nature of the cluster structure (chimeras of different types and solitary states) in the layer~\cite{Vadivasova:2020vw,Rybalova:2022wi}. 
It has also been shown that the low-frequency multiplexing noise can produce solitary states in networks of nonlocally coupled discrete-time systems~\cite{Rybalova:2022wi}. In Ref.\cite{Rybalova:2022uv} it has been shown that persistently noise-modulated parameters of local dynamics or coupling strength in a network of nonlocally coupled Lozi maps lead to reducing the domains of solitary state existence with respect to the coupling strength, while solitary states may persist in the case of randomly distributed parameters. Recently, it has been shown that additive noise in a network of nonlocally coupled logistic maps cannot only induce the appearance of chimera states but also maximize the probability of their observation within a finite range of the coupling strength and for rather strong noise \cite{Rybalova:2022wa}. It has also been established that there is  a counter-intuitive non-monotonic dependence of the chimera existence upon noise intensity, which is reminiscent of the constructive influence of noise known from coherence resonance. However, there is still unclear how the probability level of chimera observation in noisy networks of coupled oscillators can depend on the local dynamics of individual nodes as well as on the variation of coupling parameters (strength and range). Besides, it is quite interesting to uncover whether the constructive role of noise established in Ref.\cite{Rybalova:2022wa} is typical and general for networks of other nonlinear oscillators. 

In the present work we continue and extend the investigations started in Ref.\cite{Rybalova:2022wa} and systematically study the impact of additive Gaussian noise on the dynamics and chimera observation in networks of nonlocally coupled discrete-time systems. As individual elements we use the logistic map, the Henon map, and the Ricker map. We construct 2D diagrams of typical dynamical regimes which are observed in noise-free networks depending on  the local dynamics parameters and the coupling strength. After introducing additive noise, we calculate the probability of observing chimera states in the three networks when the noise intensity and the coupling strength are varied. The numerical results are summarized in the 2D distributions of the probability drawn for several selected values of the local dynamics parameters. It is shown that the chimera existence demonstrates a resonance-like dependence on the noise intensity and the coupling strength. Moreover,  there is an optimum noise level at which the interval of the coupling strength within which chimeras are observed with a high or even maximum probability is the widest. Thus, this fact constitutes the constructive role of noise in analogy with  stochastic and coherence resonance.

\section{Networks under study}

\subsection{Model equations of the network}

The object of our numerical study is a ring network of nonlocally coupled discrete-time systems, which is subjected to additive noise. The network is described by the following system of equations:
\begin{eqnarray}\label{eq:system}
x(i,n+1) &=& F(i, n)+\\\nonumber&+&\frac{\sigma}{2R}\sum_{j=i-R}^{i+R}[F(j, n)-F(i, n)]+D\xi(i,n),\\\nonumber
y(i, n+1) &=& G(i, n),
\end{eqnarray}
where $x(i, n)$ and $y(i, n)$ are dynamical variables, $i=1,2,3,\ldots,N$ numbers the elements in the ensemble and $N=1000$ is the total number of elements, $n$ denotes the discrete time. Functions $F(i, n)$ and $G(i, n)$ are defined by the right-hand sides of the equations of the respective discrete-time systems which will be given below.  The elements within the ring are coupled through a nonlocal scheme, i.e., each $i$th node is linked with $1<R<N/2$ neighbours on the left and right. The parameter $R$ denotes the coupling range and $\sigma$ is the coupling strength between the elements. 
The influence of additive noise is determined by the last term in the first equation of (\ref{eq:system}), where $\xi(i,n)$ is a Gaussian noise source, and $D$ is the noise intensity. 

\subsection{Models and dynamics of individual elements}

As individual elements in the network (\ref{eq:system}), we consider three different discrete-time systems, namely, the logistic map~\cite{Feigenbaum:1978wh,Feigenbaum:1979vj}, the Ricker~\cite{Ricker:1954wp} map, and the Henon map~\cite{Henon:1969wp}. 

\begin{figure}[h]
	\centering
\begin{tabular}{ccc}
\includegraphics[width=.33\columnwidth]{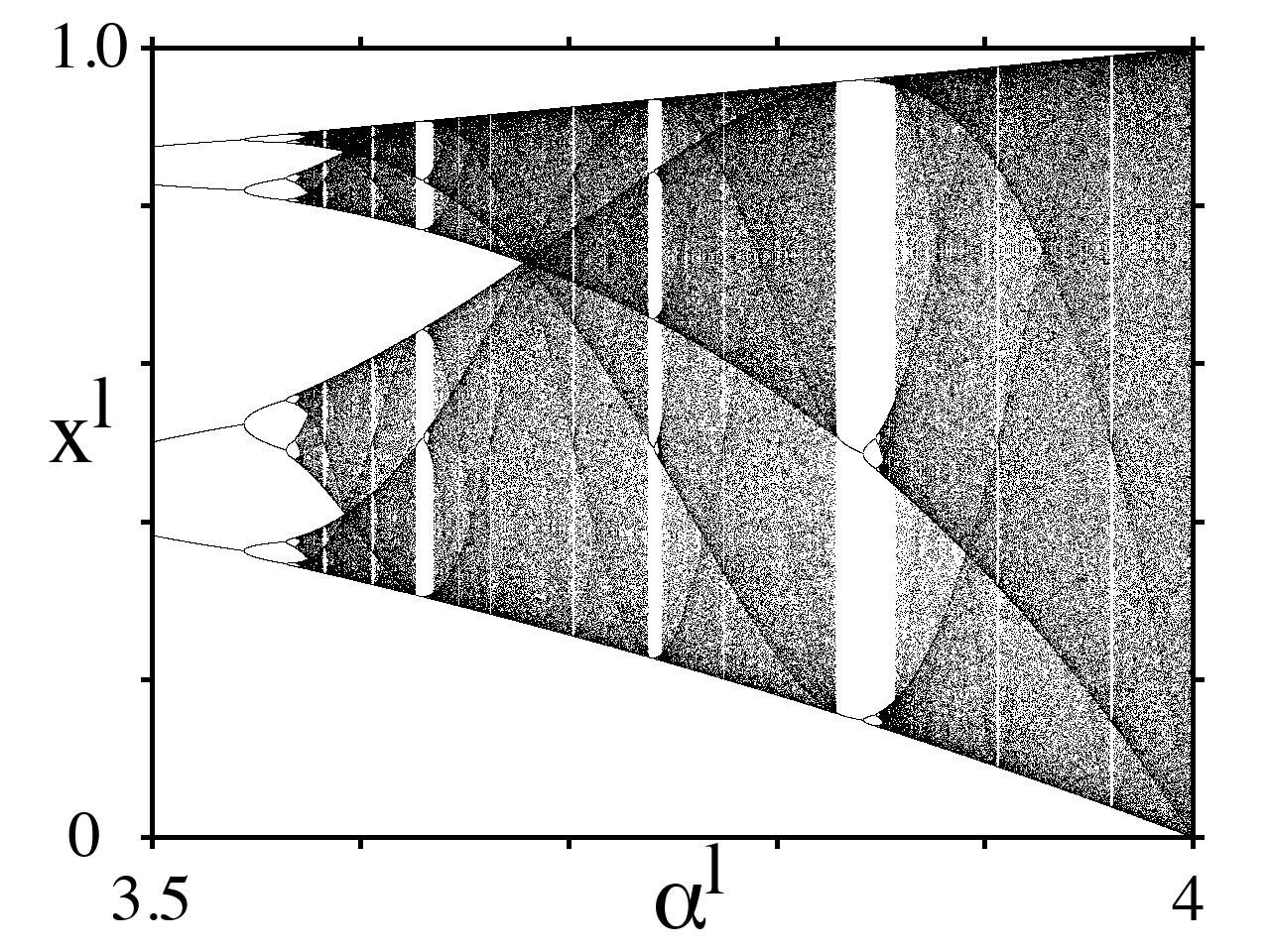} &
\includegraphics[width=.33\columnwidth]{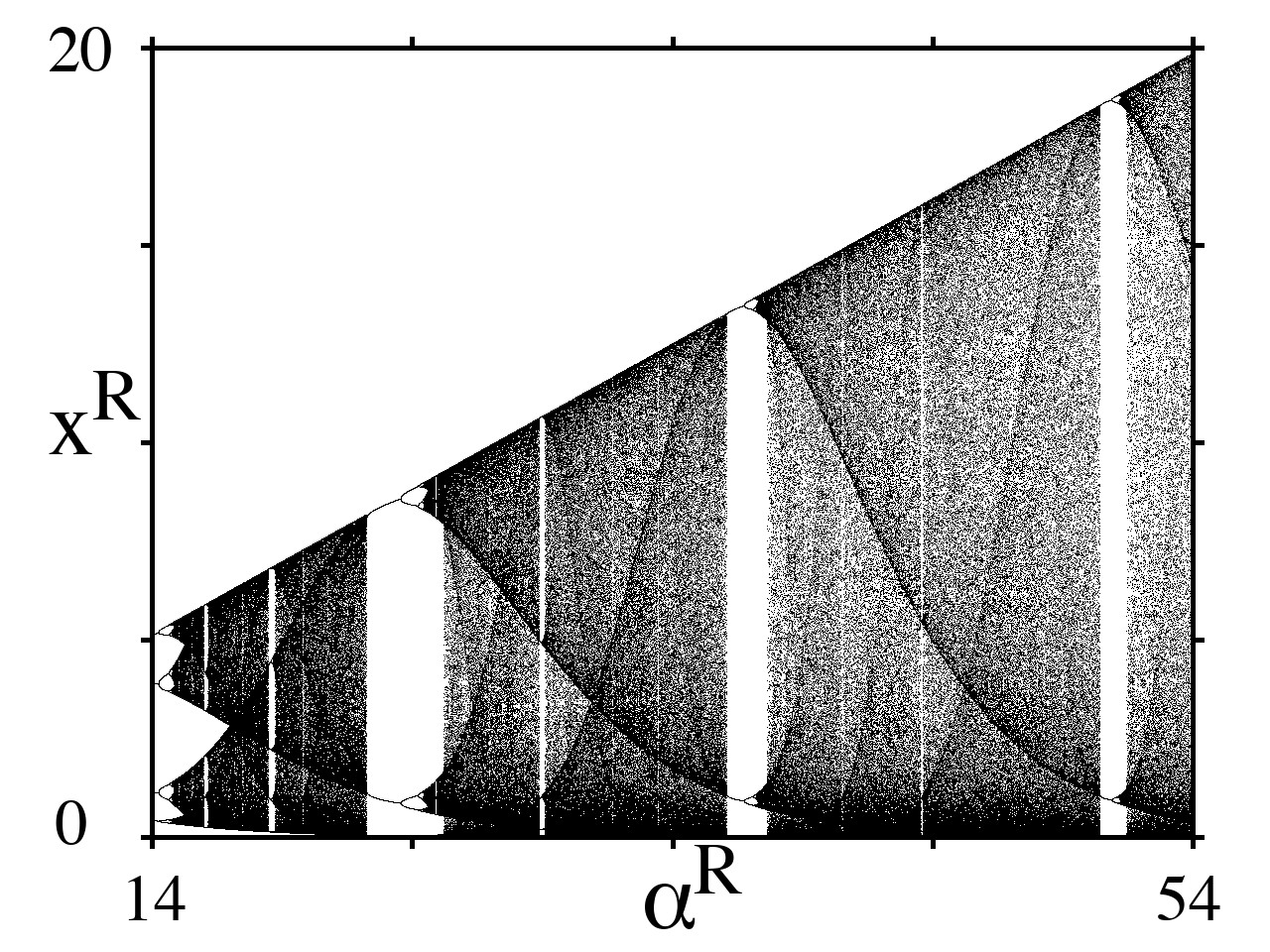}  &
\includegraphics[width=.33\columnwidth]{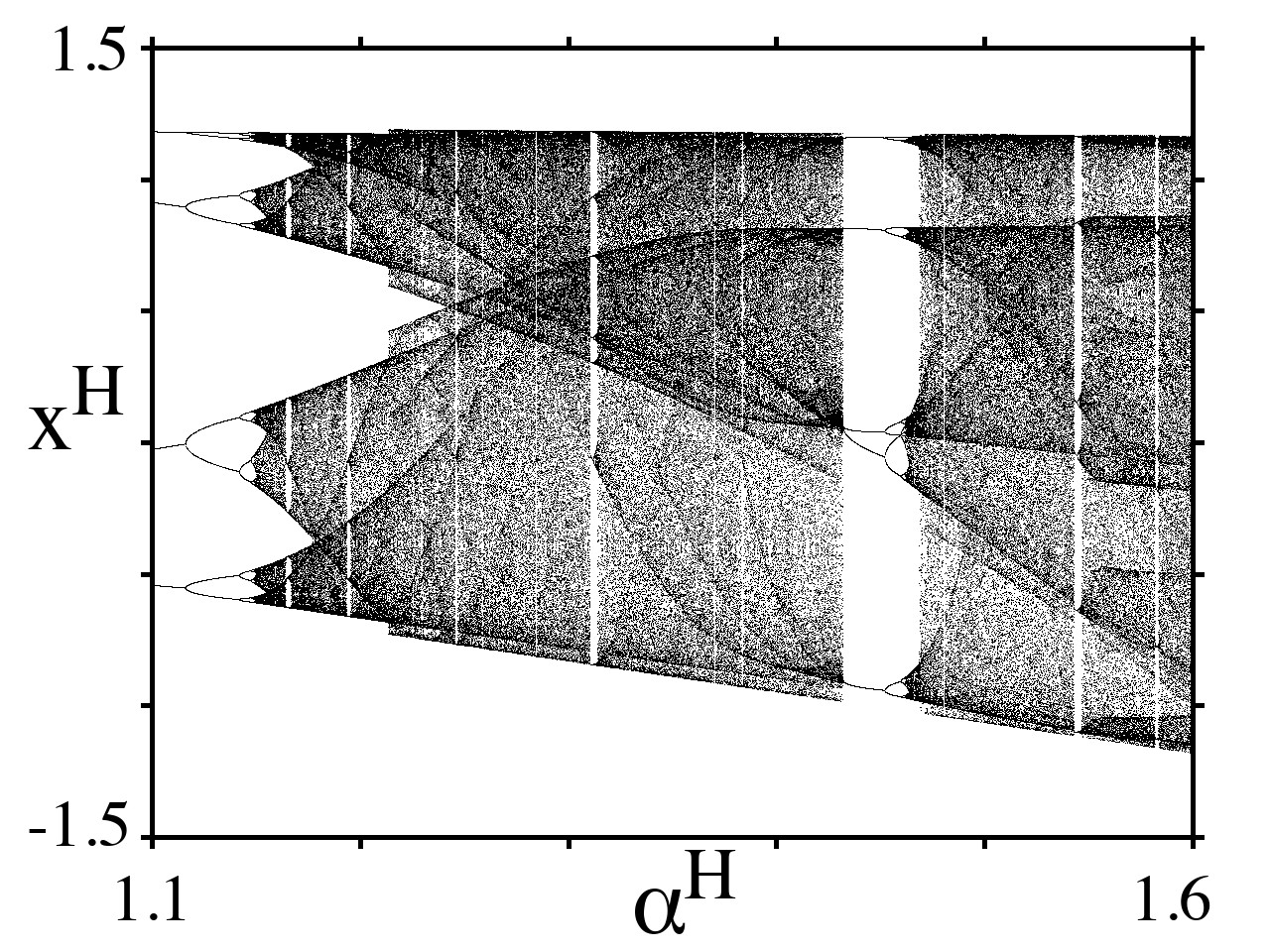}\\
\hspace{8pt} (a) & \hspace{8pt} (b) & \hspace{8pt} (c)\\
\end{tabular}
\caption{Bifurcation diagrams for isolated maps: (a) the logistic map, (b) the modified Ricker map, and (c) the Henon map  at $\beta^H=0.2$.}
	\label{fig_1}
\end{figure}

The logistic map~\cite{Feigenbaum:1978wh,Feigenbaum:1979vj} is a canonical one-dimensional map ($G(n)=0$ in (\ref{eq:system})) that is characterized by chaotic dynamics and multistability  \cite{Anishchenko:2014tw}. The logistic map is described as follows:
\begin{equation}\label{eq:logistic}
x^l(n+1)=F^l (n)=\alpha^l x^l(n) (1-x^l(n)),
\end{equation}
where $x^l(n)$ is the dynamical variable, $\alpha^l$ is the control (bifurcation) parameter. The transition to a  chaotic attractor in the logistic map occurs at $\tilde{\alpha}^{l}\approx3.57$ via a period-doubling bifurcation cascade  which is clearly seen in the bifurcation diagram  $x^l(\alpha^l)$ shown in Fig.~\ref{fig_1}(a) for the isolated map. The system trajectories diverge to infinity at  $\alpha^{l}>4.0$.

The one-dimensional Ricker map~\cite{Ricker:1954wp} ($G(n)=0$ in (\ref{eq:system})) is defined by the following equation:
\begin{eqnarray}\label{eq:ricker}
x^{R}(n+1)=F^{R} (n)=x^{R}(n)\exp\left[\alpha^{R}\left(1-\frac{x^{R}(n)}{K}\right)\right],
\end{eqnarray}
where $x^R(n)$ is the dynamical variable, $\alpha^{R}$ is the control (bifurcation) parameter and $K$ is the carrying capacity of the environment. The numerical studies have shown that in the presence of external noise, trajectories of the network of nonlocally coupled Ricker maps diverge to infinity and incoherent dynamics is not observed in the network as compared with the cases of other individual elements. Therefore, in the present work we use a modified Ricker map proposed in Ref.~\cite{Bukh:2018wa}:
\begin{eqnarray}\label{eq:mod-ricker}
x^{R}(n+1)=F^{R} (n)=\alpha^{R}|x^{R}(n)|\exp[-x^{R}(n)].
\end{eqnarray}
This map also demonstrates the transition to chaos via period-doubling bifurcations at 
$\tilde{\alpha}^{R}\approx4.77$ (Fig.~\ref{fig_1}(b)). In this case the system trajectories do not diverge to infinity for any value of $\alpha^{R}$. The bifurcation diagram for the isolated modified Ricker map (\ref{eq:mod-ricker}) (Fig.~~\ref{fig_1}(b)) shows that the maximum value of the dynamical variable 
$x^{R}$ continues to grow linearly in the whole interval of the control parameter variation. The regions of chaotic behavior alternate with periodic windows, whose width narrows down as the parameter $\alpha^{R}$ increases.   

The two-dimensional Henon map \cite{Henon:1969wp} is given by the system of equations:
\begin{eqnarray}\label{eq:henon}
x^H(n+1)&=&F^H (n)=1-\alpha^H (x^H(n))^{2} +y^H(n), \\
y^H(n+1)&=&G^H (n)=\beta^H x^H(n), \nonumber 
\end{eqnarray}
where $x^H(n)$ and $y^H(n)$ are the dynamical variables, and $\alpha^H$ and $\beta^H$ are the positive control parameters.  
The Henon map can be reduced to the logistic map as $\beta^H \rightarrow 0$. When the control parameters are varied, a period-doubling bifurcation cascade takes place and a nonhyperbolic chaotic attractor~\cite{Anishchenko:2014tw} appears in the Henon map. The corresponding bifurcation diagram $x^H(\alpha^H)$ at fixed $\beta^H=0.2$ for the isolated Henon map is drawn in Fig.~\ref{fig_1}(c). Starting with $\tilde{\alpha}^{H}\approx1.15$ the map demonstrates chaotic behavior and at 
  $\alpha^{H}>1.61$ trajectories diverge to infinity. 

It was found~\cite{Omelchenko:2011uc,Semenova:2015tt,Rybalova:2017tl,Omelchenko:2012tv} that when logistic maps or Henon maps are nonlocally coupled within a ring, the transition from complete chaotic synchronization to spatio-temporal chaos is accompanied by the appearance of amplitude and phase chimera states when the coupling strength decreases.

Due to multistability of the considered networks of chaotic maps not all of the initial conditions provide chimera states.
The spatio-temporal dynamics of each network is studied for a set of 50 to 100 different realizations of randomly distributed initial states of the dynamical variables ($x(i,0)$, $y(i,0)$) and in each case different noise realizations. We use the same noise realizations for all coupling strengths and noise intensities.

\subsection{Quantitative measure}

In our research the changes in the network dynamics are illustrated and analyzed by means of instantaneous spatial distributions of dynamical variables $x(i)$ at a fixed time $n$ (snapshots) and spatio-temporal diagrams $x(i,n)$ of the studied network. 
The evolution of complex structures observed in the networks is also estimated quantitatively by using the cross-correlation coefficient between the network elements \cite{Vadivasova:2016td}:  
\begin{equation}\label{eq:correlation}
 C_{1,i}=\frac{\langle\tilde{x}(1,n) \tilde{x}(i,n)\rangle}{\sqrt{\langle(\tilde{x}(i,n))^2\rangle \langle (\tilde{x}(i,n))^2\rangle}},~~~i=2,3,\dots,N, \\
\end{equation}
where $\tilde{x} = x(n) - \langle x(n) \rangle$, $ \langle x(n) \rangle $ is the average of the  $x$ variable over $T = 50000$ iterations. The cross-correlation coefficient (\ref{eq:correlation}) indicates the degree of correlation between the first and all the other nodes of the network and takes the values within the interval $[-1,+1]$. Plotting spatial distributions of $C_{1,i}$ values ($i=2,3,\ldots,N$) enables one to diagnose the dynamics of the studied network and the type of observed spatio-temporal structures. The boundary values $C_{1,i}=-1$ and $C_{1,i}=+1$ correspond to complete anti-phase and in-phase synchronization, respectively, while  $C_{1,i}=0$ relates to the incoherent behavior of the elements (desynchronization). In the regime of a phase chimera, values of the cross-correlation coefficient can alternate irregularly between $+1$ and $-1$ within an incoherent cluster\cite{Vadivasova:2016td}.
 In the case of partial phase shift, values of $C_{1,i}$ lie between [0,1] (or [-1,0]), and this shift (these $C_{1,i}$  values) indicates the presence of phase chimeras and solitary nodes. The aforementioned peculiarities of the cross-correlation coefficient allow us to clearly distinguish phase chimeras and solitary states while simulating the network dynamics.

\section{Results}

Our results show that the presence of additive noise of certain intensities can induce the appearance and observation of chimera states in the studied networks  within a sufficiently wide range of the nonlocal coupling strength $\sigma$. This $\sigma$-interval can be the widest at  a certain optimum noise level at which chimera states are observed with a high probability for a large number of different realizations of  randomly distributed initial conditions. This phenomenon demonstrates a constructive role of  noise in analogy with the effects of stochastic~\cite{Benzi:1981us,Benzi:1982wv} and coherence~\cite{Pikovsky:1997vf,Lindner:1999tv} resonance and may be referred to as {\it chimera resonance}.

\subsection{Network of nonlocally coupled logistic maps}\label{sec_ring_logistic}

We start with considering the spatio-temporal dynamics of the ring network of nonlocally coupled logistic maps. In our simulation we fix the coupling range at $R=320$ and vary the local dynamics parameter $\alpha^l\in[3.5,4]$ and the coupling strength $\sigma\in [0.15,0.55]$.

\begin{figure}[ht]
	\centering
\begin{tabular}{cc}
\includegraphics[width=.47\columnwidth]{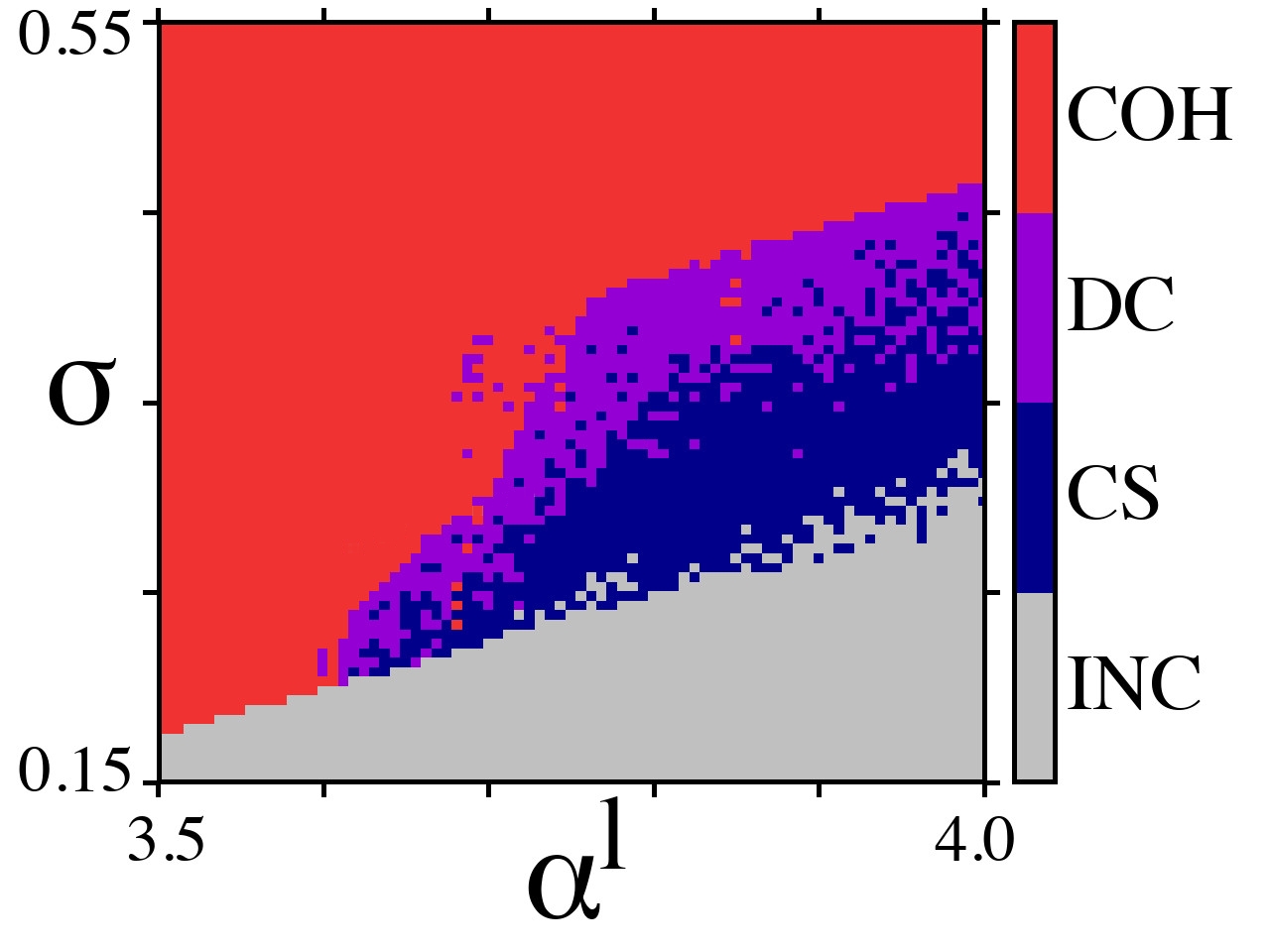} &
\includegraphics[width=.47\columnwidth]{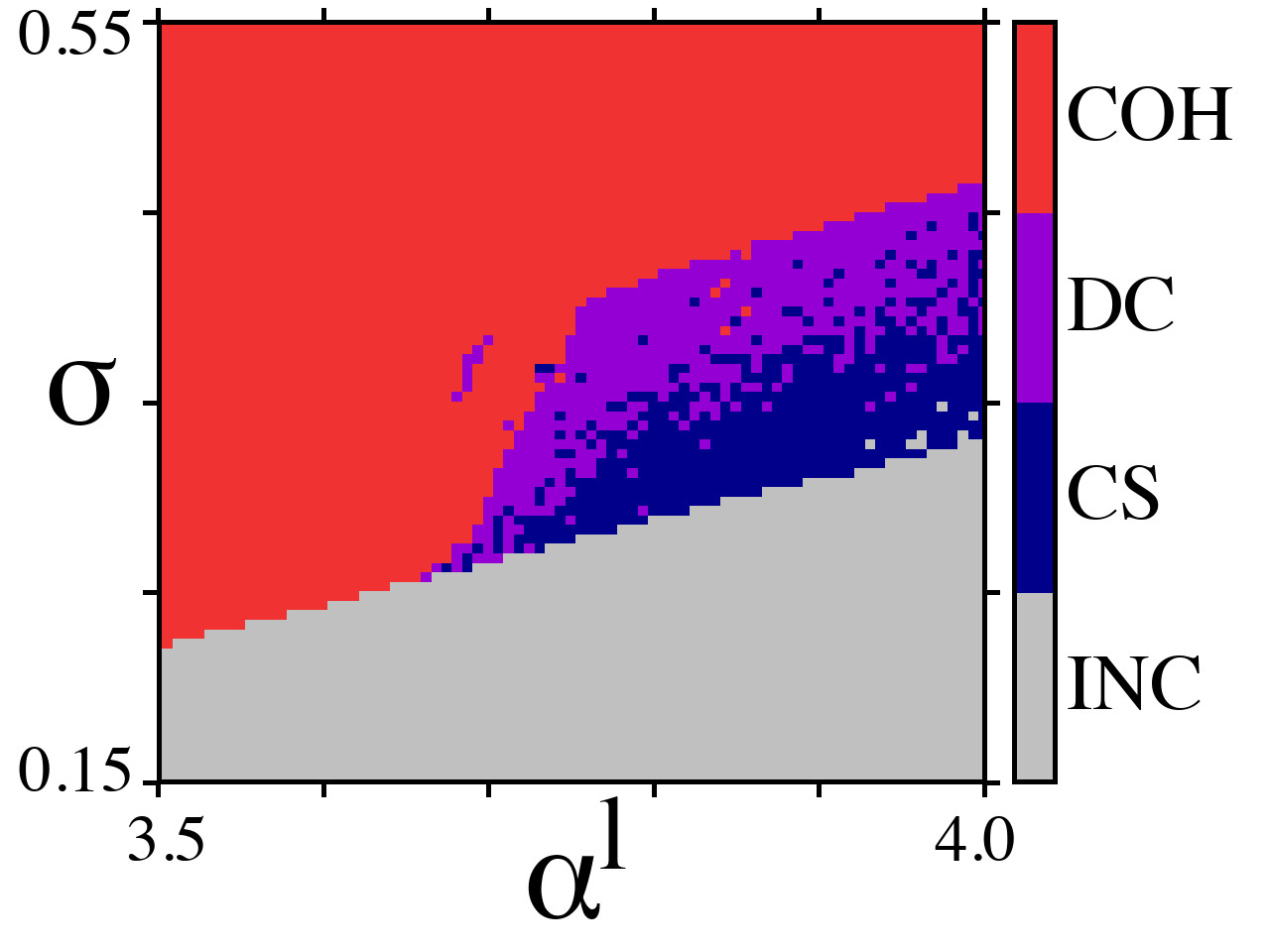}  \\
\hspace{8pt} (a) & \hspace{8pt} (b)\\
\end{tabular}
\begin{tabular}{cc}
\includegraphics[width=.47\columnwidth]{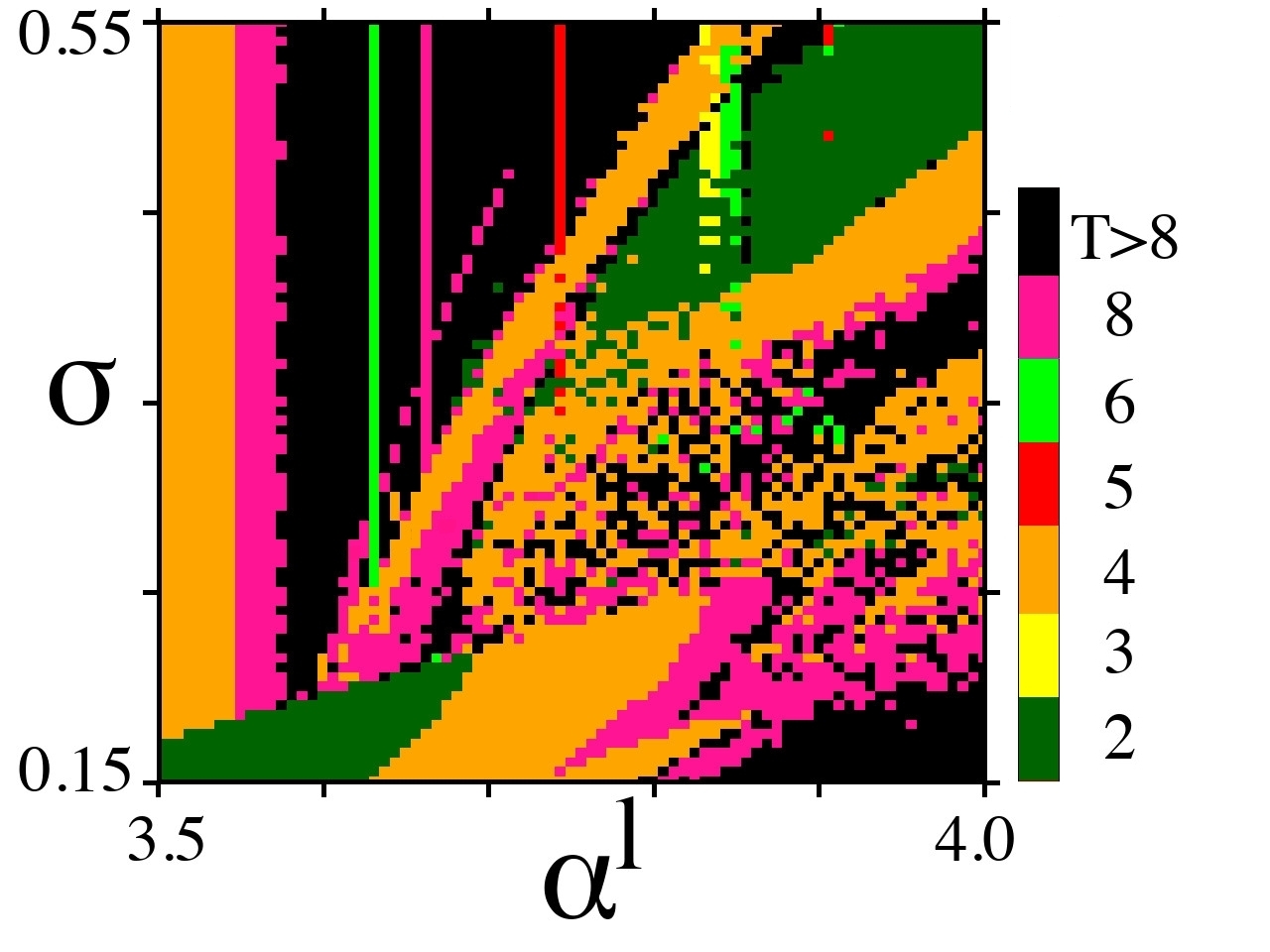} &
\includegraphics[width=.47\columnwidth]{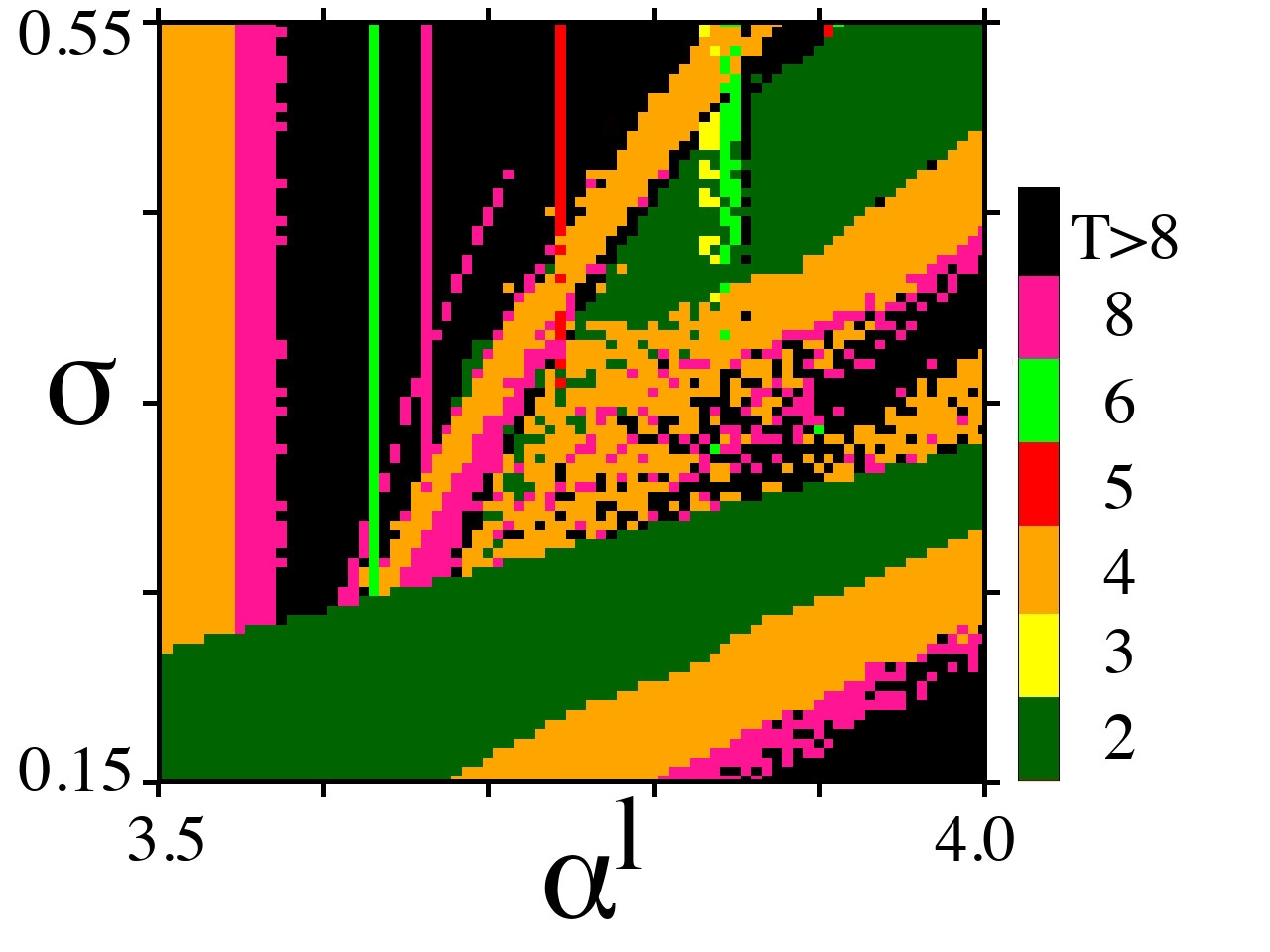}  \\
\hspace{8pt} (c) & \hspace{8pt} (d)\\
\end{tabular}
\caption{2D diagrams of spatio-temporal regimes (a,b) and of corresponding temporal dynamics (c,d) for the noise-free network of nonlocally coupled logistic maps in the ($\alpha^{l},\sigma$) parameter plane  for two different realizations of initial conditions randomly distributed within the interval $[0,1]$.  COH is coherence or complete synchronization between elements, DC corresponds to snapshots with profile discontinuities, CS is chimera states, and INC is incoherence. The color scale in (c,d) indicates the period of temporal dynamics. Other parameters: $R=320$, $N=1000$, $D=0$.}
	\label{fig_2}
\end{figure}

\begin{figure}[ht]
	\centering
\begin{tabular}{cccc}
\includegraphics[width=.23\columnwidth]{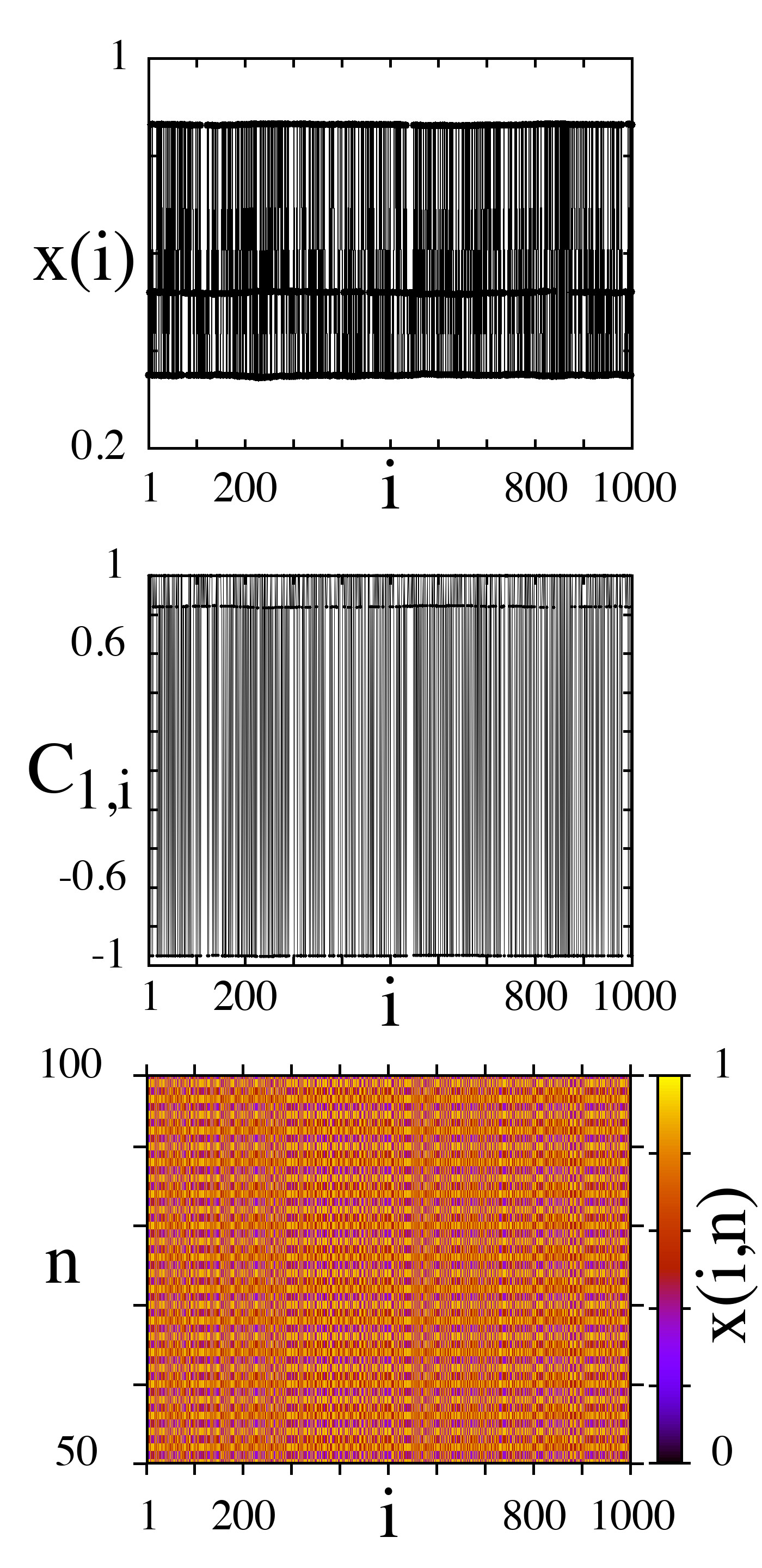} &
\includegraphics[width=.23\columnwidth]{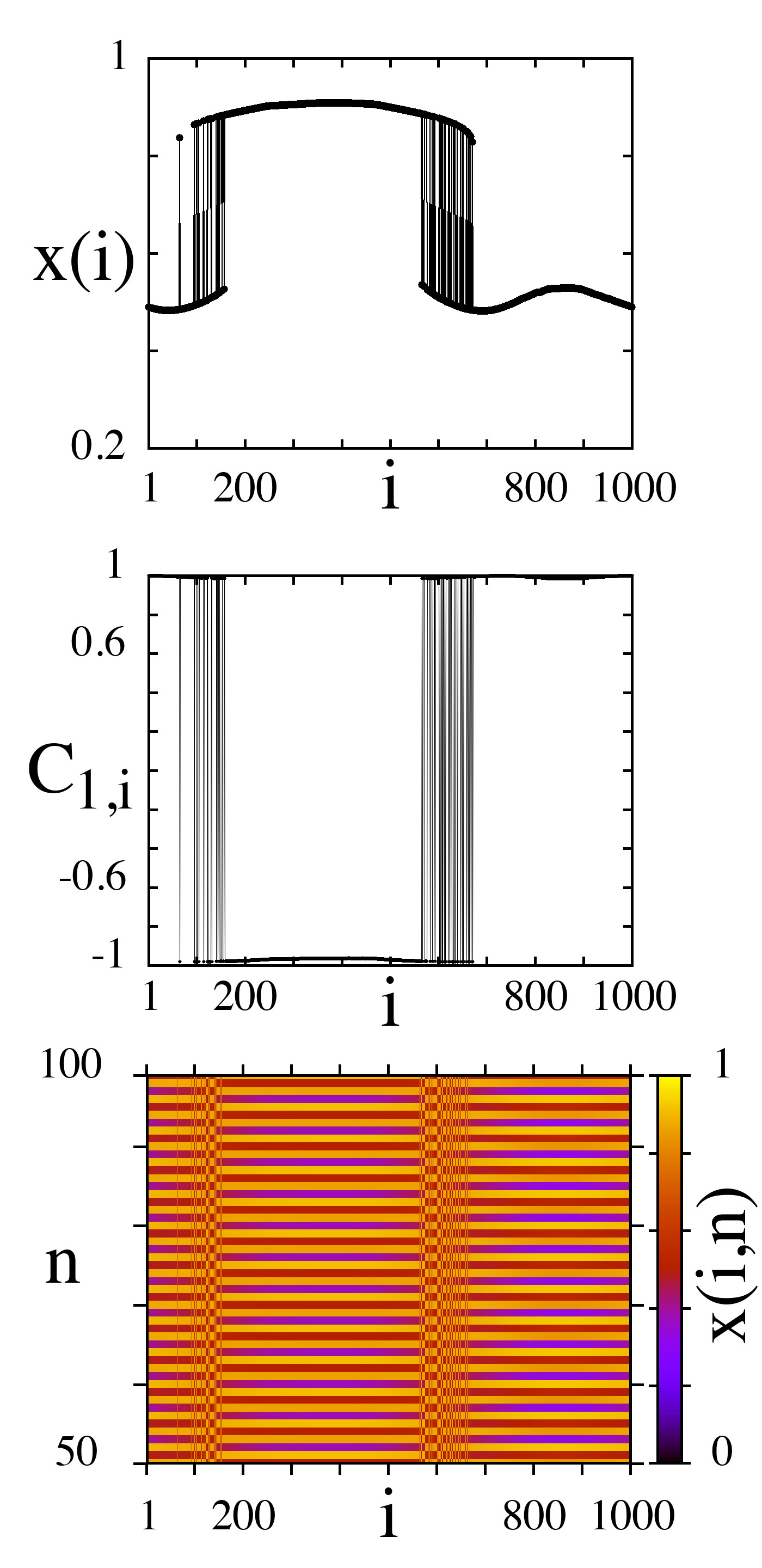}  &
\includegraphics[width=.23\columnwidth]{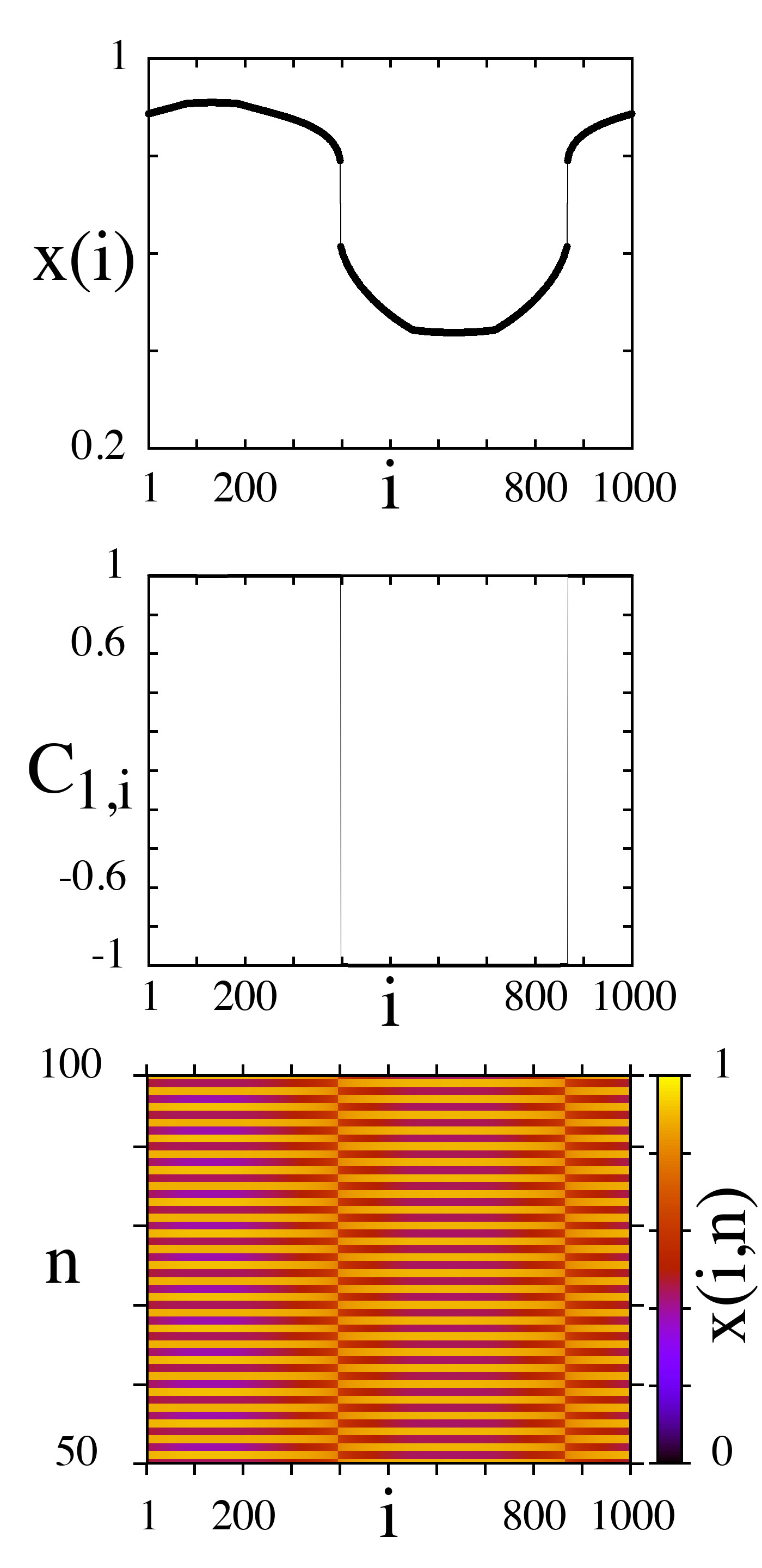} &
\includegraphics[width=.23\columnwidth]{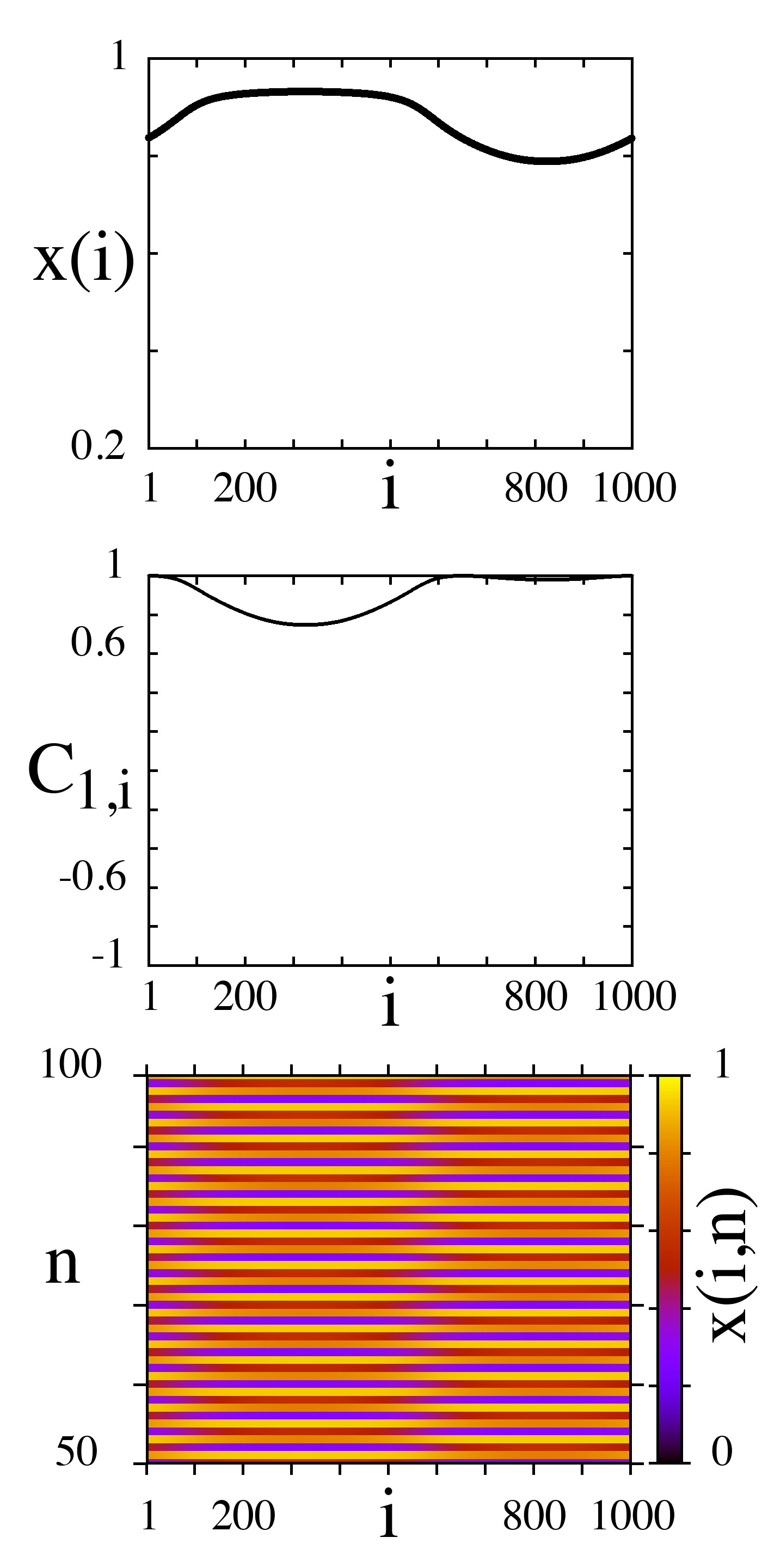}  \\
\hspace{8pt} (a) & \hspace{8pt} (b) & \hspace{8pt} (c) & \hspace{8pt} (d)\\
\end{tabular}
\caption{Exemplary snapshots of the $x(i)$ variables (upper row), spatial distributions of $C_{1,i}$ (\ref{eq:correlation})  (middle row), and space-time diagrams $x(i,n)$ (lower row) for four typical regimes in the noise-free network of nonlocally coupled logistic maps (Fig.~\ref{fig_2}(a),(b)): (a) incoherence (INC) for $\sigma=0.16$, (b) chimera state (CS) for $\sigma=0.29$, (c) snapshots with profile discontinuities (DC) at $\sigma=39$, and (d) coherence (COH) at $\sigma=0.47$. Other parameters:  $\alpha^{l}=3.8$, $R=320$, $N=1000$, $D=0$.}
	\label{fig_3}
\end{figure}

We first study the noise-free case when $D=0$ in (\ref{eq:system}). Figure~\ref{fig_2}(a),(b) shows diagrams of dynamical regimes in the logistic map network in the $(\alpha^{l},\sigma)$ parameter plane for two different realizations of randomly distributed initial conditions. Four regions can be distinguished in the diagrams. The red (COH) and grey (INC) regions correspond to coherent and incoherent dynamics of the network, respectively. The network dynamics is characterized by snapshots with profile discontinuities inside the violet region (DC) and chimera states are observed within the dark-blue region (CS). 
Exemplary snapshots of the network dynamics, spatial distributions of the cross-correlation coefficient (\ref{eq:correlation}) and space-time diagrams are presented in Fig.~\ref{fig_3} for each of the aforementioned regions. 

 The regime diagrams reflect their qualitative similarity but there are some quantitative differences in the boundaries between the regions with coherent and incoherent dynamics and between the regions with coherent dynamics and snapshots with profile discontinuities (for $\alpha^l <3.7$).  The $\alpha^l$ range in the regime diagrams (Fig.~\ref{fig_2}(a),(b)) can be divided into two subranges each corresponding to a different transition from incoherence to coherence as the coupling strength $\sigma$ increases. In the first case, when $\alpha^l <3.6$ (Fig.~\ref{fig_2}(a)) or $\alpha^l <3.65$ (Fig.~\ref{fig_2}(b)), there is a direct transition to the coherent dynamics of all network elements (Fig.~\ref{fig_3}(d)), which occurs for sufficiently weak coupling. In this case the values of the control parameter $\alpha^l$ correspond to local periodic or weakly chaotic behavior of individual nodes in time (Fig.~\ref{fig_1}(a)). 

Within the second subrange, $\alpha^l>3.6$ (Fig.~\ref{fig_2}(a)) or $\alpha^l>3.65$ (Fig.~\ref{fig_2}(b)), the transition to coherence is accompanied by the appearance of chimera states (Fig.~\ref{fig_3}(b)) and profiles with discontinuities (Fig.~\ref{fig_3}(c)) when $\sigma$ increases. A small region of coherence (red color) can be noted in the regime diagrams at $\alpha^{l} \approx 3.85$ near the boundary of the region of snapshots with profile discontinuities (violet color). This "coherent window" corresponds to the existence of a periodic window in the isolated logistic map (Fig.~\ref{fig_1}(a)). 
Besides, there is a small finite region of snapshots with profile discontinuities  within the coherence region for $\alpha^{l} \in [3.68,3.71]$ (Fig.~\ref{fig_2}(a),(b)). Thus, two ways of transition to coherence can be distinguished for the second subrange of the $\alpha^l$ variation: (1) "incoherence  $\rightarrow$ chimera state $\rightarrow$ snapshot with profile discontinuities $\rightarrow$ coherence"$,$ (2) 
"incoherence $\rightarrow$ chimera state $\rightarrow$ snapshot with profile discontinuities $\rightarrow$ coherence ("coherent window")  $\rightarrow$ snapshot with profile discontinuities $\rightarrow$ coherence". As a consequence and it will be shown further, there are two different impacts of additive noise on the probability of observing chimera states. 

In order to get insight into the temporal dynamics of nonlocally coupled logistic maps we calculate and plot two-parameter diagrams of oscillation period distributions, which correspond to the regime diagrams shown in Fig.~\ref{fig_2}(a),(b). The obtained distributions of regular and chaotic dynamics of all network elements in time are presented in Fig.~\ref{fig_2}(c),(d). 
It is seen that within the coherence region (red color in Fig.~\ref{fig_2}(a),(b)) the temporal dynamics of the network can be either regular with different periods or irregular depending on the value of $\alpha^l$. The incoherent spatial structure of the network (grey color in  Fig.~\ref{fig_2}(a),(b)) is characterized by periodic and irregular dynamics of the nodes in time and period-doubling bifurcations occur in the network as the parameter $\alpha^l$ increases (Fig.~\ref{fig_2}(c),(d)). Inside the regions of snapshots with profile discontinuities and of chimera states (violet and dark-blue colors in Fig.~\ref{fig_2}(a),(b), respectively), subsequent period-doubling bifurcations are realized in time when the coupling strength $\sigma$ is decreased \cite{Omelchenko:2011uc}. Note that if the phase chimera \cite{Omelchenko:2011uc,Bogomolov:2017wq} is observed in the network, the nodes demonstrate regular dynamics in time with periods $T=2,4,8$. In the case of amplitude chimera \cite{Bogomolov:2017wq}, the temporal behavior of the network elements is chaotic.

We now introduce additively a Gaussian noise source $D\xi(i,n)$ into the logistic map network (\ref{eq:system}) and analyze how the observation of chimera states depends on the noise intensity, the local dynamics parameter and the coupling parameters. In order to get statistically significant results, we use 50 or 100 different pairs of realizations of initial conditions randomly distributed in the interval $[0,1]$ and noise realizations. 

\begin{figure}[ht]
	\centering
\begin{tabular}{cc}
\includegraphics[width=.45\columnwidth]{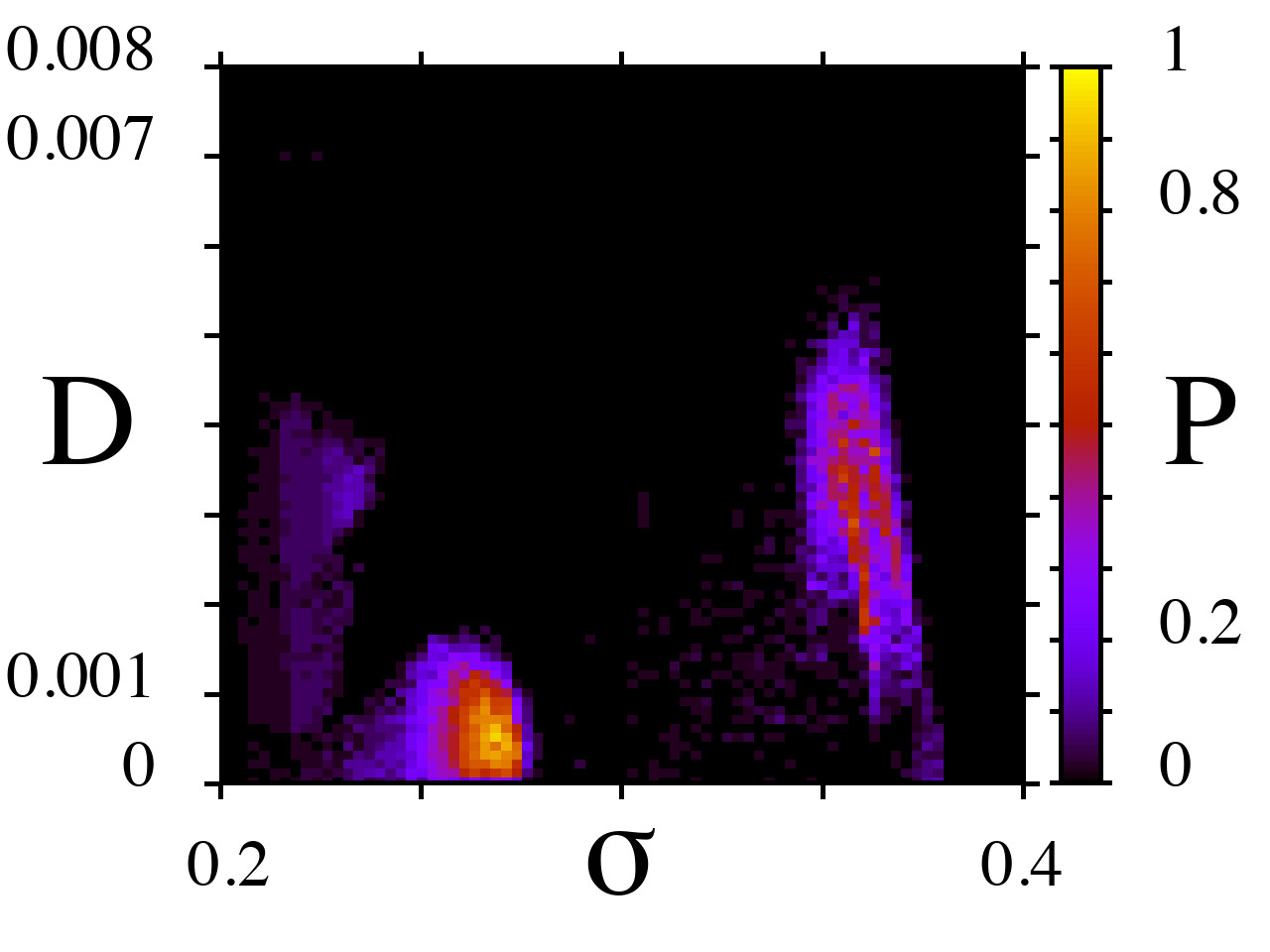} &
\includegraphics[width=.45\columnwidth]{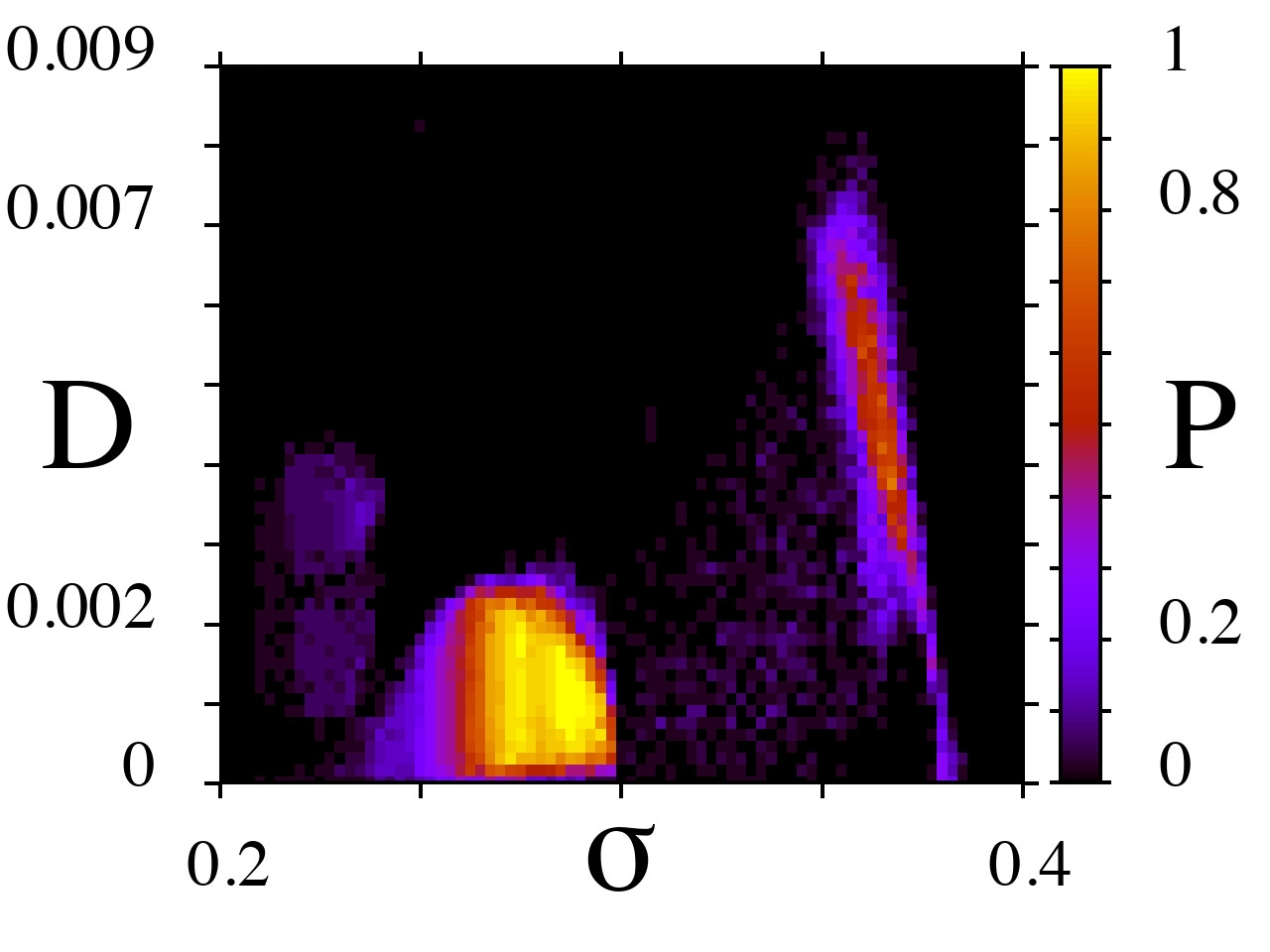}  \\
\hspace{8pt} (a) & \hspace{8pt} (b)\\
\end{tabular}
\begin{tabular}{cc}
\includegraphics[width=.45\columnwidth]{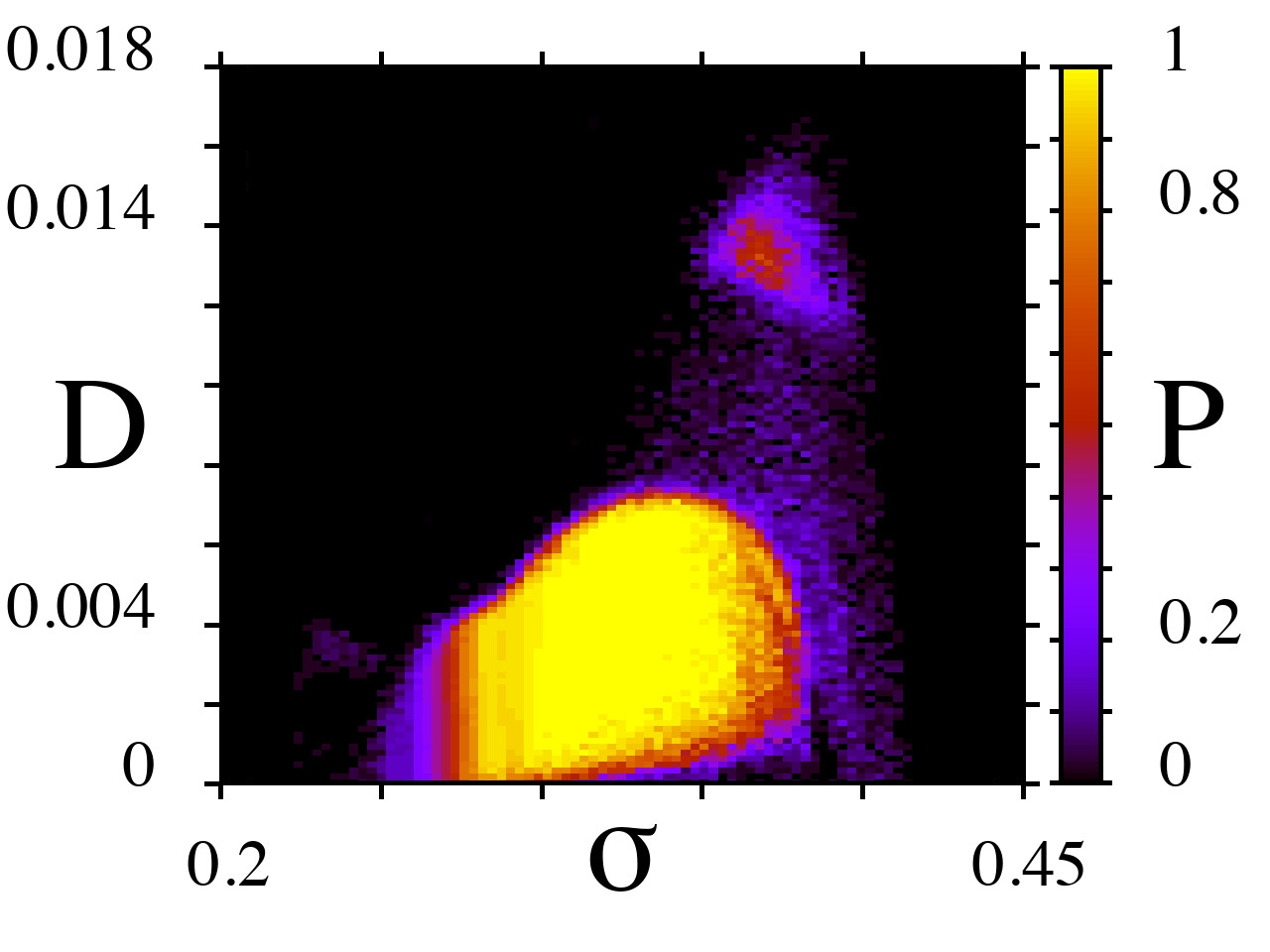} &
\includegraphics[width=.45\columnwidth]{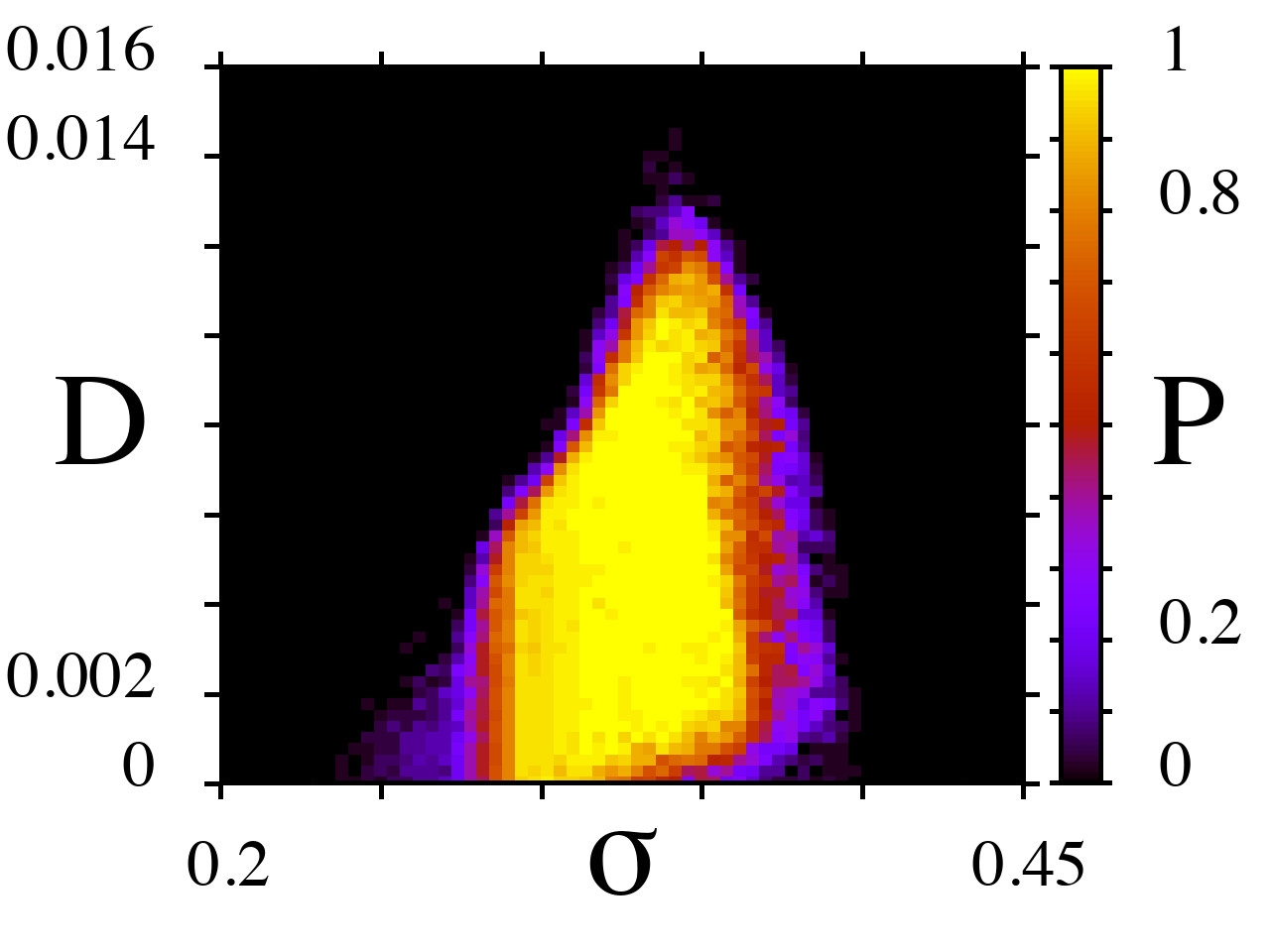}  \\
\hspace{8pt} (c) & \hspace{8pt} (d)\\
\end{tabular}
\caption{Distribution diagrams for the probability $P$ of observing chimera states  in the ($\sigma,D$) parameter plane in the logistic map network  for different values of the local dynamics parameter $\alpha^l$:  (a) $\alpha^{l}=3.69$, (b) $\alpha^{l}=3.7$, (c) $\alpha^{l}=3.75$, and (d) $\alpha^{l}=3.8$. The diagrams are plotted using $M=50$ different pairs of realizations of random initial conditions  and noise realizations. Other parameters:  $R=320$, $N=1000$}
	\label{fig_4}
\end{figure}

Figure~\ref{fig_4} shows distribution diagrams for the probability of observing chimera states in the logistic map network in the ($\sigma,D$) parameter plane for four different values of the local dynamics parameter $\alpha^l$. 
The quantity $P$ is the normalized number of initial realizations $P = K/M$, where $K$ is the
number of initial sets for which chimeras arise in the network and $M$ is the total number of initial realizations used.
The first two diagrams (Fig.~\ref{fig_4}(a),(b)) correspond to the $\alpha^l$ values for which the coherent window is observed within the region of snapshots with profile discontinuities (see Fig.~\ref{fig_2}(a),(b)). In this case the probability distributions have two regions with a maximal (or very close to it) probability separated by a zero-probability region (Fig.~\ref{fig_4}(a),(b)).
Note that in the noise-free case, chimera states exist mainly at small values of the coupling strength: around  $\sigma \approx 0.26$ for $\alpha^{l}=3.69$ and $\sigma \approx 0.27$ for $\alpha^{l}=3.7$, while only profiles with discontinuities are observed for larger values of $\sigma$. As follows from the distribution diagrams, even the low-intensity additive noise  not only extends the existing interval with a high probability of observing chimeras (yellow and orange colors in Fig.~\ref{fig_4}(a),(b)) but also induces the appearance of chimera states (with a non-zero probability) in the strong coupling range. With increasing noise intensity, the regions of chimera existence in the weak coupling range decrease and then disappear when  $D \approx 0.0017$ for $\alpha^{l}=3.69$ (Fig.~\ref{fig_4}(a)) and $D \approx 0.0027$ for $\alpha^{l}=3.7$ (Fig.~\ref{fig_4}(b)). The noise-induced chimera observed in the strong coupling range appears to be more stable towards the additive noise and exists up to $D \approx 0.0056$ for $\alpha^{l}=3.69$ (Fig.~\ref{fig_4}(a)) and $D \approx 0.0081$ for $\alpha^{l}=3.7$ (Fig.~\ref{fig_4}(b)). 
As the local dynamics parameter $\alpha^{l}$ increases further, the region of non-zero probability expands with respect to both the coupling strength and the noise intensity. However,  a small region with $P\approx0.5$ can still exist in the strong coupling range when $\alpha^{l}$ lies near the coherent window (Fig.~\ref{fig_4}(c)).

For the values of $\alpha^l$ when there is no coherent window in the region with profile discontinuities (Fig.~\ref{fig_2}(a),(b)), the probability distribution for the observation of chimera states occupies a single region in the ($\sigma$, $D$) parameter plane. A typical distribution is shown in Fig.~\ref{fig_4}(d), it is seen that chimeras can be observed in the noisy network with a maximum and non-zero probability (yellow region in Fig.~\ref{fig_4}(d)) within a rather wider interval of the coupling strength $\sigma$. This $\sigma$-interval is gradually narrowing as $D$ increases up to $D\approx 0.014$ for $\alpha^{l}=3.8$ (Fig.~\ref{fig_4}(d)). Besides, for weak noise, the region with high probability expands towards larger values of the coupling strength $\sigma$. 

As follows from the presented diagrams (Fig.~\ref{fig_4}(c),(d)), the width of the noise intensity $D$ range of a non-zero probability changes as the coupling strength increases within the $\sigma$-interval corresponding to the observation of chimeras with a high  probability. This $D$-range first gradually expands, achieves its maximum at a certain value of $\sigma$, e.g., at $\sigma\approx 0.33$ for $\alpha^l=3.75$ (Fig.~\ref{fig_4}(c)) and at $\sigma\approx 0.35$ for $\alpha^l=3.8$ (Fig.~\ref{fig_4}(d)) and then gradually decreases when $\sigma$ approaches the right boundary of the $\sigma$-interval.  

It can also be noticed from the distribution diagrams (Fig.~\ref{fig_4}(c),(d)) that there is a certain optimum noise level ($D_{\rm opt}$) at which the width of the $\sigma$-interval corresponding to the high probability of observing chimeras ($P>0.95$) is the largest. In our cases, at $D_{\rm opt}\approx0.0028$ $\sigma \in [0.28,0.36]$ for $\alpha^{l}=3.75$ (Fig.~\ref{fig_4}(c)) and  at $D_{\rm opt}=0.0032$ $\sigma \in [0.29,0.36]$ for $\alpha^{l}=3.8$ (Fig.~\ref{fig_4}(d)).  Thus, the $\sigma$-interval can be significantly increased by appropriately tuning the additive noise intensity to a certain non-vanishing value. Such an effect demonstrates a constructive role of additive noise, which is clearly manifested and typical for the phenomena of stochastic~\cite{Benzi:1981us,Benzi:1982wv} and coherence~\cite{Pikovsky:1997vf,Lindner:1999tv} resonance. In this context, the revealed peculiarity of the influence of noise on the observation of chimera states  may be called {\it chimera resonance}. This resonance-like effect is not an exclusive feature of the dynamics of the noisy network  of nonlocally coupled logistic maps but, as will be described below, is typically realized in networks of other chaotic discrete-time systems in the presence of additive noise. 
  
\begin{figure}[ht]
	\centering
\begin{tabular}{ccc}
\includegraphics[width=.3\columnwidth]{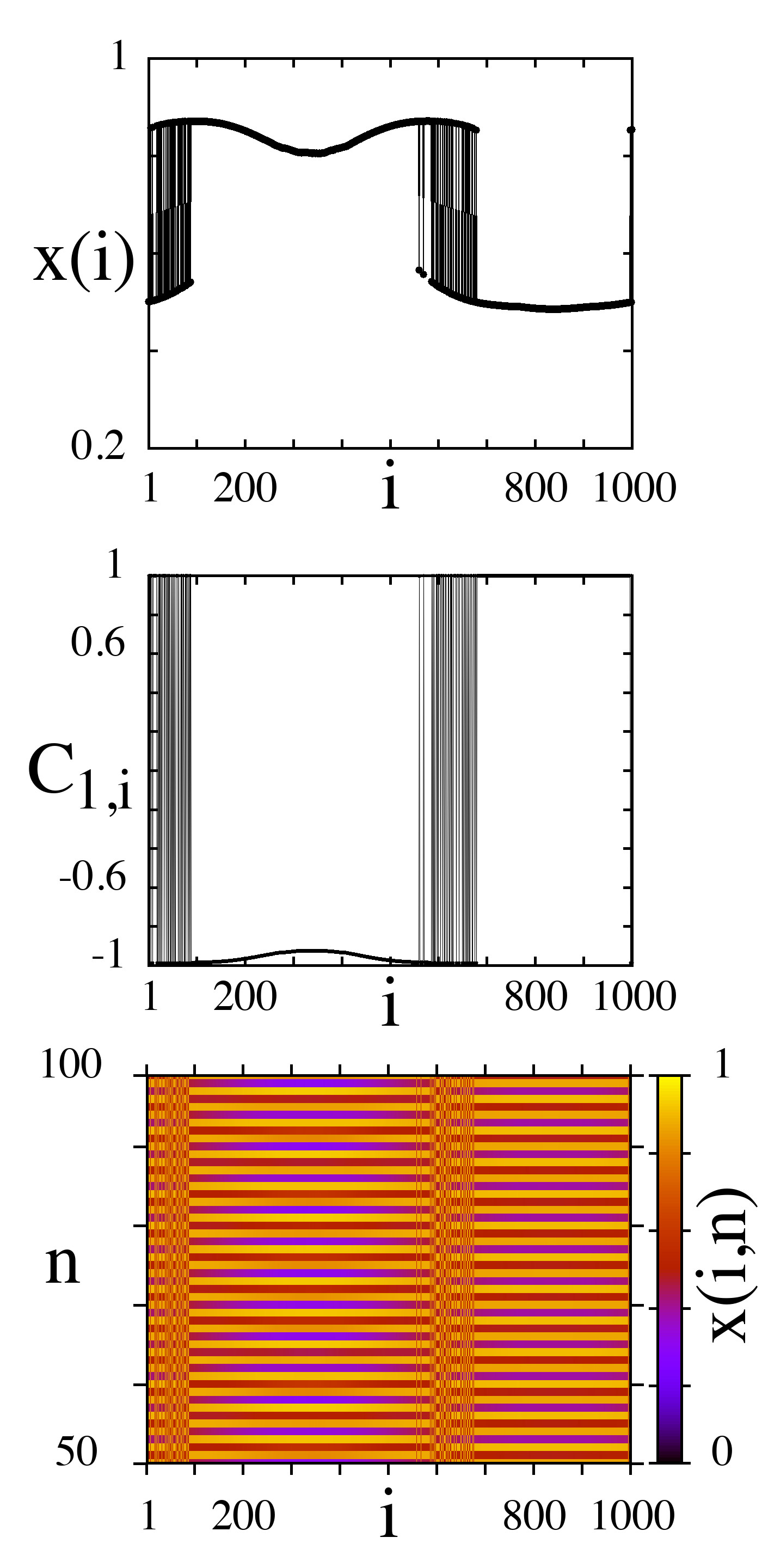} &
\includegraphics[width=.3\columnwidth]{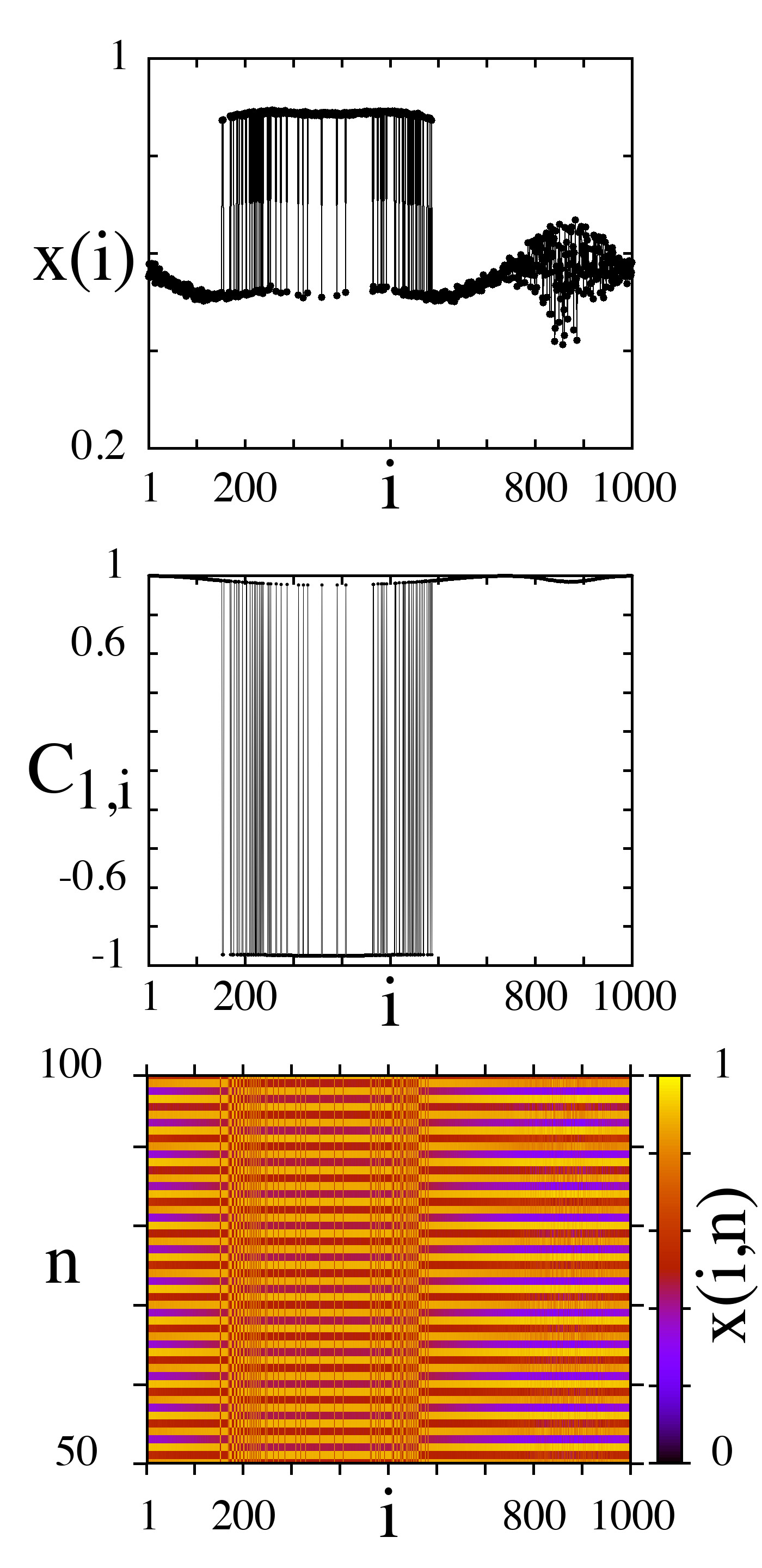}  &
\includegraphics[width=.3\columnwidth]{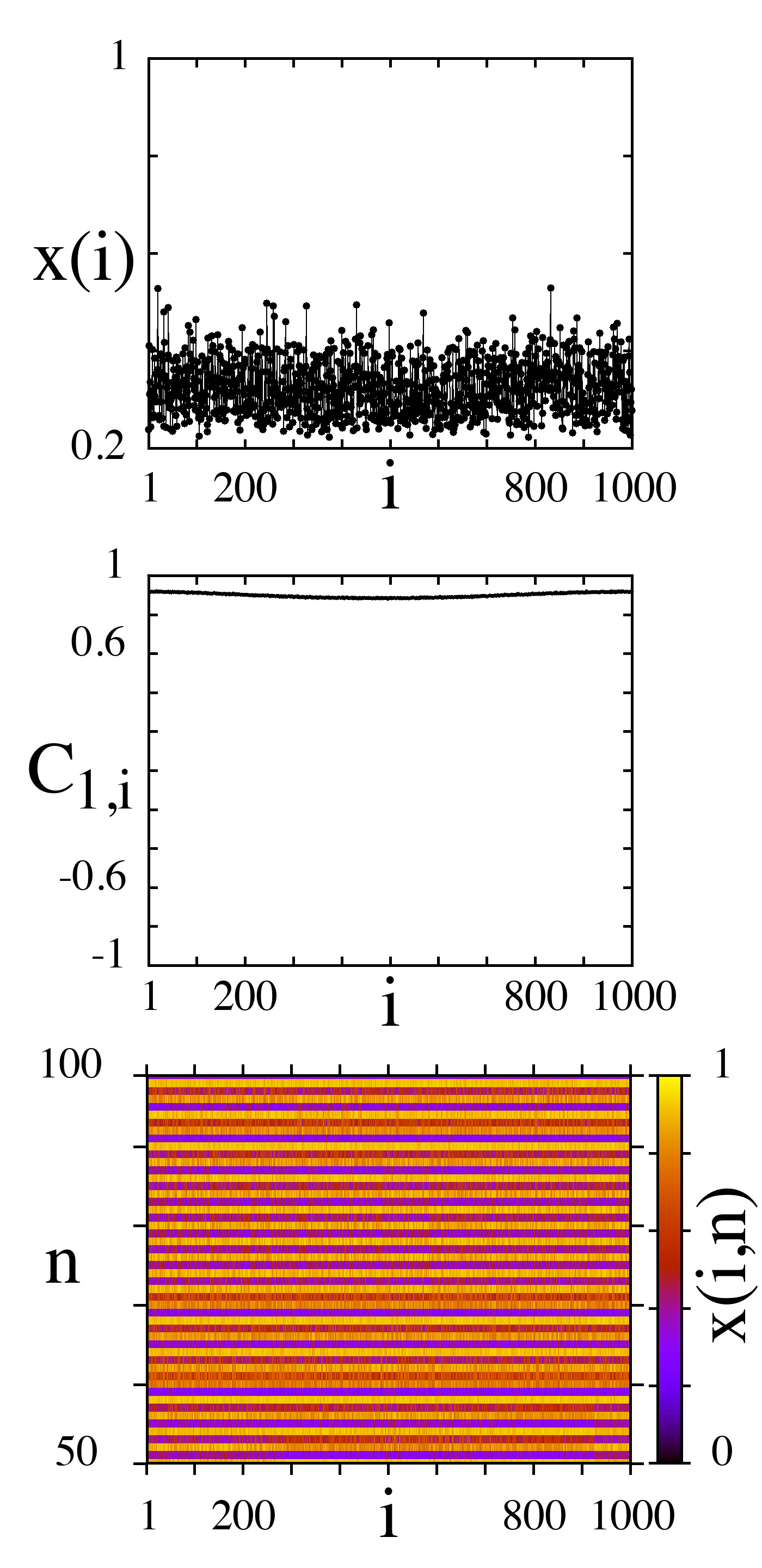}  \\
\hspace{0pt} (a) & \hspace{0pt} (b) & \hspace{0pt} (c)\\
\end{tabular}
\begin{tabular}{ccc}
\includegraphics[width=.3\columnwidth]{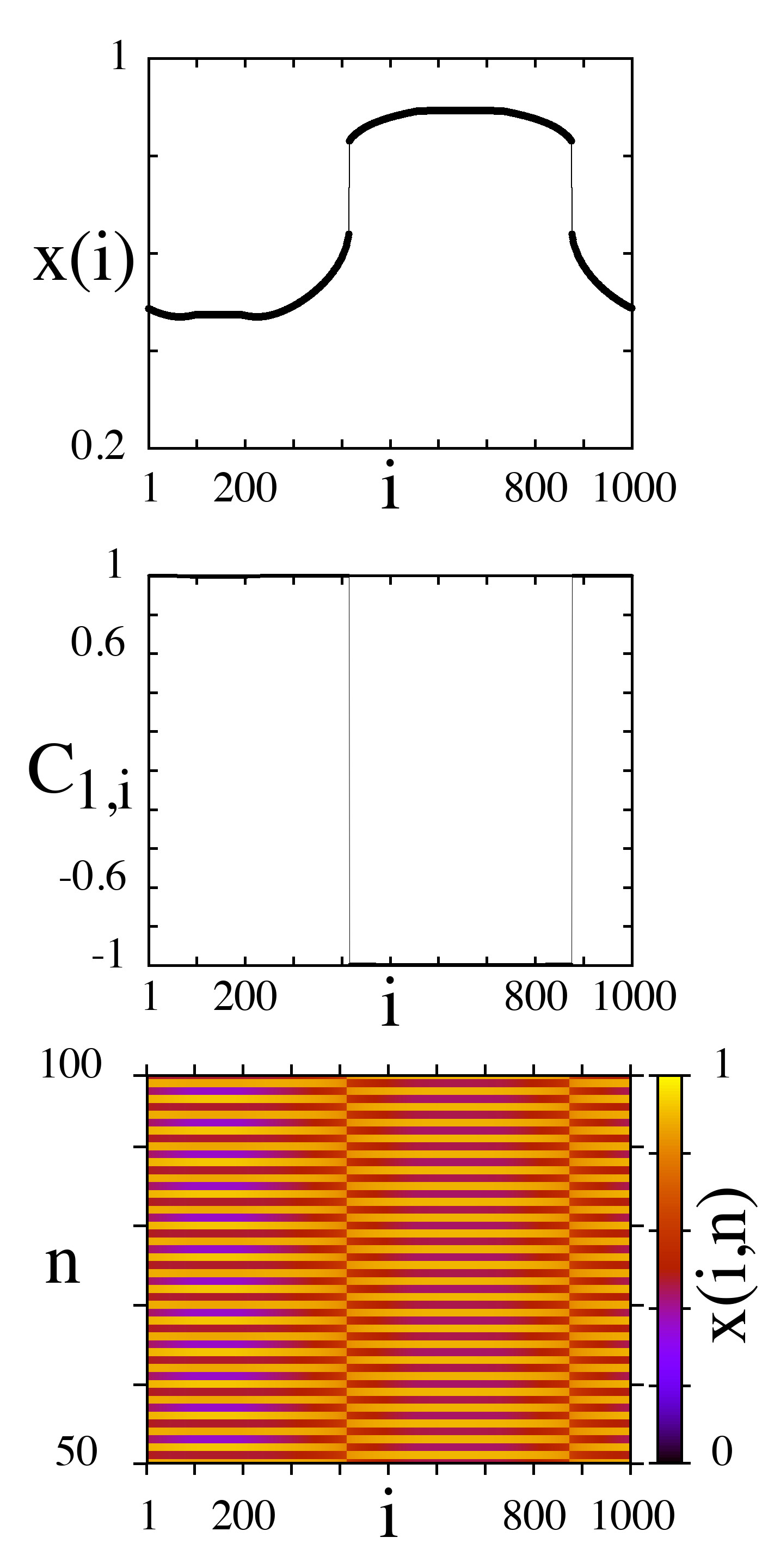} &
\includegraphics[width=.3\columnwidth]{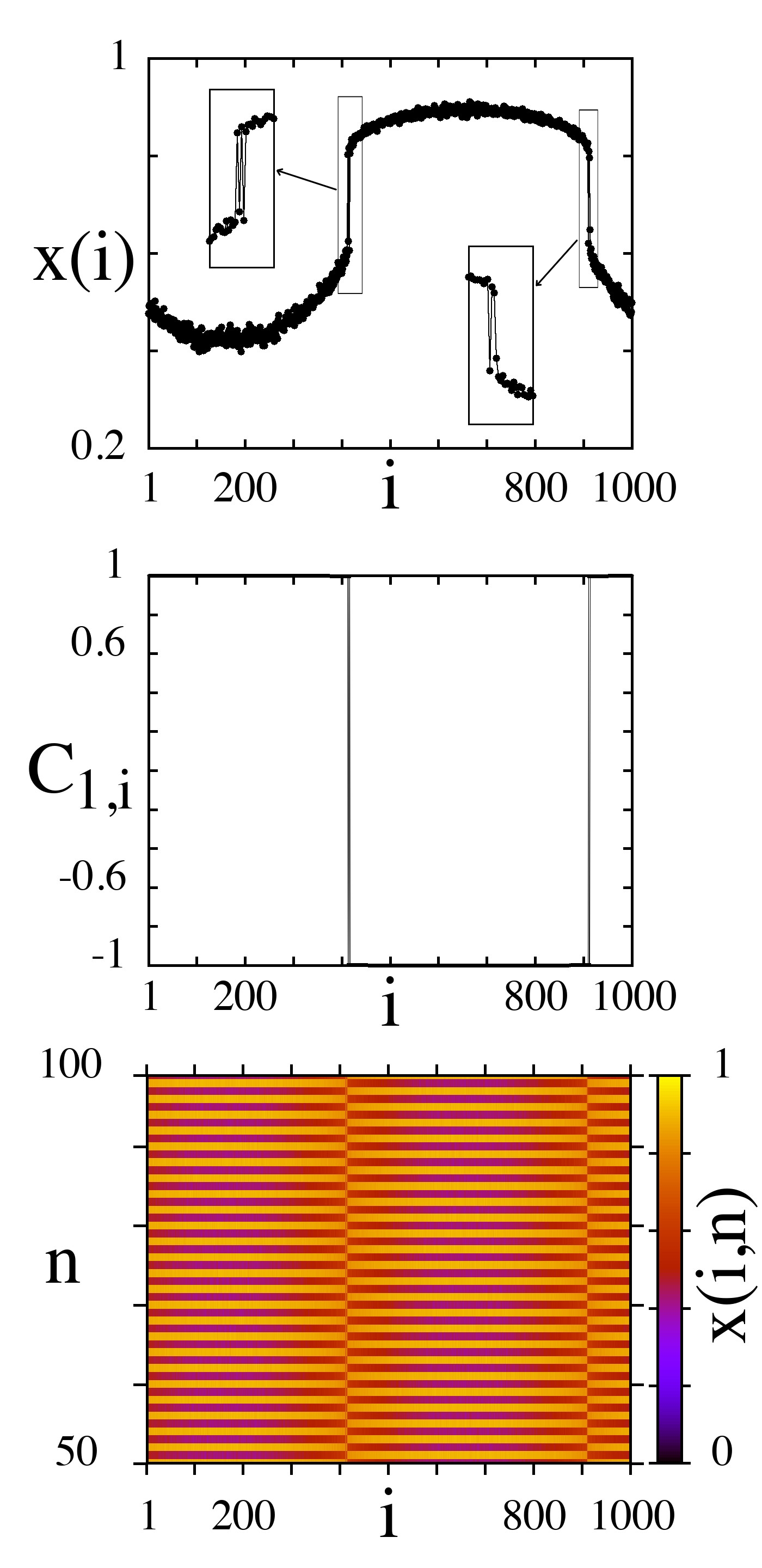}  &
\includegraphics[width=.3\columnwidth]{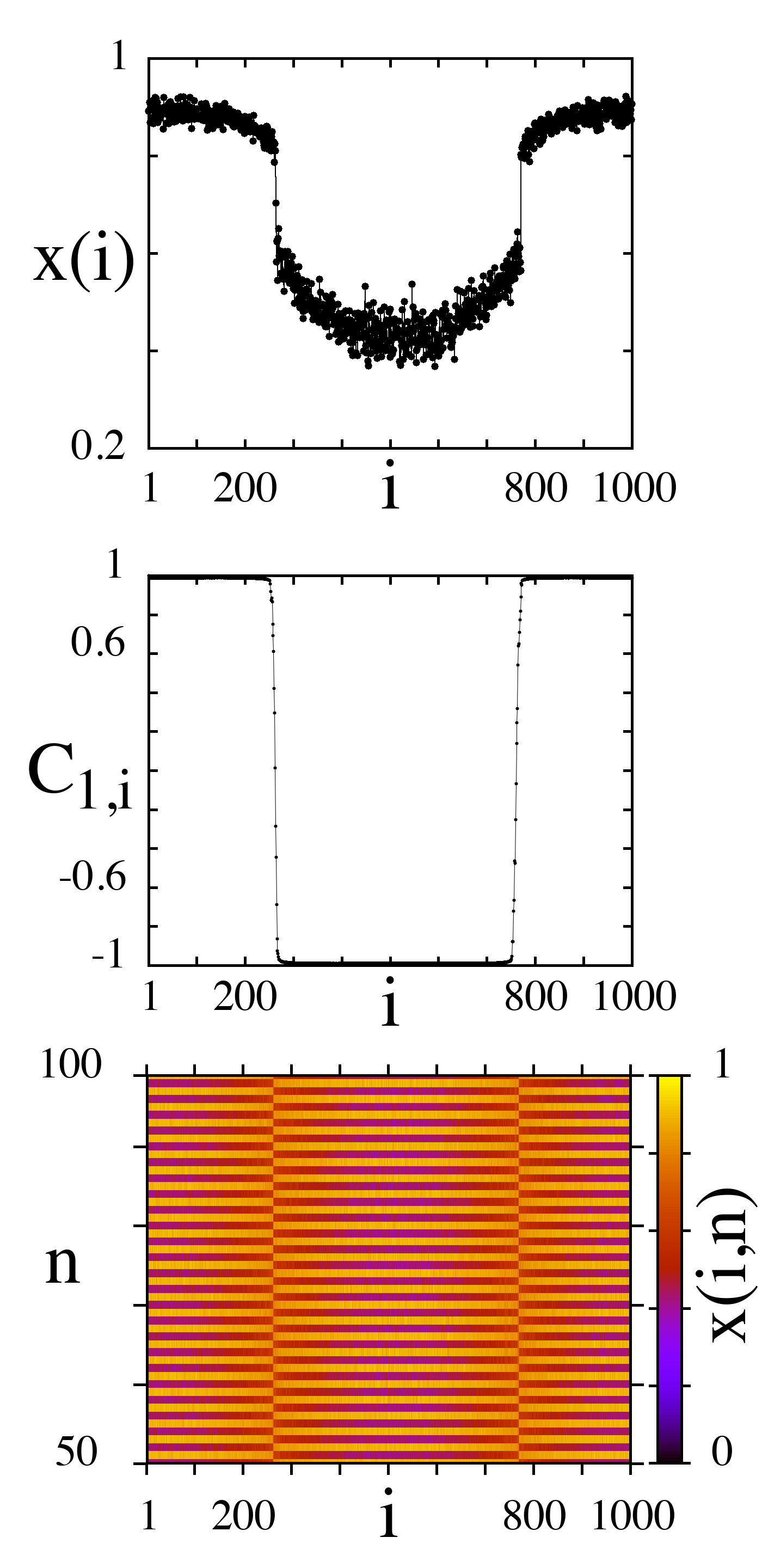} \\
 \hspace{0pt} (d) & \hspace{0pt} (e) & \hspace{0pt} (f)\\
\end{tabular}
\caption{Snapshots of the $x(i)$ variables (upper row), spatial distributions of the cross-correlation coefficient (middle row), and spatial distributions of the cross-correlation coefficient (lower row) for the logistic map network for different values of the noise intensity and for weak (a)-(c) and strong (d)-(f) coupling:  (a) $D=0$, $\sigma=0.280$, (b) $D=0.00165$, $\sigma=0.280$, (c) $D=0.0055$, $\sigma=0.280$, (d) $D=0$, $\sigma=0.380$, (e) $D=0.004$, $\sigma=0.380$, (f) $D=0.0112$, $\sigma=0.380$. Other parameters: $\alpha^{l}=3.8$, $R=320$, $N=1000$. The insets in (e) top row show blow-ups.}
	\label{fig_5}
\end{figure}

The evolution of the logistic map network dynamics both without and in the presence of noise with increasing intensity is illustrated in Fig.~\ref{fig_5} for weak (Fig.~\ref{fig_5}(a)-(c)) and strong (Fig.~\ref{fig_5}(d)-(f)) coupling between the nodes. In the first case (at $\sigma=0.28$), the additive noise first induces the expansion of incoherence clusters of a phase chimera (compare Fig.~\ref{fig_5}(a) and (b)) and appearance of  an amplitude chimera (Fig.~\ref{fig_5}(b), $750<i<1000$) and a solitary state chimera (Fig.~\ref{fig_5}(b), $270<i<410$). The further increase of noise intensity leads to the incoherent dynamics (Fig.~\ref{fig_5}(c)), i.e., the chimera state is destroyed. 
A more interesting impact of noise is observed for strong coupling, e.g., for $\sigma=0.38$ (Fig.~\ref{fig_5}(d)-(f)) when the noise-free network dynamics is characterized by a snapshot with profile discontinuities (Fig.~\ref{fig_5}(d)). Weak additive noise causes the oscillators located near the profile discontinuities to jump to the other coherent branch and thus induces the formation of incoherence cluster, i.e., noise promotes the appearance of a chimera  state (Fig.~\ref{fig_5}(e)).  With a higher noise intensity, the spatial profile is smeared out (in spite of slight noise-induced fluctuations of amplitudes of the network nodes), which is clearly seen in the distribution of cross-correlation coefficients (Fig.~\ref{fig_5}(f)).

\begin{figure*}[ht]
	\centering
\begin{tabular}{ccc}
\includegraphics[width=.67\columnwidth]{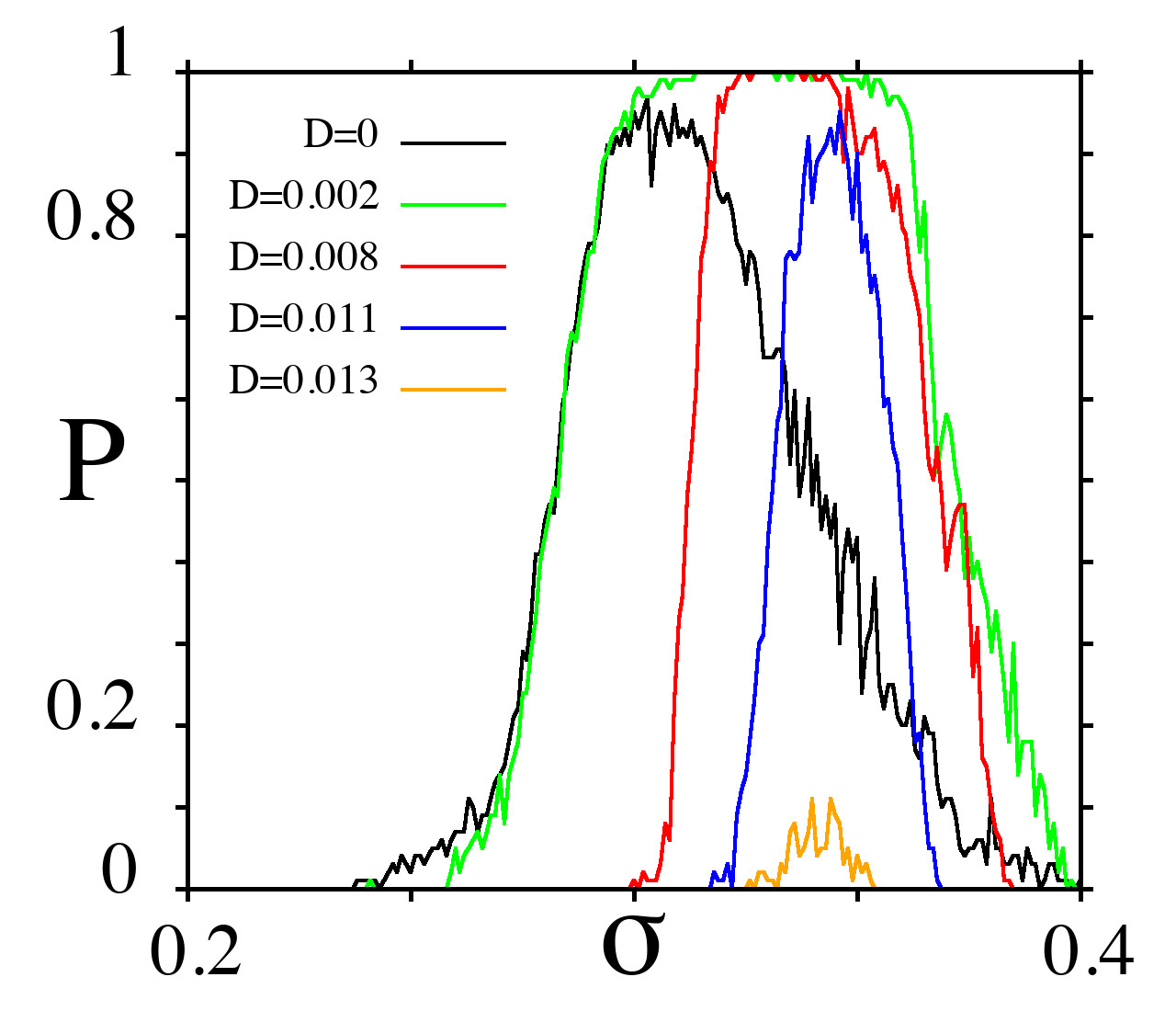} &
\includegraphics[width=.67\columnwidth]{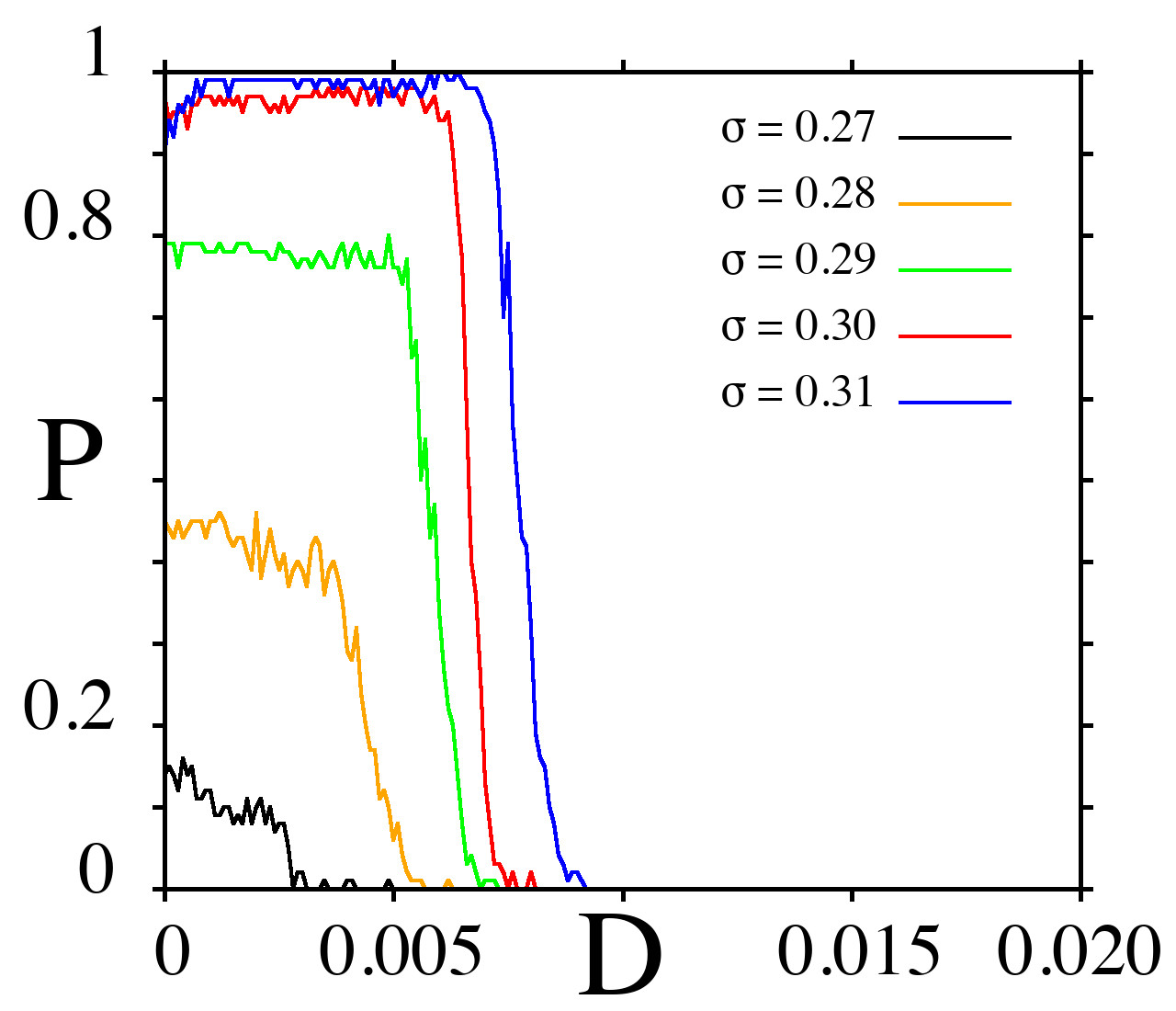}  &
\includegraphics[width=.67\columnwidth]{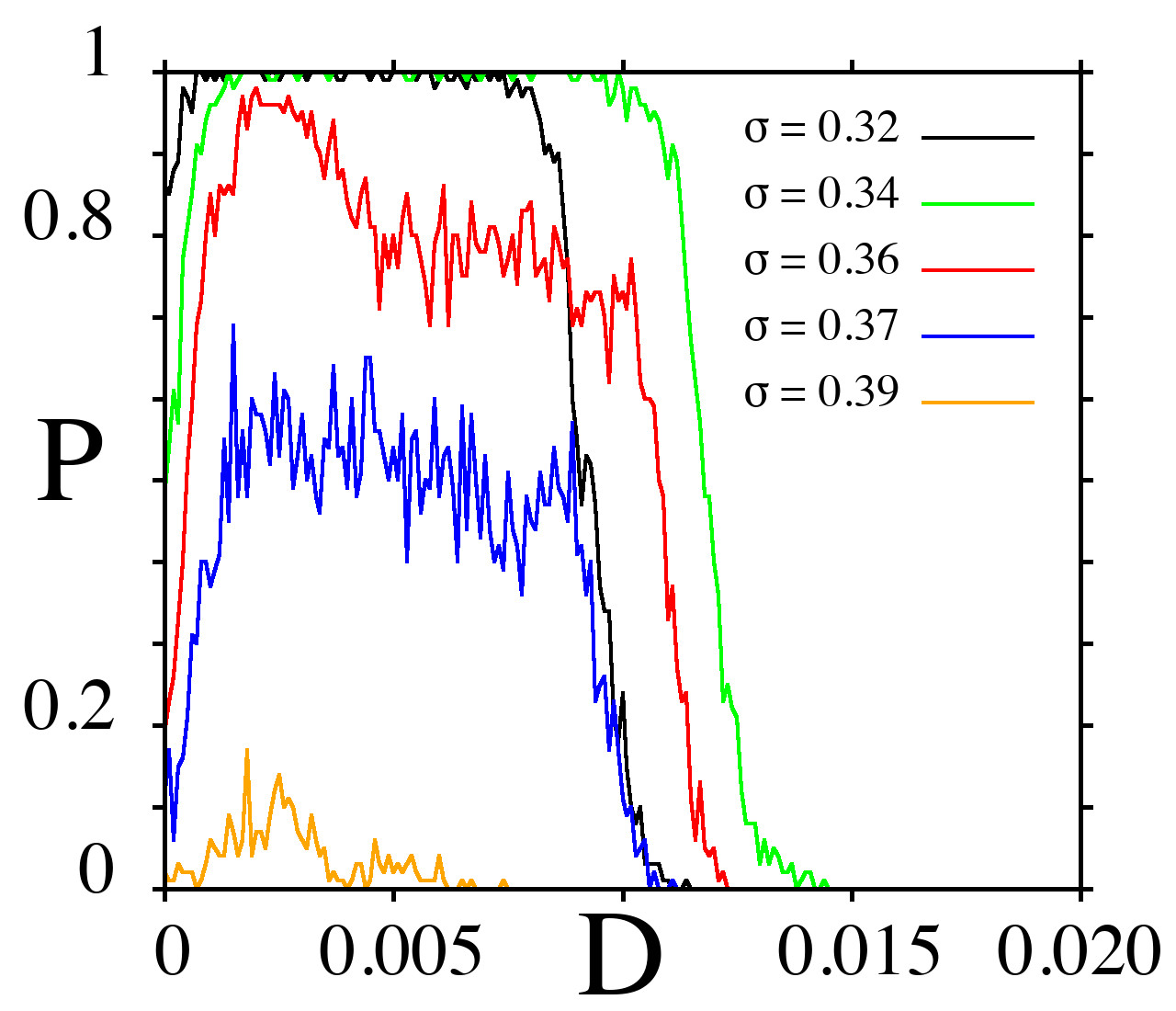} \\
\hspace{8pt} (a) & \hspace{8pt} (b) & \hspace{8pt} (c)\\
\end{tabular}
\caption{Dependences of the probability $P$ of observing chimera states in the logistic map network (a) on the coupling strength   $\sigma$ for five different values of the noise intensity $D$, (b) and (c) on the noise intensity $D$ for different values of 
$\sigma$. The graphics are plotted using $M=100$ different pairs of realizations of random initial conditions randomly distributed in the interval $[0,1]$ and noise realizations. Other parameters: $\alpha^{l}=3.8$, $R=320$, $N=1000$}
	\label{fig_6}
\end{figure*}

\begin{figure*}[ht]
	\centering
\begin{tabular}{ccc}
\includegraphics[width=.67\columnwidth]{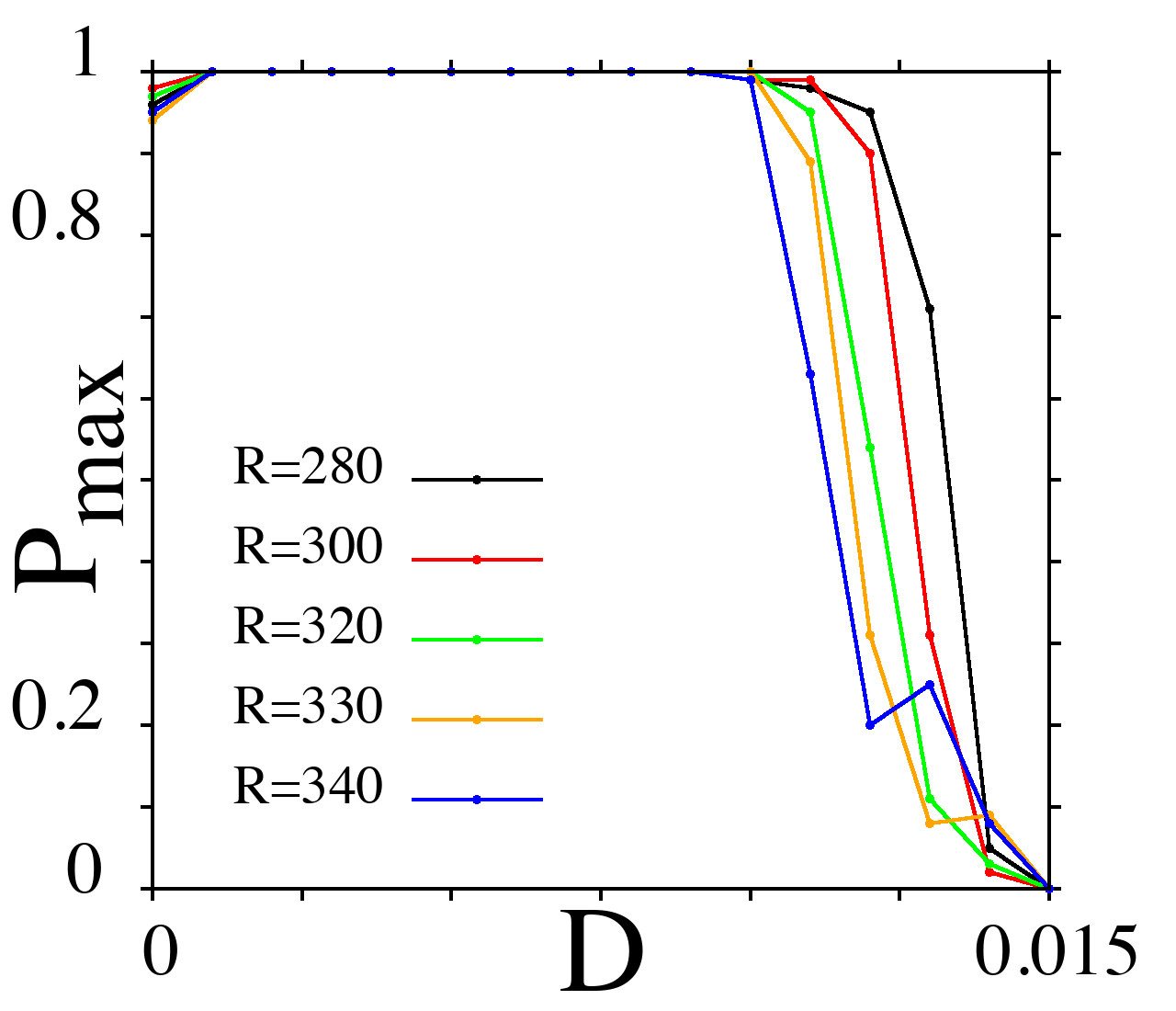}  &
\includegraphics[width=.67\columnwidth]{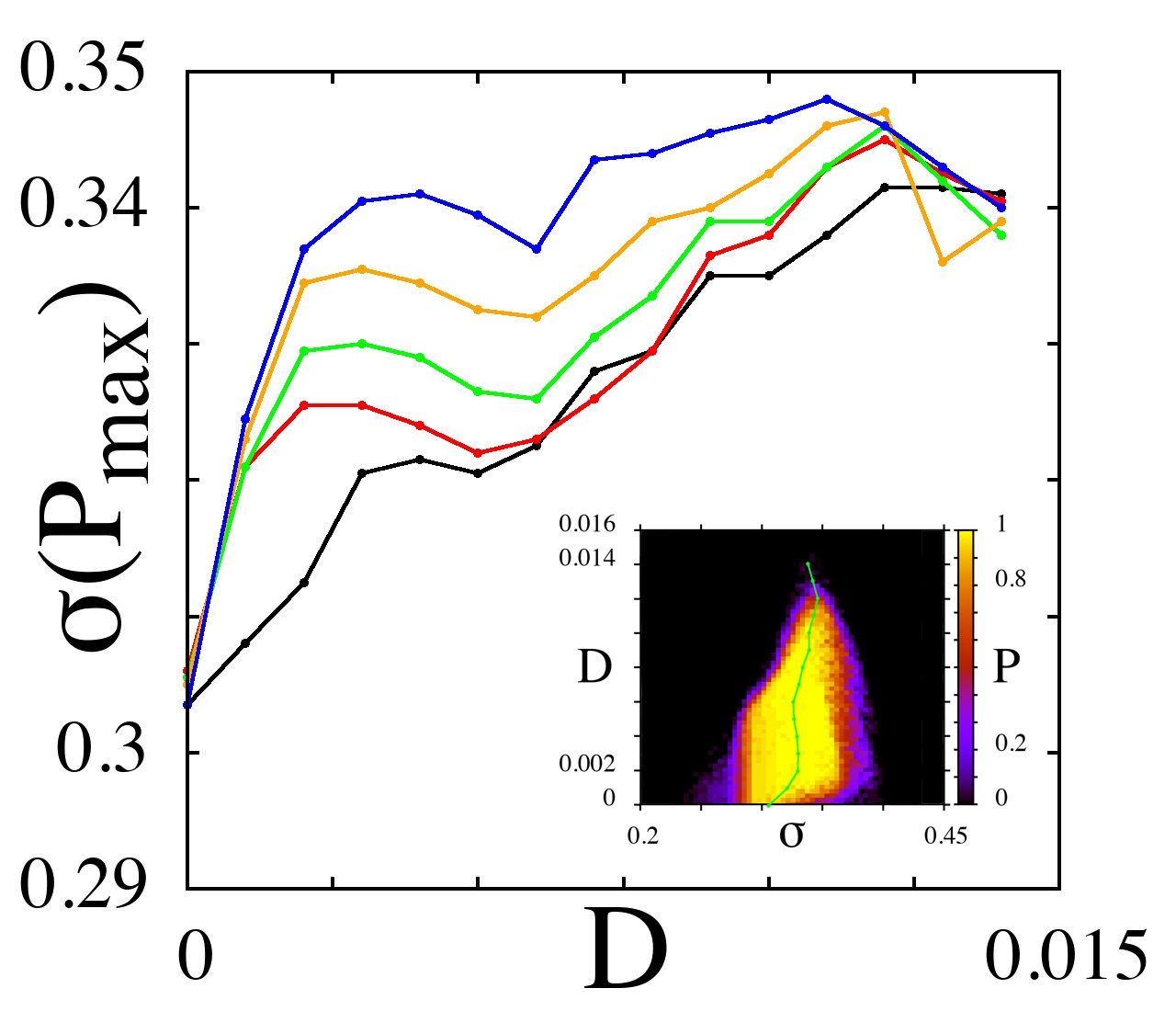} &
\includegraphics[width=.67\columnwidth]{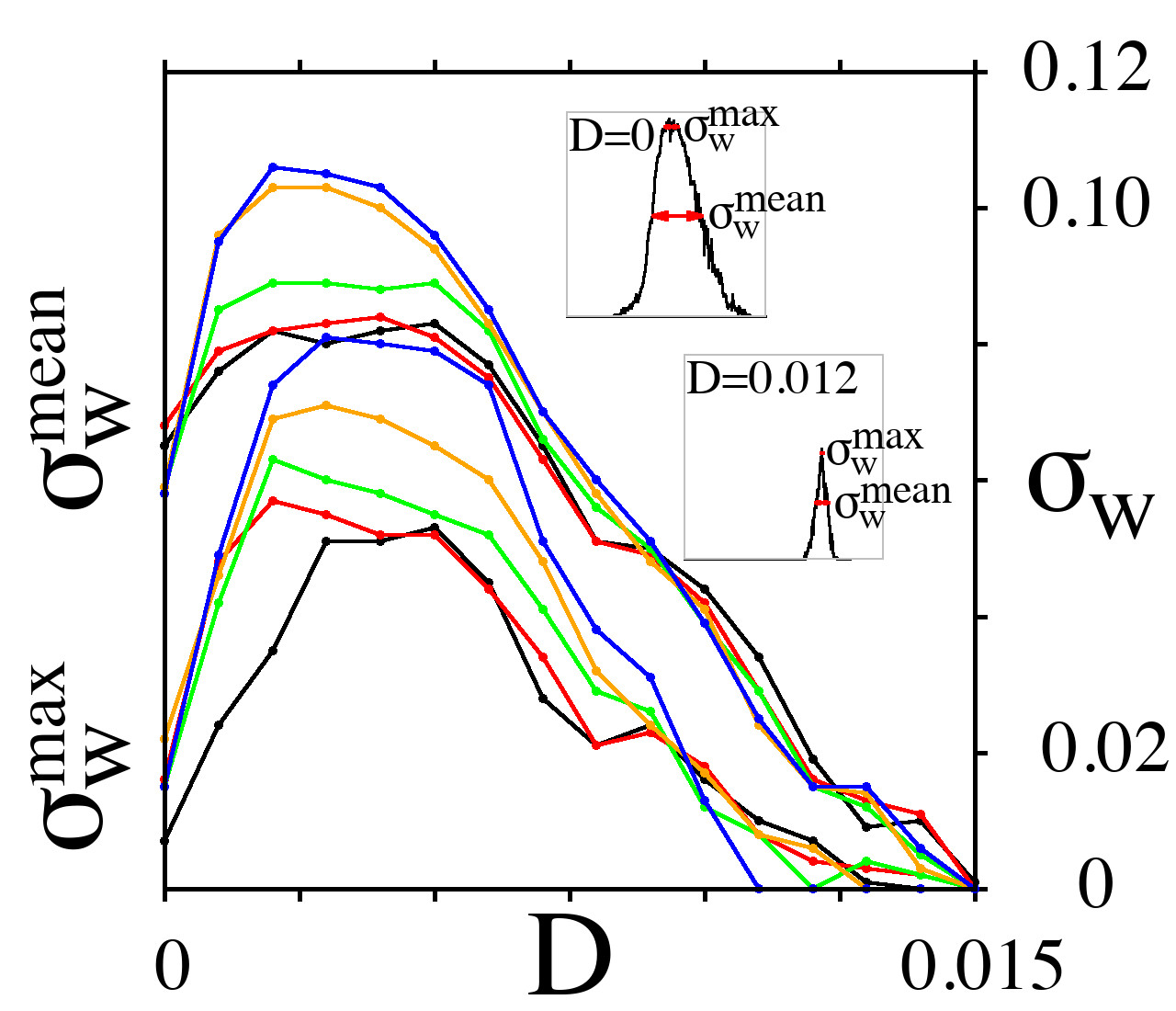}  \\
\hspace{8pt} (a) & \hspace{8pt} (b) & \hspace{8pt} (c)\\
\end{tabular}
\caption{(a) The maximum probability of chimera observation  $P_{\rm max}$ versus the noise intensity $D$ for five different values of the coupling range $R$, (b) the coupling strength $\sigma(P_{\rm max})$, at which the maximum chimera  probability is observed, versus the noise intensity $D$ for five different values of $R$ (see the legend in (a)), and  (c) the width of the $\sigma$-interval at the mean probability level (upper five curves $\sigma_{\rm w}^{\rm mean}$) and at the level of 0.95 of the maximum  (lower five curves $\sigma_{\rm w}^{\rm max}$) versus the noise intensity $D$ for five different values of $R$ (see the legend in (a)). The probability levels ($\sigma_{\rm w}^{\rm mean}$ and $\sigma_{\rm w}^{\rm max}$)  are schematically marked in the two resonance curves $P(\sigma)$ shown in the insets in (c). The graphics are plotted using $M=100$ different pairs of realizations of random initial conditions randomly distributed in the interval $[0,1]$ and noise realizations. Other parameters: $\alpha^{l}=3.8$, $N=1000$}
	\label{fig_7}
\end{figure*}

We now consider in more detail how the probability $P$ of observing chimeras changes when the noise intensity and the coupling parameters (strength $\sigma$ and range $R$) are varied and the local dynamics parameter is fixed at $\alpha^{l}=3.8$. The dependence of  $P$ versus $\sigma$ is presented in Fig.~\ref{fig_6}(a) for five different values of the noise intensity $D$. It is clearly seen that in all cases the plots have a resonant-like shape. When $D=0$, a non-zero probability exists within the interval  $\sigma \in [0.238,0.401]$, however, its maximum value $P=1$ is never reached. 
With introducing additive noise of a very low intensity,  the $\sigma$-interval within which the chimeras are observed can slightly decrease but there appears a finite $\sigma$-range in which the probability of chimera existence is maximum ($P\approx 1$) (the plot for $D=0.002$ in  Fig.~\ref{fig_6}(a)). A further increase in the noise intensity essentially narrows the $\sigma$-interval of the high probability of chimera observation (the cases of $D=0.008$, $D=0.011$, $D=0.013$ in Fig.~\ref{fig_6}(a)). Besides, both the location of the  $\sigma$-interval and the resonant value of $\sigma$ corresponding to the maximum probability shift towards larger values of coupling strength $\sigma$. 

The 2D diagrams of probability distributions (Fig.~\ref{fig_4}) can be analyzed using selected sections of the parameter $\sigma$. For our calculations the diagram for $\alpha^l=3.8$ presented in Fig.~\ref{fig_4}(d) is used. Figure~\ref{fig_6}(b),(c) shows dependences of the probability $P$ on the noise intensity $D$ for 10 different values of $\sigma$. In this case it is more convenient to divide the entire $\sigma$-interval of chimera observation into two subintervals corresponding to weak ($\sigma\in[0.23,0.31]$, Fig.~\ref{fig_6}(b)) and strong  ($\sigma\in(0.31,0.4]$, Fig.~\ref{fig_6}(c)) coupling. In the case of weak coupling, there is a finite range with respect to $D$ where the probability $P$ is almost constant, for example, this is $D\in[0,0.0055]$ for $\sigma=0.29$ (green curve in Fig.~\ref{fig_6}(b)). For larger $\sigma$, e.g., $\sigma=0.30,~0.31$ (red and blue curves in Fig.~\ref{fig_6}(b), respectively), the value of $P$ increases and can achieve its maximum value ($P\approx1$) which remains unchanged over a rather wide range with respect to $D$. The probability begins to rapidly vanish after a certain value of $D$ which depends on the coupling strength. The described  peculiarities are also observed for the case of strong coupling (Fig.~\ref{fig_6}(c)).  It is clearly seen that there is a certain (resonant) value of $\sigma$ for which  the $D$-range of the high probability of observing chimeras is the widest (green line at $\sigma=0.34$ in Fig.~\ref{fig_6}(c) for which $D\in[0.001,0.0105]$)). Comparing all the plots shown in 
Fig.~\ref{fig_6}(b) and (c) one can conclude that there are coupling strengths (e.g., $\sigma=0.32$, $0.34$, $0.36$, $0.37$) for which additive noise with even a very low noise intensity ($D\leqslant0.001$) can significantly increase the probability of observing chimeras even up to 1, while without noise this probability is very small.  

In order to verify the generality of the revealed effect of chimera resonance, we have carried out numerical simulations for several values of the coupling range $R$ in the logistic map network. Calculations were performed using 100 different pairs of random initial conditions realizations and noise realizations and the obtained results are presented in Fig.~\ref{fig_7}(a)-(c). The dependence  of the maximum probability $P_{\rm max}$ on the noise intensity $D$ is shown in Fig.~\ref{fig_7}(a) for five different values of $R$ (see legend in the figure). 
Note that without noise ($D=0$) $P_{\rm max}$ does not exceed $0.98$ (for $R=300$) for all the considered cases of $R$ and only the introduction of additive noise into the network can increase the maximum value of $P$ to 1. This absolute maximum is preserved within a finite and rather wide $D$-range $[0.001, 0.011]$. The coupling strength $\sigma$ corresponding to the maximum probability of chimera observation grows as the noise intensity $D$ increases (Fig.~\ref{fig_7}(b)) and this is valid for all selected values of $R$. 
However, when the noise intensity exceeds the value of $0.014$, the chimera states are destroyed. Thus we cannot speak about  $\sigma(P_{\rm max})$ since the probability of observing chimeras is equal to 0. As a consequence, all the curves shown in Fig.~\ref{fig_7}(b) terminate.
 It is worth noting that the dependence of $\sigma(P_{\rm max})$ is a non-monotonic  function of $D$ and a dip  in the curves  shown in Fig.~\ref{fig_7}(b) occurs around $D\approx 0.006$. In order to get insight into this peculiarity, we include the distribution diagram shown in Fig.~\ref{fig_4}(d) as the inset in Fig.~\ref{fig_7}(b), where the green curve shows how the value of $\sigma(P_{\rm max})$ changes as $D$ increases (this green curve relates to $R=320$ and fully corresponds to the green curve in the main picture). It is seen that at first $\sigma(P_{\rm max})$ increases monotonically for $D\leq0.002$ and then slightly decreases when $0.002<D<0.006$. It can be mentioned that the shape  of the probability distribution significantly changes at $D =0.006$ (see the inset in Fig.~\ref{fig_7}(b)). As a consequence, when $D>0.006$, the coupling strength $\sigma(P_{\rm max})$ again continues to grow monotonically as a function of $D$ and vanishes abruptly at the noise level which relates to the disappearance of chimera states.

The effect of chimera resonance is well illustrated in Fig.~\ref{fig_7}(c) that shows the impact of additive noise of different intensity on the width of the $\sigma$-interval at the mean probability level (upper five curves $\sigma_{\rm w}^{\rm mean}$) and at the level of $0.95$ of the maximum probability (lower five curves $\sigma_{\rm w}^{\rm max}$) for five different values of the coupling range $R$. The probability levels are schematically marked in the two resonant curves $P(\sigma)$ shown in the insets in Fig.~\ref{fig_7}(c). 
The presented dependences have a resonant-like form and give evidence that there is a certain optimum noise level $D$ (different for different $R$) at which the width of the $\sigma$-interval corresponding to the maximum probability of observing chimeras is the largest. It can also be noted that the width of the largest $\sigma_{\rm w}^{\rm mean}$ and $\sigma_{\rm w}^{\rm max}$ intervals decreases as the coupling range $R$ decreases. The maximum width of both $\sigma$-intervals is observed for $R=340$  and the minimum -- for $R=280$ (blue and black curves in Fig.~\ref{fig_7}(c), respectively).

\subsection{Network of nonlocally coupled Henon maps}\label{sec_ring_henon}

\begin{figure}[ht]
	\centering
\begin{tabular}{cc}
\includegraphics[width=.45\columnwidth]{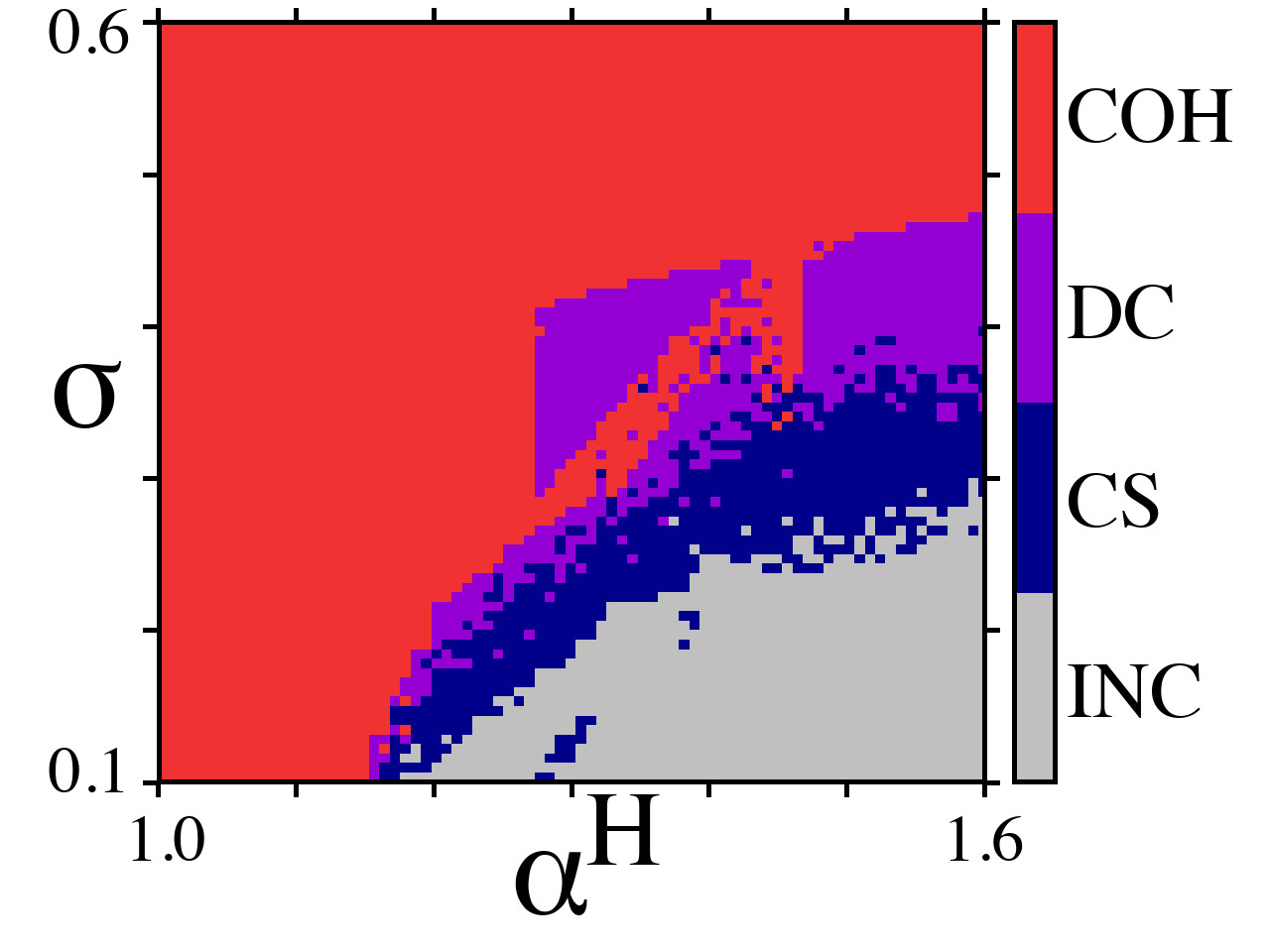} &
\includegraphics[width=.45\columnwidth]{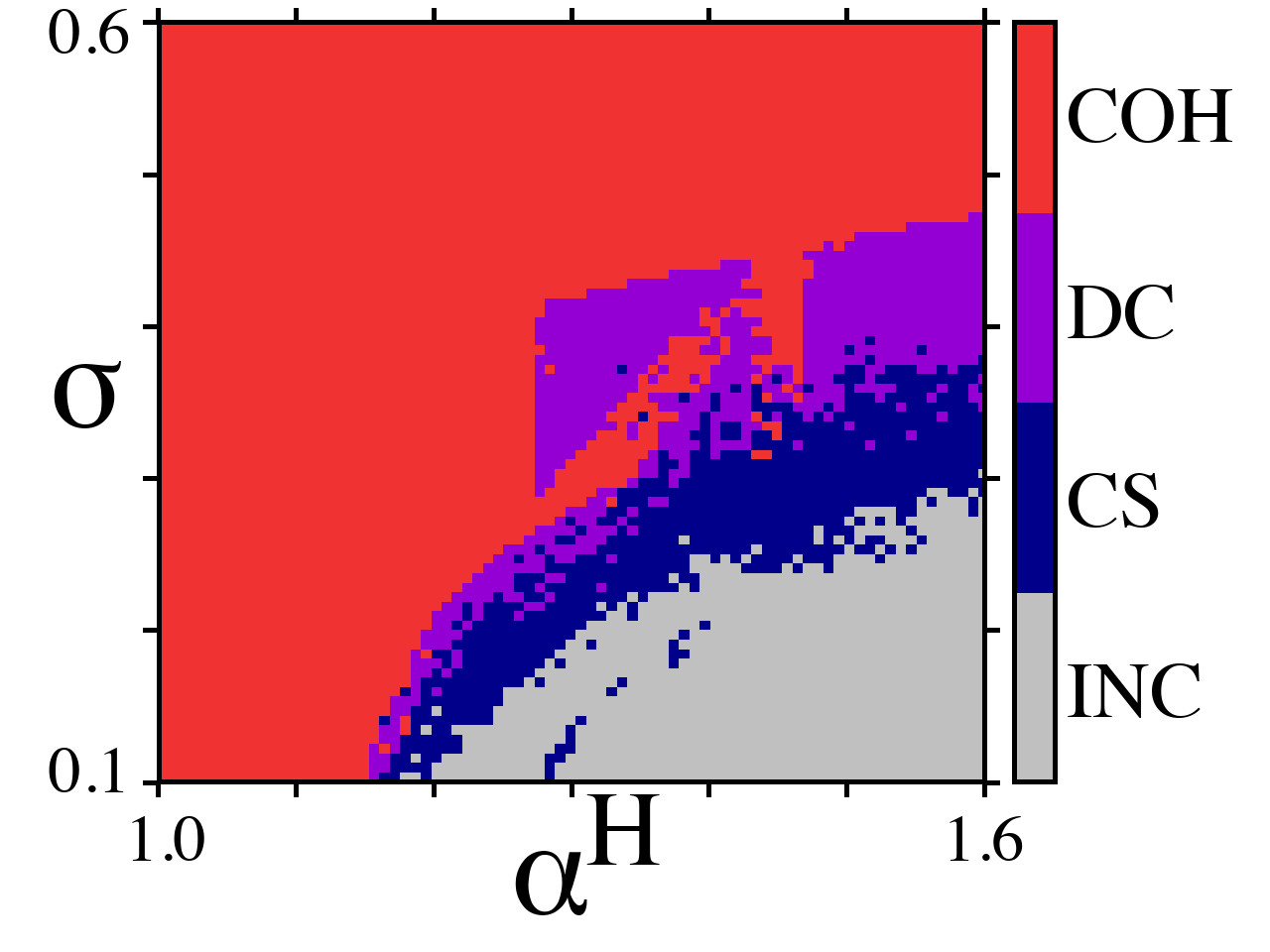}  \\
\hspace{8pt} (a) & \hspace{8pt} (b)\\
\end{tabular}
\begin{tabular}{cc}
\includegraphics[width=.45\columnwidth]{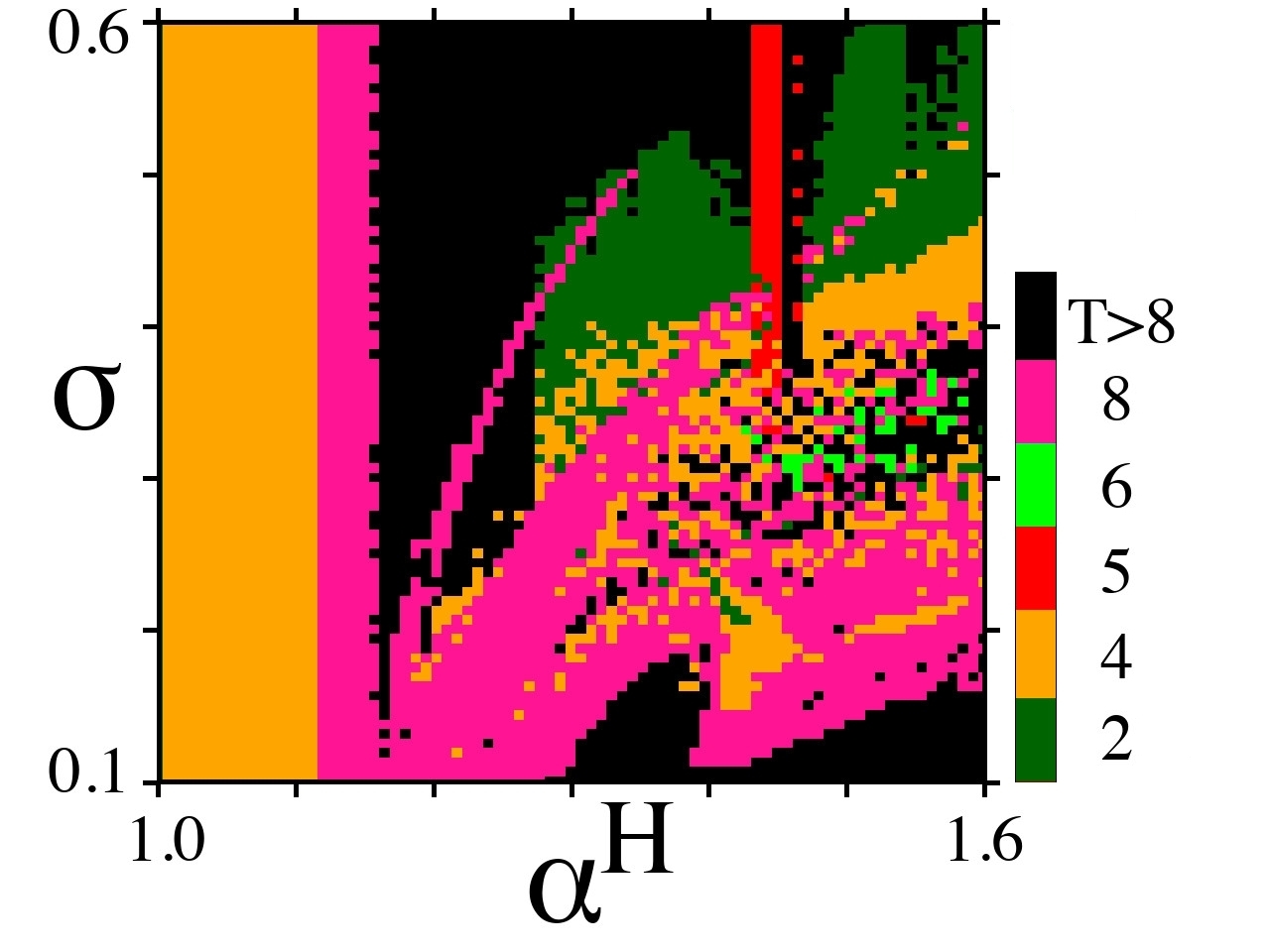} &
\includegraphics[width=.45\columnwidth]{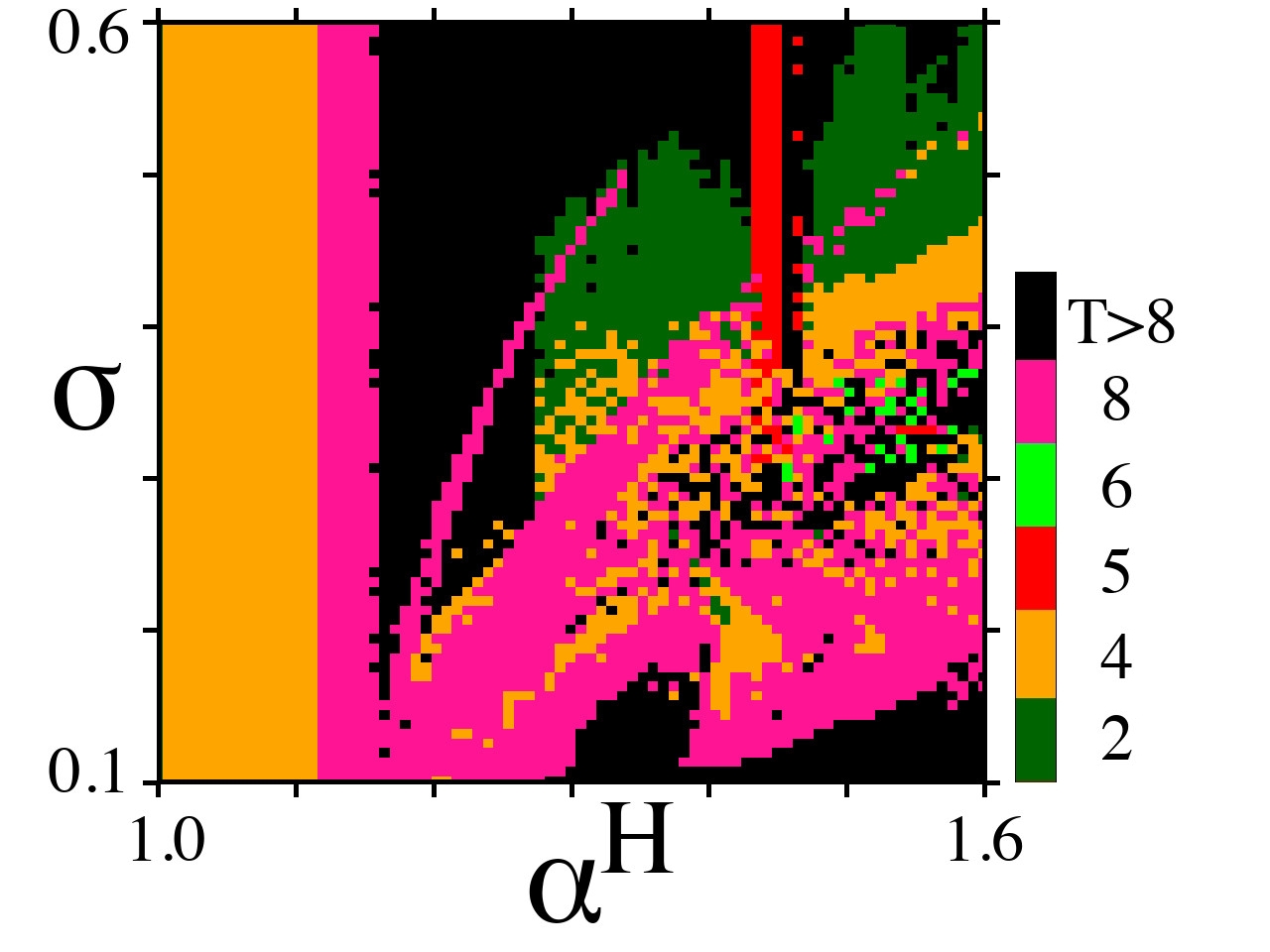}  \\
\hspace{8pt} (c) & \hspace{8pt} (d)\\
\end{tabular}
\caption{2D diagrams of spatio-temporal regimes (a,b) and of corresponding temporal dynamics (c,d) for the noise-free network of nonlocally coupled Henon maps in the ($\alpha^{H},\sigma$) parameter plane  for two different realizations of initial conditions randomly distributed in  the intervals $x(i,0)\in [-0.5,0.5]$ and  $y(i,0)\in [-0.15,0.15]$.  COH is coherence or complete synchronization between elements, DC corresponds to snapshots with profile discontinuities, CS is chimera states, and INC is incoherence. The color scale in (c,d) indicates the period of temporal dynamics. Other parameters: $\beta^{H}=0.2$, $R=320$, $N=1000$, $D=0$.}
	\label{fig_8}
\end{figure}

We now proceed to analyze the impact of additive noise on the probability of observing chimera states in a network of nonlocally coupled Henon maps (\ref{eq:henon}). As in the previous section, we first construct 2D diagrams of dynamical regimes (Fig.~\ref{fig_8}(a),(b)) which  typically exist in the noise-free network and the corresponding diagrams of temporal behavior (Fig.~\ref{fig_8}(c),(d)) in the ($\alpha^{H}, \sigma$) parameter plane for two different realizations of initial conditions randomly distributed  in the intervals $x(i,0)\in [-0.5,0.5]$ and  $y(i,0)\in [-0.15,0.15]$. We fix 
$\beta^{H}=0.2$ (\ref{eq:henon}) and the nonlocal coupling range $R=320$.

As in the case of the logistic map network (Sec.~\ref{sec_ring_logistic}), the network of nonlocally coupled Henon maps demonstrates four typical dynamical regimes when $\alpha^{H}$ and $\sigma$ are varied (Fig.~\ref{fig_8}(a),(b)). When $\alpha^H \in [1.0,1.15]$, only coherent dynamics is observed in the Henon map network for the whole range of $\sigma\in[0.1,0.6]$. In this case the elements oscillate periodically in time with $T=4$ and $T=8$, i.e., there is a period doubling in time~\cite{Omelchenko:2011uc,Omelchenko:2012tv} as $1.0 < \alpha^H<1.15$ (Fig.~\ref{fig_8}(c),(d)). Such temporal behavior is in full accordance with the dynamics of the uncoupled Henon map (Fig.~\ref{fig_1}(c)). 
A transition from incoherence to coherence occurs through the existence of chimera states when $\alpha^H \in [1.15,1.6]$. With this, two different routes of the transition can be clearly distinguished with increasing coupling strength, namely, with the presence of "coherent window" ($1.27<\alpha^{H}<1.43$) and without it ($\alpha^{H}<1.27$ and $\alpha^{H}>1.43$) (Fig.~\ref{fig_8}(a),(b)). As in the case of the logistic map ensemble, the elements of the Henon map ring demonstrate periodic dynamics mainly with $T=8$ within the "coherent window" (Fig.~\ref{fig_8}(c),(d)), and the coherent window is much wider than in the logistic map. 
Inside the coherence region located above the region of snapshots with profile discontinuities at $\sigma>0.42$, there are 2- and 8-periodic oscillations, while at $\sigma>0.48$ the temporal dynamics of the elements becomes practically chaotic (black region in Fig.~\ref{fig_8}(c),(d)). Note that there is a wide periodic window within the interval $1.432<\alpha^{H}<1.46$ for the isolated Henon map (Fig.~\ref{fig_1}(c)). Within the same  range of  $\alpha^H$ variation the Henon map network demonstrates an earlier transition to coherent dynamics with respect to $\sigma$ than for larger values of $\alpha^H$ (Fig.~\ref{fig_8}(a),(b)).

\begin{figure}[ht]
	\centering
\begin{tabular}{cc}
\includegraphics[width=.45\columnwidth]{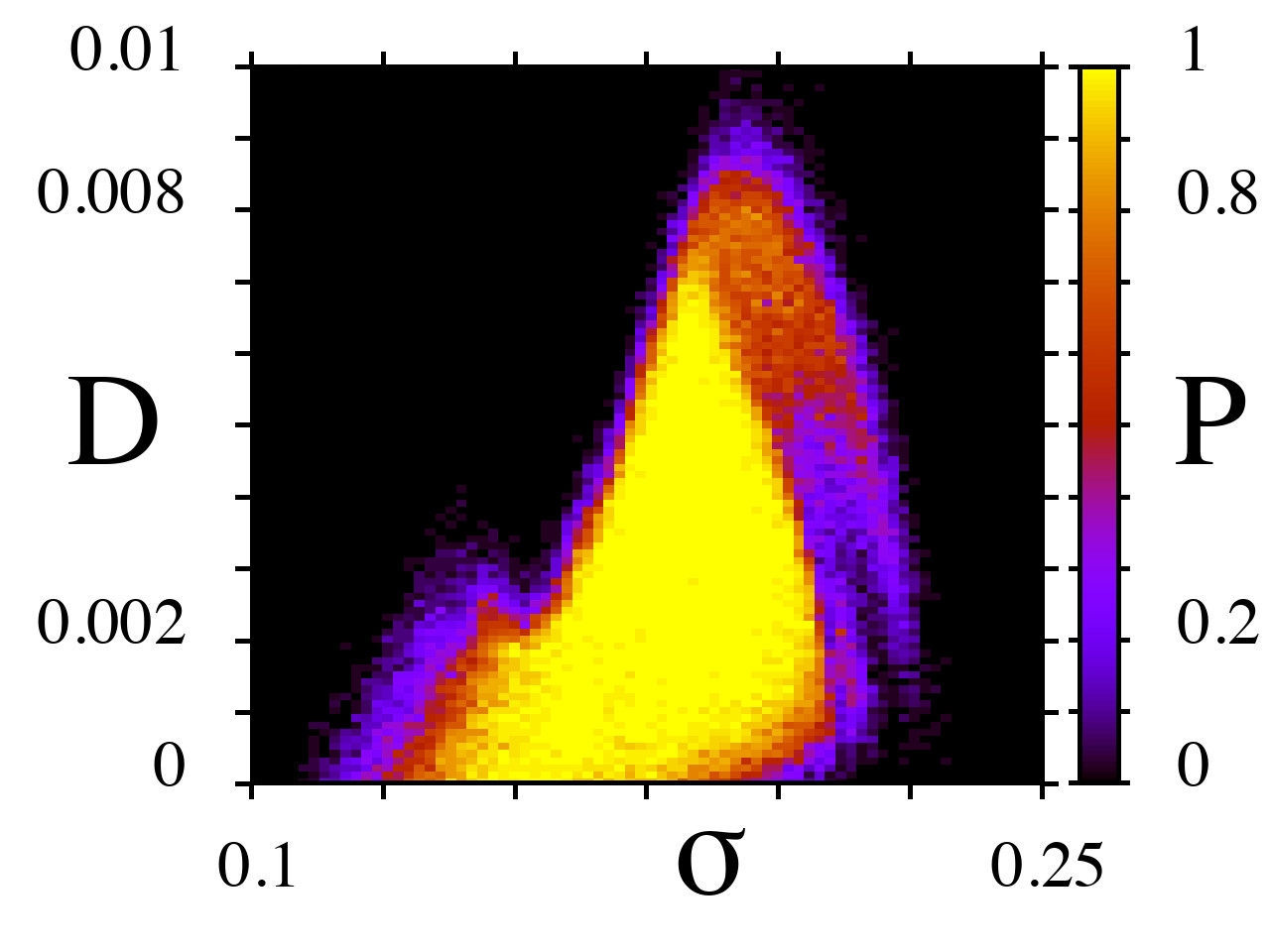} &
\includegraphics[width=.45\columnwidth]{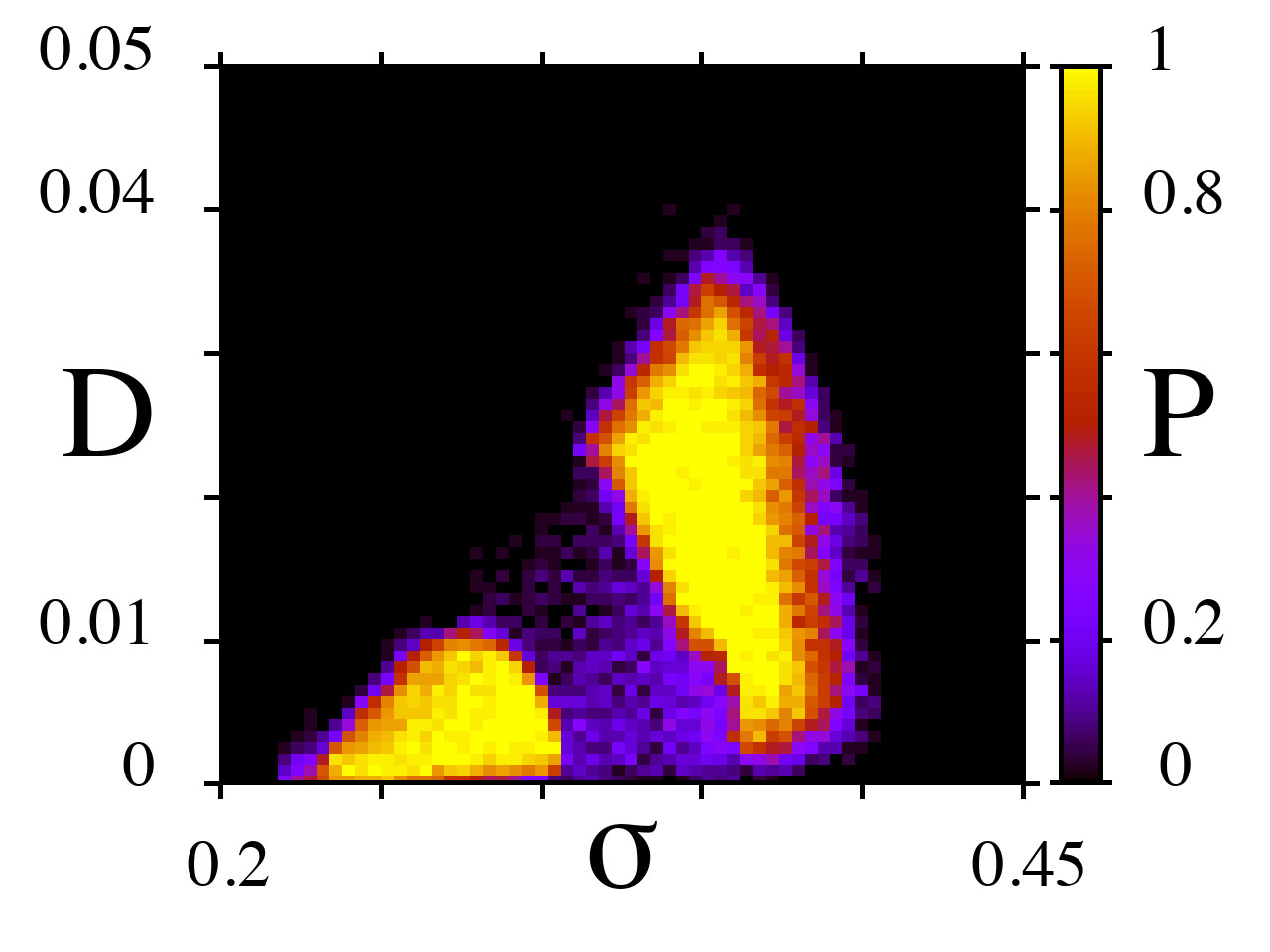}  \\
\hspace{8pt} (a) & \hspace{8pt} (b)\\
\end{tabular}
\begin{tabular}{cc}
\includegraphics[width=.45\columnwidth]{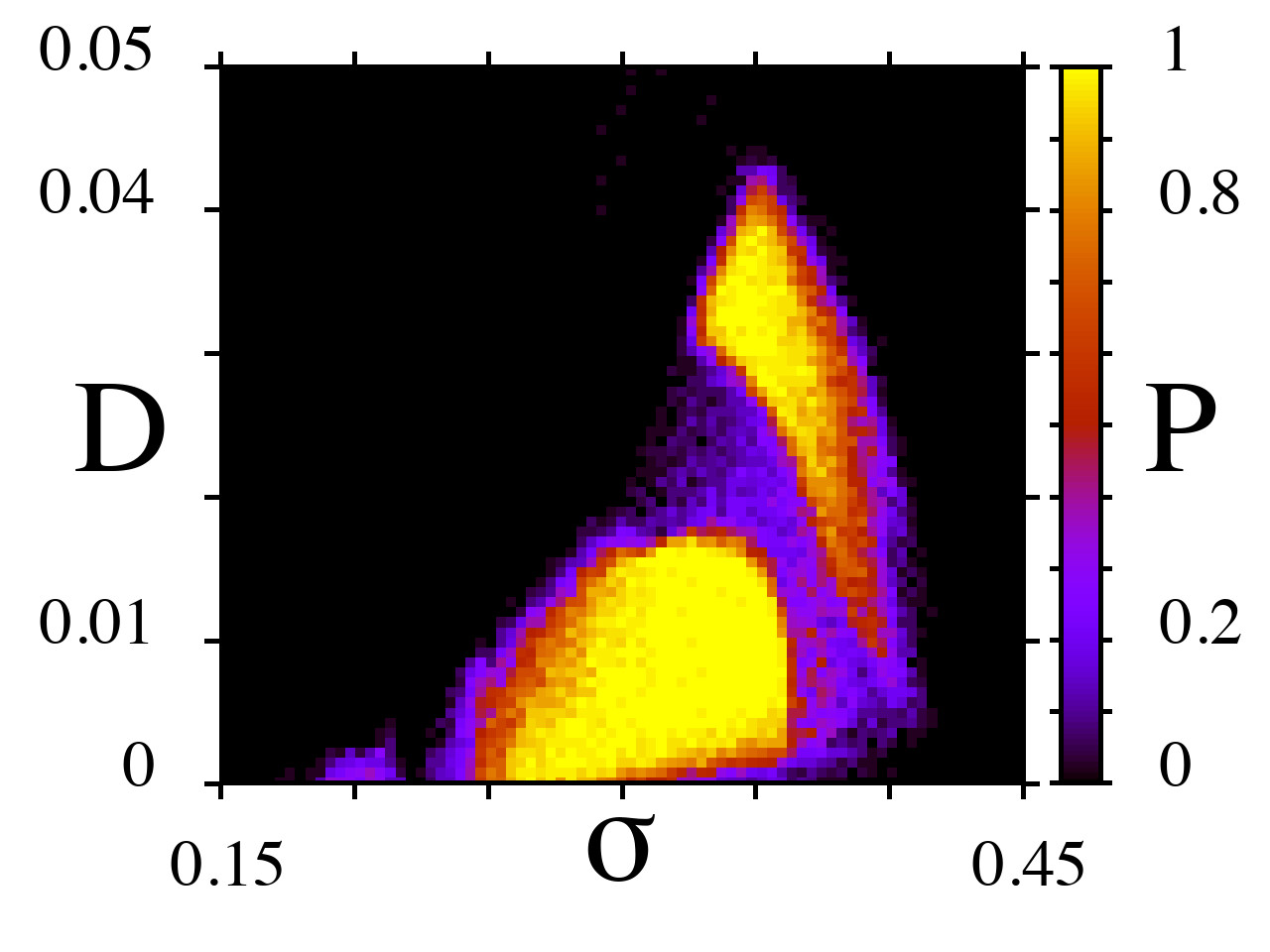} &
\includegraphics[width=.45\columnwidth]{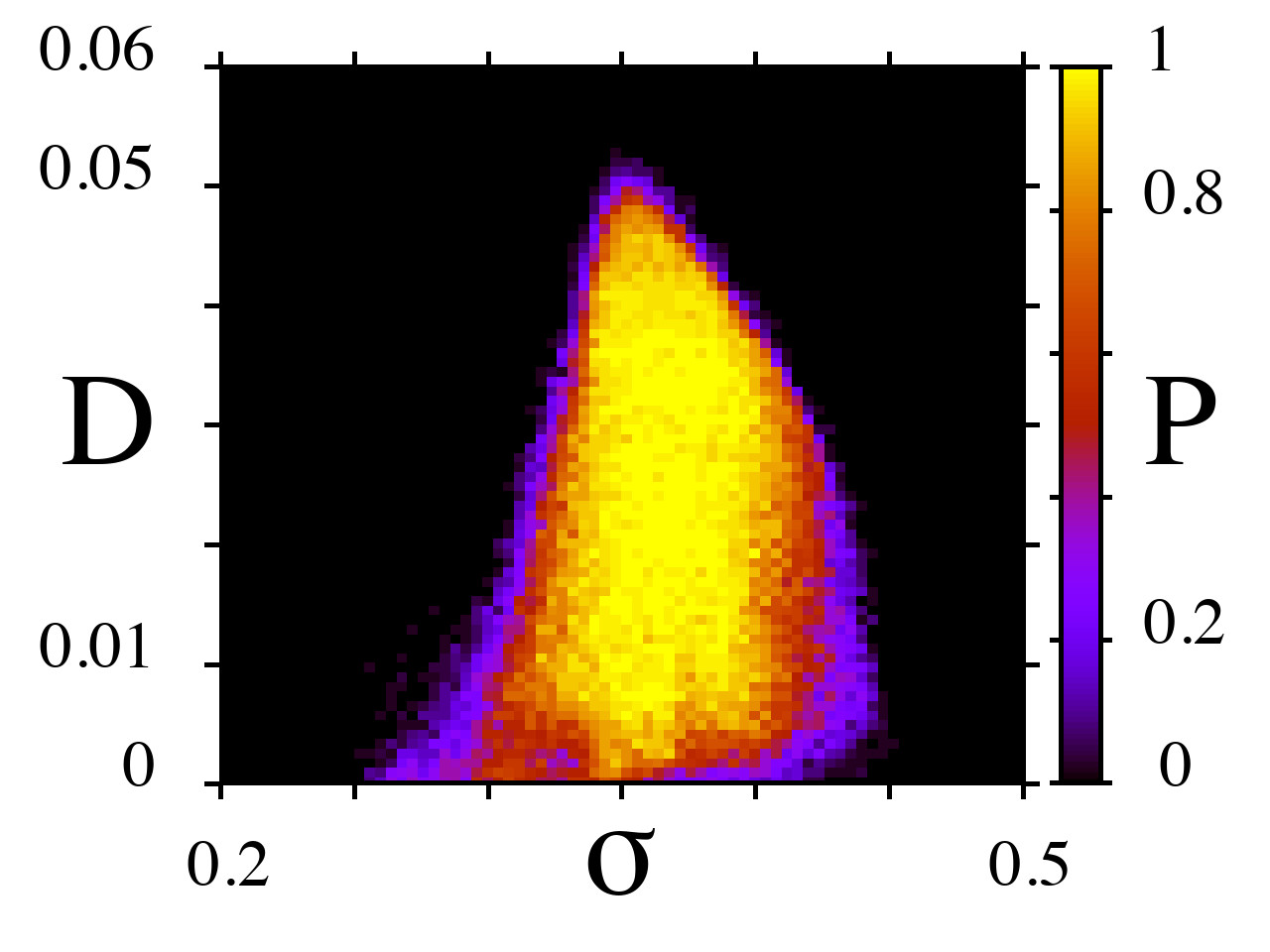}  \\
\hspace{8pt} (c) & \hspace{8pt} (d)\\
\end{tabular}
\caption{Distribution diagrams for the probability  $P$ of observing chimera states  in the ($\sigma,D$) parameter plane for the Henon map network  for different values of the local dynamics parameter $\alpha^H$: (a) $\alpha^{H}=1.22$, (b) $\alpha^{H}=1.35$, (c) $\alpha^{H}=1.4$, and (d) $\alpha^{H}=1.6$. The diagrams are plotted using $M=50$ different pairs of realizations of random initial conditions  and noise realizations. Other parameters: $\beta^{H}=0.2$, $R=320$, $N=1000$.}
	\label{fig_9}
\end{figure}

We now add external noise to the Henon map network and study its influence on the probability of appearing and existing chimera states as the noise intensity $D$ and the coupling strength $\sigma$ are varied. Figure~\ref{fig_9} shows distribution diagrams for the probability of observing chimera states in the Henon map network in the ($\sigma,D$) parameter plane for four different values of the local dynamics parameter $\alpha^H$. 
As is seen from the diagrams of spatio-temporal regimes (Fig.~\ref{fig_8}(a),(b)), each of the two routes of ``incoherence-coherence'' transition occupies a rather wide range with respect to the parameter $\alpha^{H}$. Our simulations show that this fact can lead to the appearance of one  (Fig.~\ref{fig_9}(a),(d)) or two  (Fig.~\ref{fig_9}(b),(c)) regions with a high probability $P$ of chimera observation. For a small value of $\alpha^{H}$  and for any noise intensity $D$, chimera states can exist only for weak coupling (Fig.~\ref{fig_9}(a)).  When $\alpha^{H}$ increases, a second region corresponding to the high probability of chimera observation appears in the region of larger values of the  coupling strength  (Fig.~\ref{fig_9}(b),(c)). With this, the two-region probability distributions are more extended in size than in the case of the logistic map network (compare Fig.~\ref{fig_9}(b),(c) and Fig.~\ref{fig_4}(a),(b)). As is seen from Fig.~\ref{fig_9}(b),(c), the two regions of the high probability are separated by a ``channel'' (the violet region) which is related to a low probability of chimera existence.  The width of this channel decreases as $\alpha^{H}$ successively increases and eventually (at $\alpha^{H}=1.6$) there is a single region of high probability of observing chimeras (Fig.~\ref{fig_9}(d)), as in the case of $\alpha^{H}=1.22$ (Fig.~\ref{fig_9}(a)). 
However,  for large values of $\alpha^{H}$, this region is shifted towards the range of strong coupling, and the resonant value of $\sigma$ at which chimeras are observed within a sufficiently wide range of the noise intensity also becomes larger ($\sigma \approx 0.35$ for $\alpha^{H}=1.6$) as compared with the case of small values of $\alpha^{H}$ ($\sigma \approx 0.195$ for $\alpha^{H}=1.22$). Note that at $\alpha^{H}=1.6$ chimeras are observed within a significantly wider region with respect to the noise intensity $D$, than at $\alpha^{H}=1.22$ (compare Figs.~ \ref{fig_9}(a) and (d)). We suppose that these peculiarities may be caused by a highly-developed chaotic dynamics observed in the isolated Henon map at $\alpha^{H}=1.6$.

The probability distributions of chimera observation obtained for the noisy network of nonlocally coupled Henon maps also demonstrate and verify the effect of chimera resonance. There is a certain optimum noise level at which the $\sigma$-interval corresponding to the high probability of existing chimeras is the largest. For example, at $D_{\rm opt} \approx 0.0022$ $\sigma\in[0.158,0.207]$ for $\alpha^{H}=1.22$ (Fig.~\ref{fig_9}(a)) and  at $D_{\rm opt} \approx 0.002$ and $\sigma\in[0.338,0.401]$ for $\alpha^{H}=1.6$ (Fig.~\ref{fig_9}(d)). Note that in the case of two-region (bimodal) probability distributions there exist two different optimum noise intensities which are related to the two widest $\sigma$-intervals.

\subsection{Network of nonlocally coupled modified Ricker maps}\label{sec_ring_ricker}

\begin{figure}[ht]
	\centering
\begin{tabular}{cc}
\includegraphics[width=.45\columnwidth]{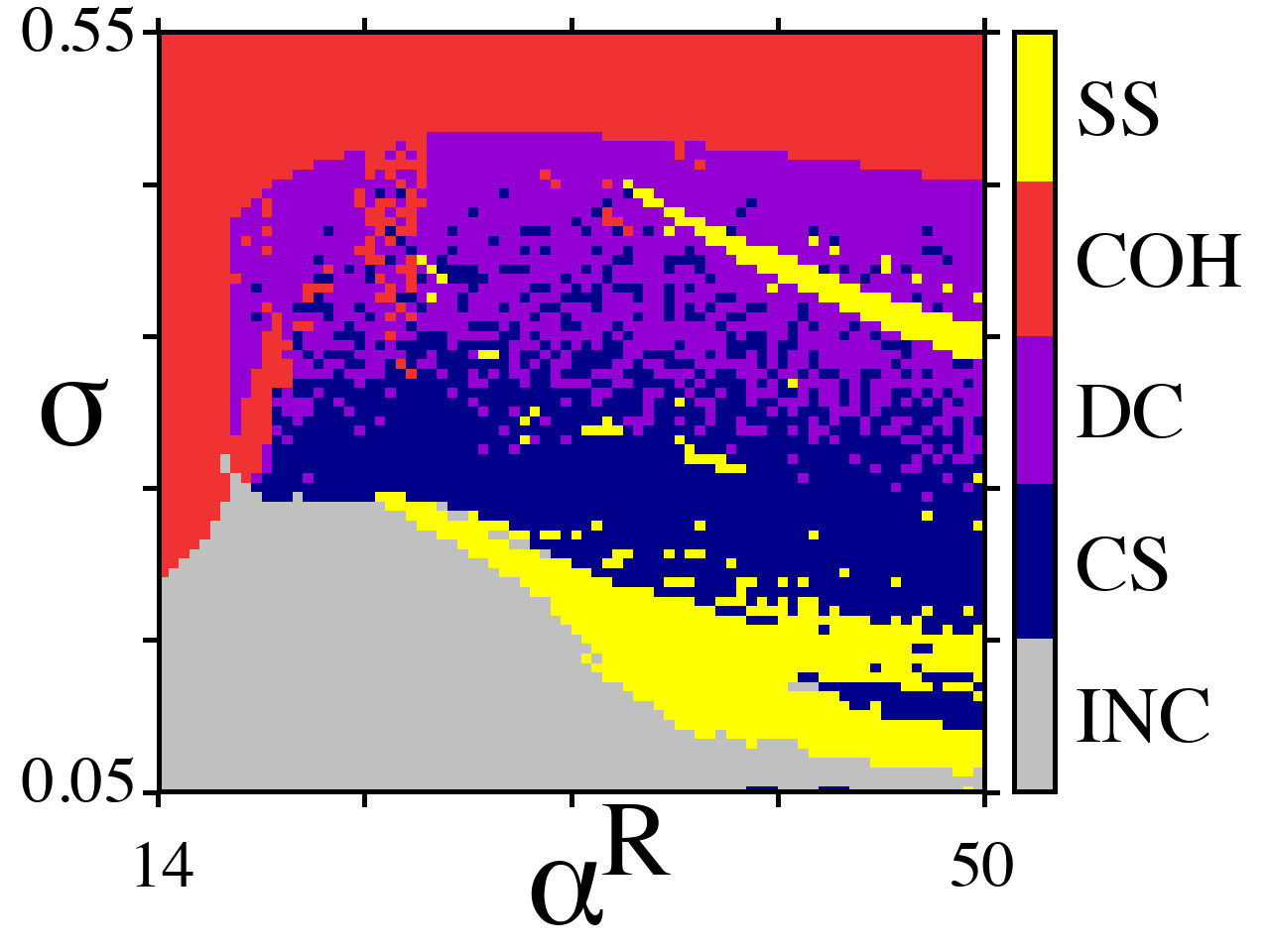} &
\includegraphics[width=.45\columnwidth]{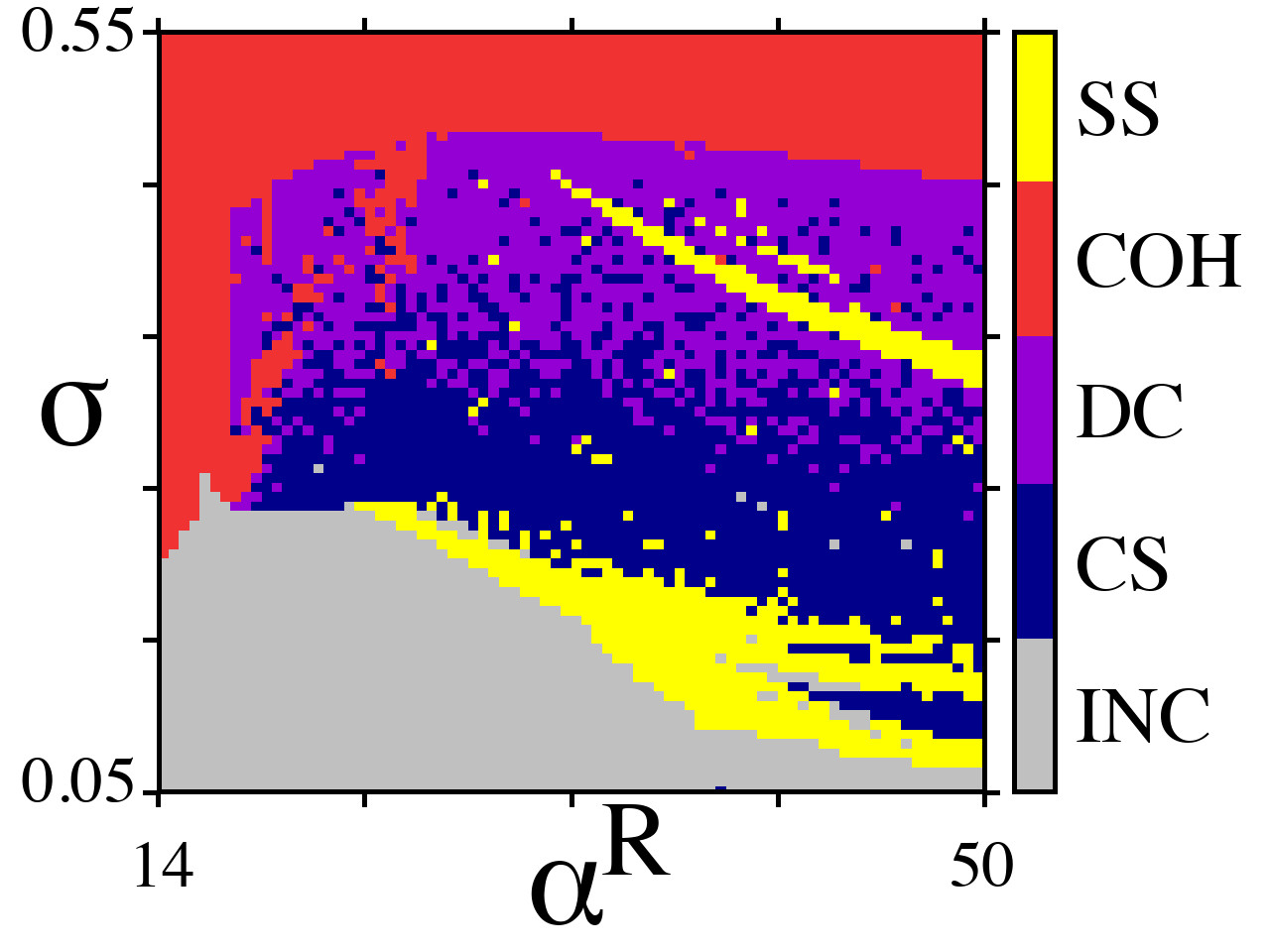}  \\
\hspace{8pt} (a) & \hspace{8pt} (b)\\
\end{tabular}
\begin{tabular}{cc}
\includegraphics[width=.45\columnwidth]{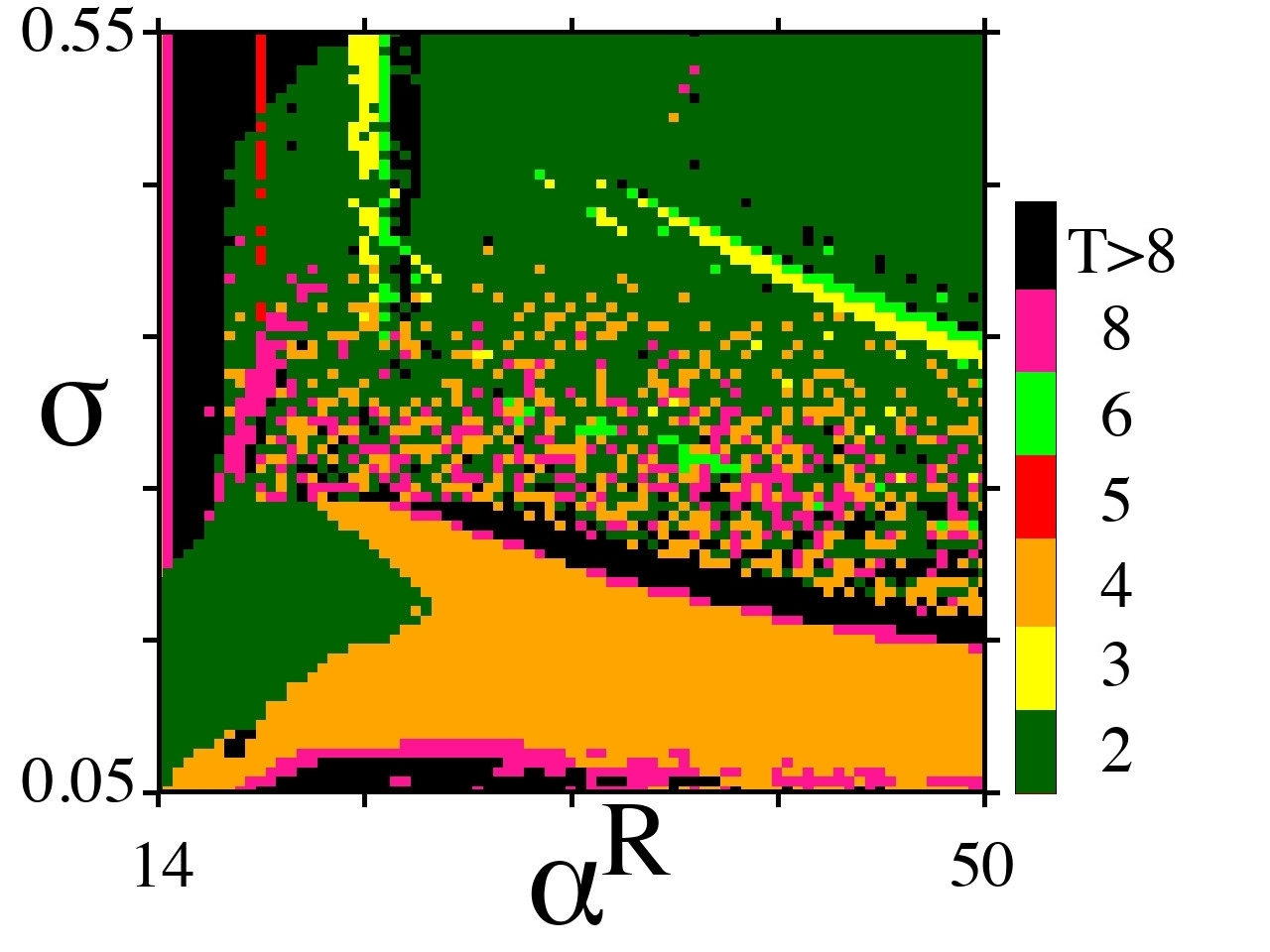} &
\includegraphics[width=.45\columnwidth]{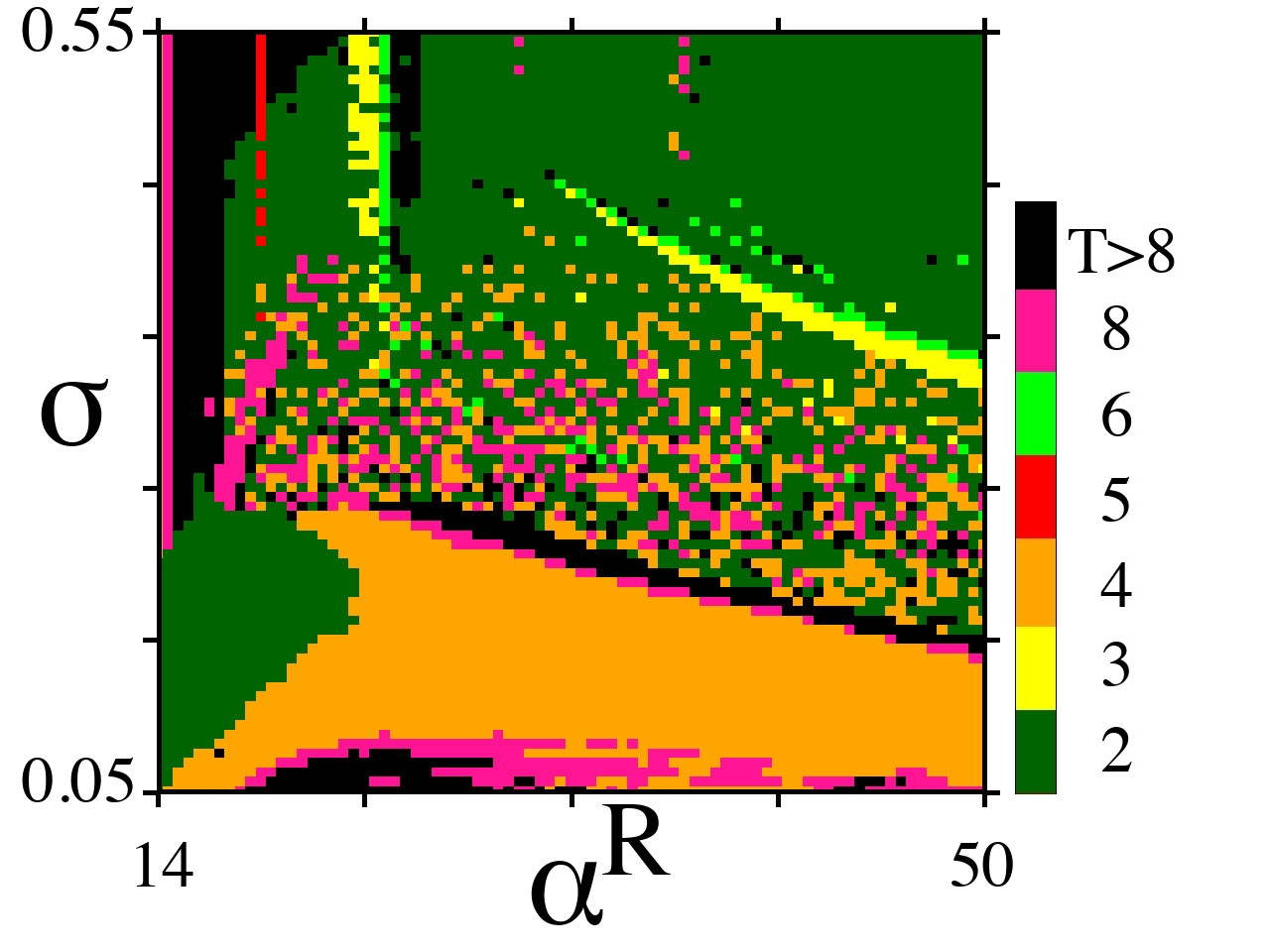}  \\
\hspace{8pt} (c) & \hspace{8pt} (d)\\
\end{tabular}
\caption{2D diagrams of spatio-temporal regimes (a,b) and temporal dynamics (c,d) for the noise-free network of nonlocally coupled modified Ricker maps in the ($\alpha^{R},\sigma$) parameter plane  for two different realizations of initial conditions randomly distributed within the interval $[-1,1]$.  COH is coherence or complete synchronization between elements, DC corresponds to snapshots with profile discontinuities, SS is solitary states, CS is chimera states, and INC is incoherence. The color scale in (c,d) indicates the period of temporal dynamics. Other parameters: $R=320$, $D=0$, $N=1000$.}
	\label{fig_10}
\end{figure}

Finally, we analyze numerically the dynamics of the third network of nonlocally coupled modified Ricker maps (\ref{eq:mod-ricker}) both without and in the presence of additive noise (\ref{eq:system}). Our calculations show that this network can exhibit both chimeras and solitary states when the local dynamics parameter $\alpha^{R}$ is varied. In analogy with the previously considered networks (Secs.~\ref{sec_ring_logistic} and \ref{sec_ring_henon}), we construct 2D diagrams of spatio-temporal regimes and temporal dynamics distributions in the  ($\alpha^{R},\sigma$) parameter plane for the noise-free network of Ricker maps. The results are shown in Fig.~\ref{fig_10}. Along with the same four regions with typical regimes (incoherence, coherence, snapshots with profile discontinuities, and chimera states) which are also observed in the logistic map and Henon map networks, two new regions appear in the ($\alpha^{R},\sigma$) plane (yellow-colored regions in Fig.~\ref{fig_10}(a),(b)), which correspond to the existence of solitary states at weak and strong coupling. 

\begin{figure}[ht]
	\centering
\begin{tabular}{cccc}
\includegraphics[width=.23\columnwidth]{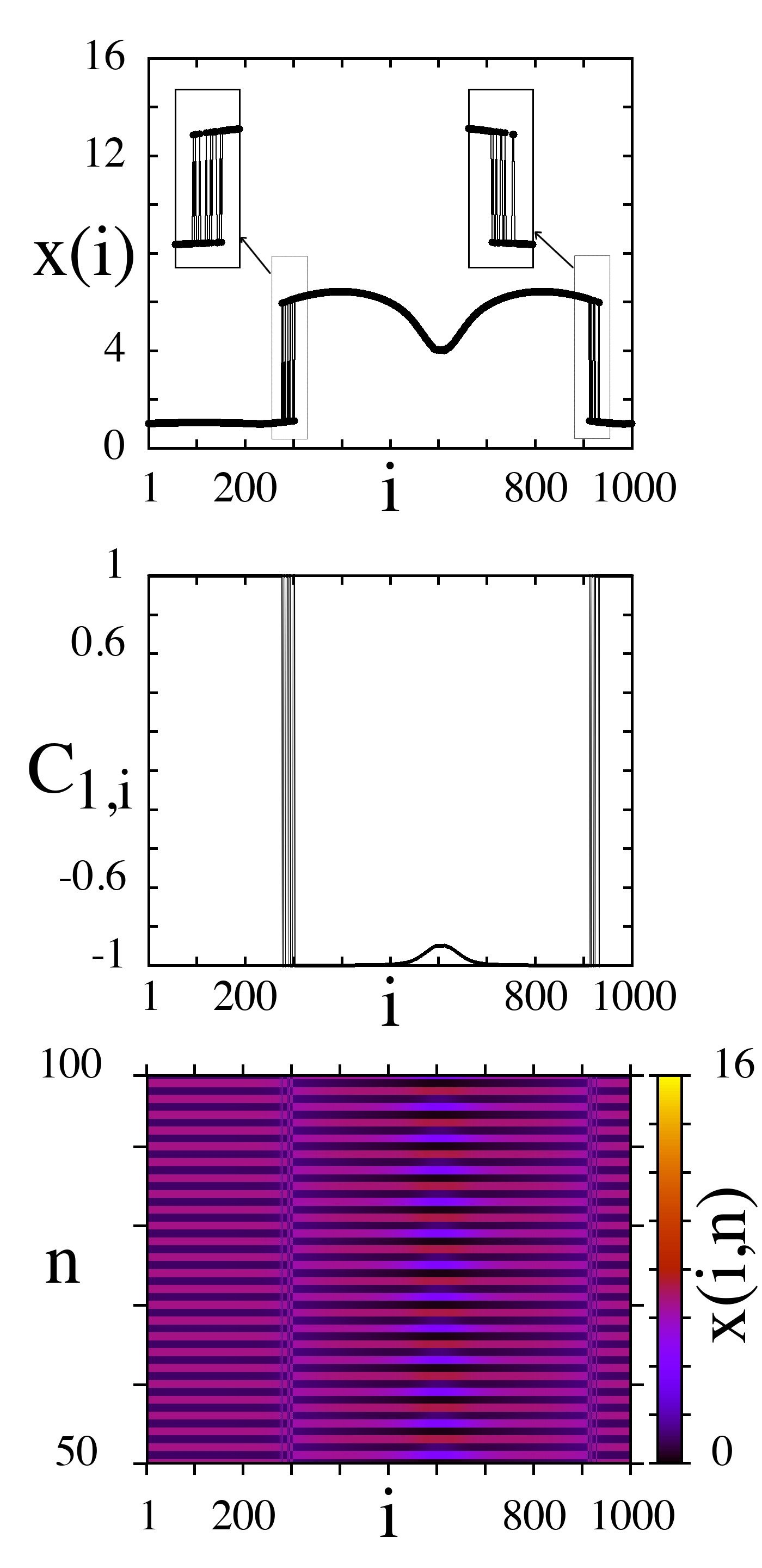} &
\includegraphics[width=.23\columnwidth]{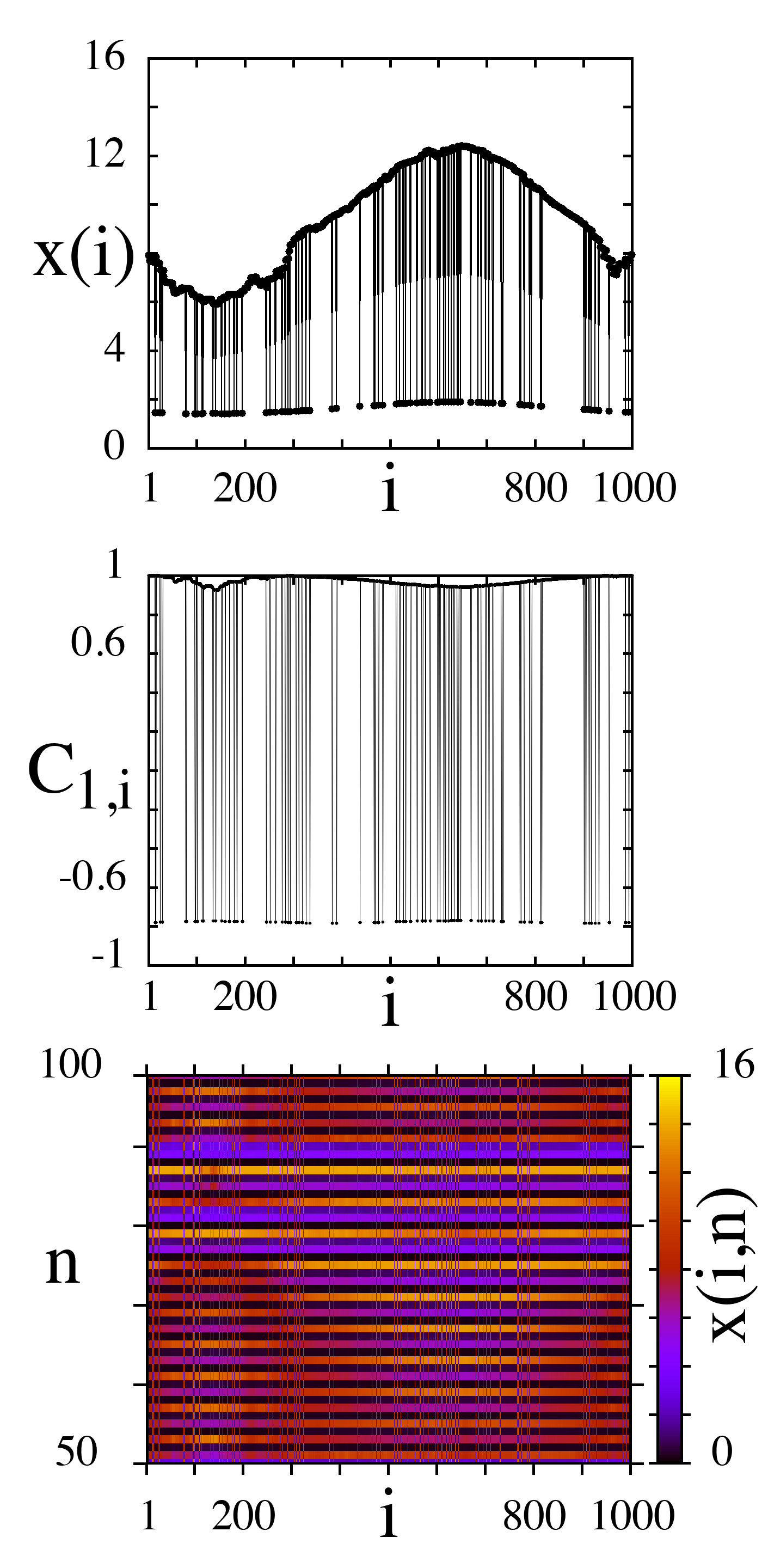}  &
\includegraphics[width=.23\columnwidth]{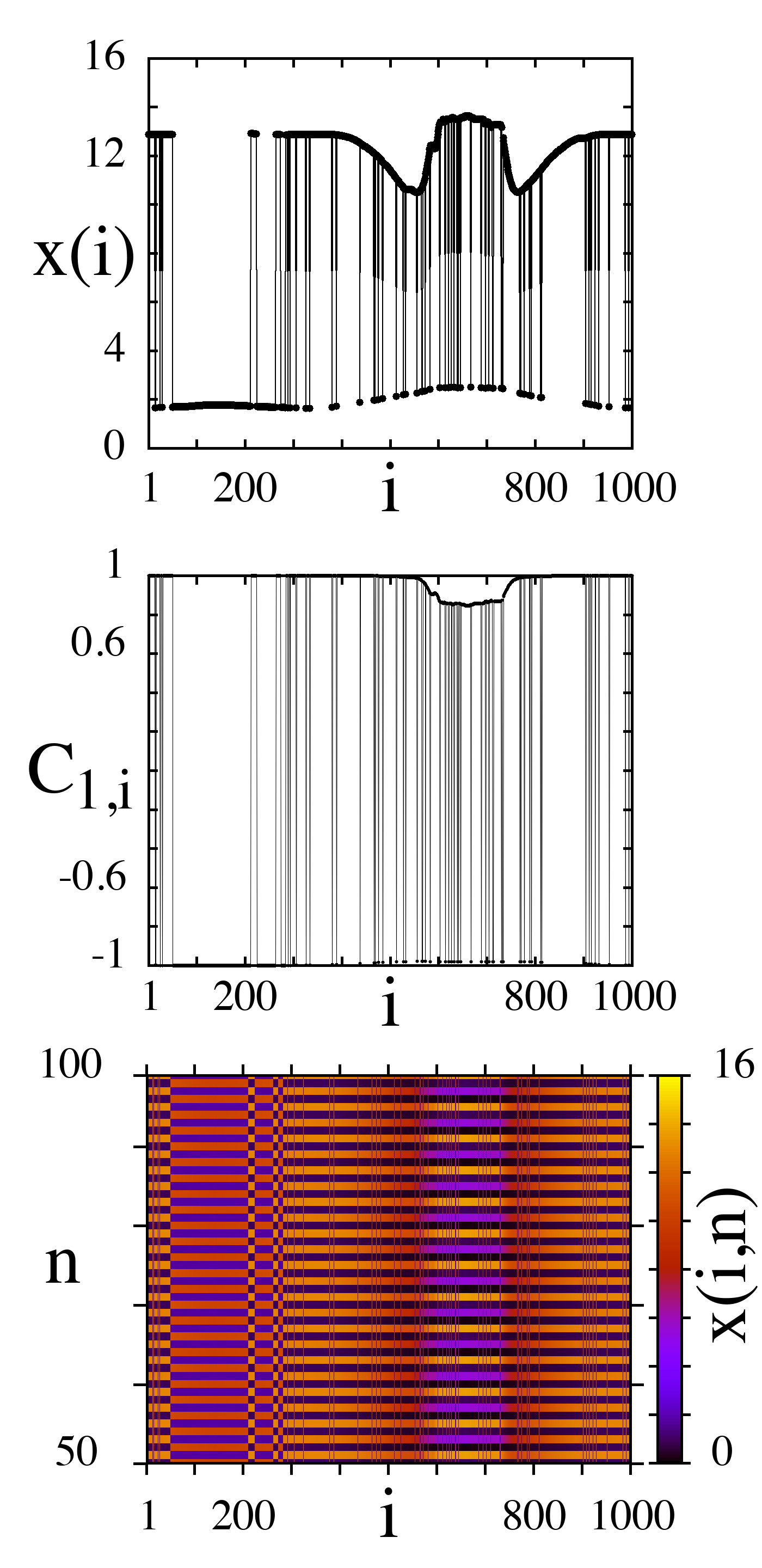} &
\includegraphics[width=.23\columnwidth]{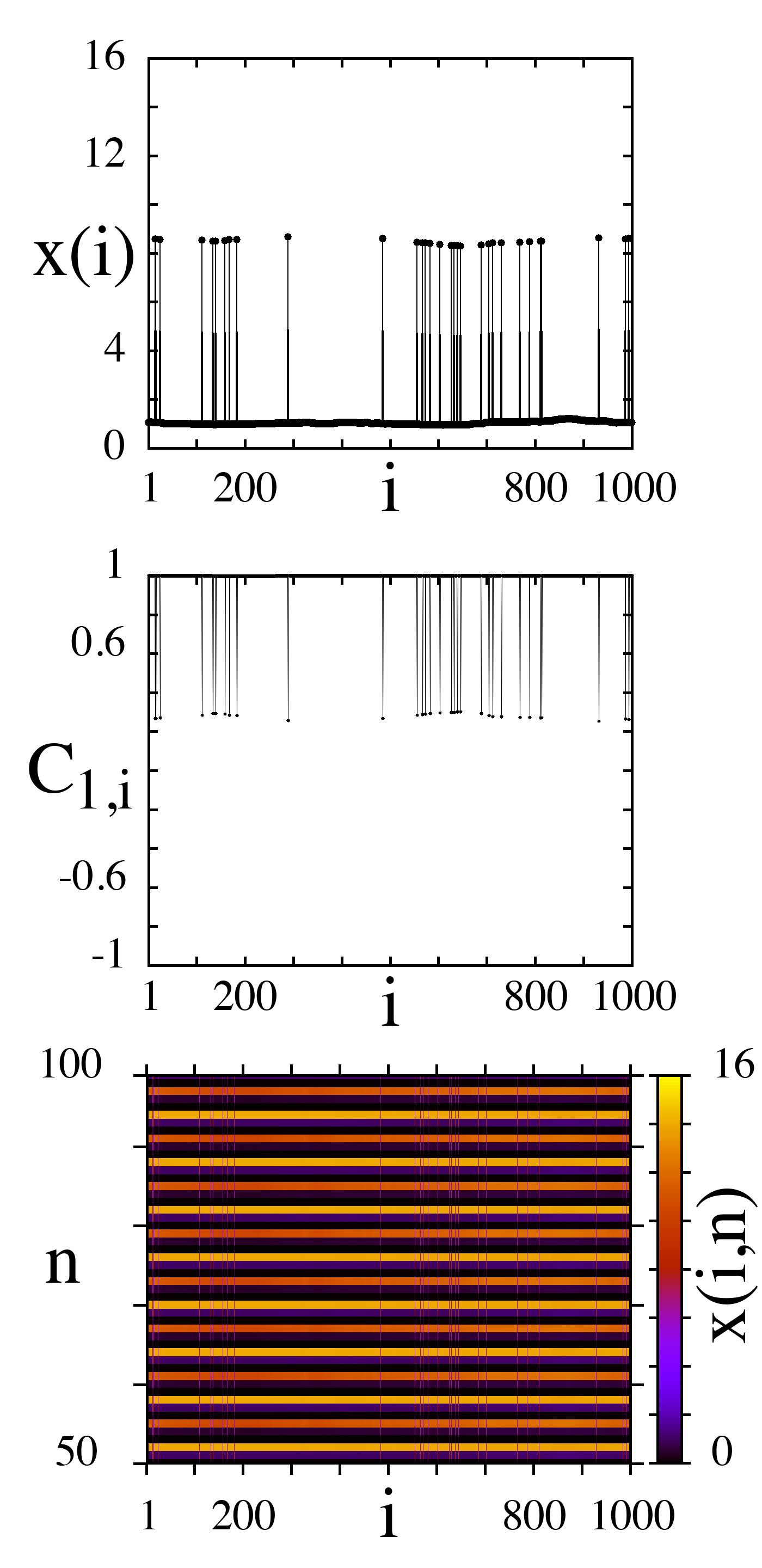}  \\
\hspace{8pt} (a) & \hspace{8pt} (b) & \hspace{8pt} (c) & \hspace{8pt} (d)\\
\end{tabular}
\caption{Snapshots of the $x(i)$ variables (upper row), spatial distributions of the cross-correlation coefficient (middle row), and space-time diagrams $x(i,n)$ (lower row) for different values of the local dynamics parameter $\alpha^{R}$ and the coupling strength $\sigma$ in the noise-free network of modified Ricker maps:  (a) $\alpha^{R}=20.3$, $\sigma=0.2875$ (chimera state), (b) $\alpha^{R}=38.3$, $\sigma=0.1875$ (solitary state), (c) $\alpha^{R}=38.3$, $\sigma=0.2$ (coexistence of chimera and solitary state), (d) $\alpha^{R}=38.3$, $\sigma=0.41875$ (solitary state). Other parameters:  $R=320$, $D=0$, $N=1000$. The insets in (a) top row show blow-ups.}
	\label{fig_11}
\end{figure}

A chimera state regime is exemplified in Fig.~\ref{fig_11}(a) by a snapshot of the $x(i)$ variables, a spatial distribution of the cross-correlation coefficient (\ref{eq:correlation}), and a space-time diagram $x(i,n)$. The chimera states observed in the Ricker map network and the logistic map network differ only in the amplitude of oscillations of individual elements. 

As can be seen from the diagrams in Fig.~\ref{fig_10}(a),(b), the $\alpha^R$ range  can be divided into two subranges each related to a different route of the transition from incoherence to coherence as the coupling strength $\sigma$ increases. The abrupt change to coherence occurs within the first subrange ($\alpha^R <17.2$). The second subrange ($\alpha^R >17.2$), in which the chimera states are observed in the Ricker map network, corresponds to the transition to coherence through both the presence of "coherent window" ($17.2<\alpha^{R}<22$) and without it $\alpha^{R}>22$). However, it is worth noting that the second route is also accompanied by the appearance of solitary states (before the appearance of chimera states) which are observed in the Ricker map network at decreasing  values of the coupling strength as the parameter $\alpha^{R}$ increases (Fig.~\ref{fig_10}(a),(b)). 

Solitary states begin to appear in the Ricker map network at  $\alpha^{R}>23.4$ (Fig.~\ref{fig_10}(a),(b)). The solitary nodes demonstrate regular dynamics in time with periods $T=4$, $T=8$, and $T>8$ for weak coupling and with $T=3$ and $T=6$ for strong coupling (Fig.~\ref{fig_10}(c),(d)). Solitary state regimes are exemplified in Fig.~\ref{fig_11}(b),(d) for weak and strong coupling, respectively.
Note that at the transition from solitary states to chimeras with increasing coupling strength the coexistence of both regimes can be observed (Fig.~\ref{fig_11}(c)) but in our work we classify this regime as chimera states. 

\begin{figure}[ht]
	\centering
\begin{tabular}{cc}
\includegraphics[width=.45\columnwidth]{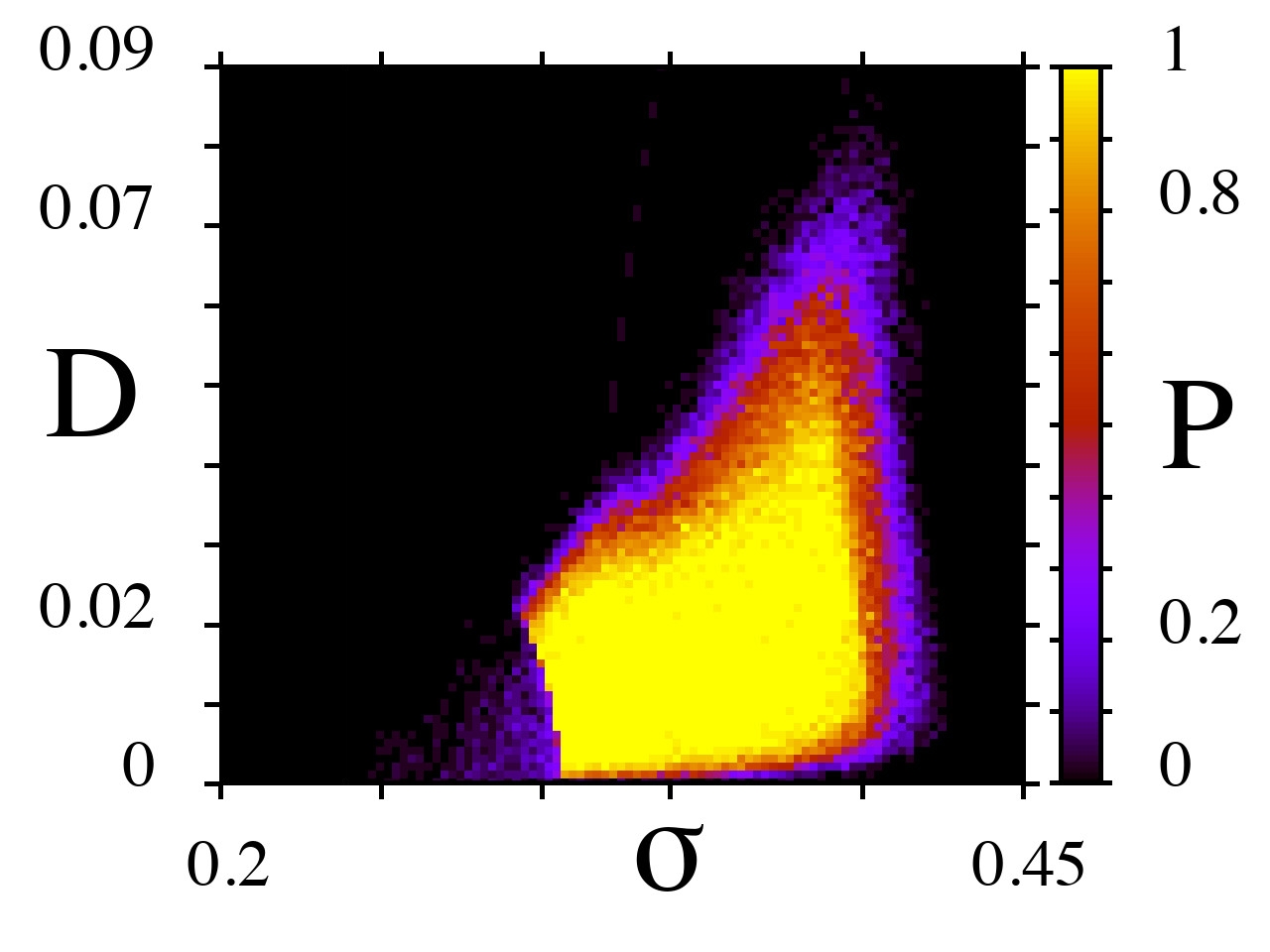} &
\includegraphics[width=.45\columnwidth]{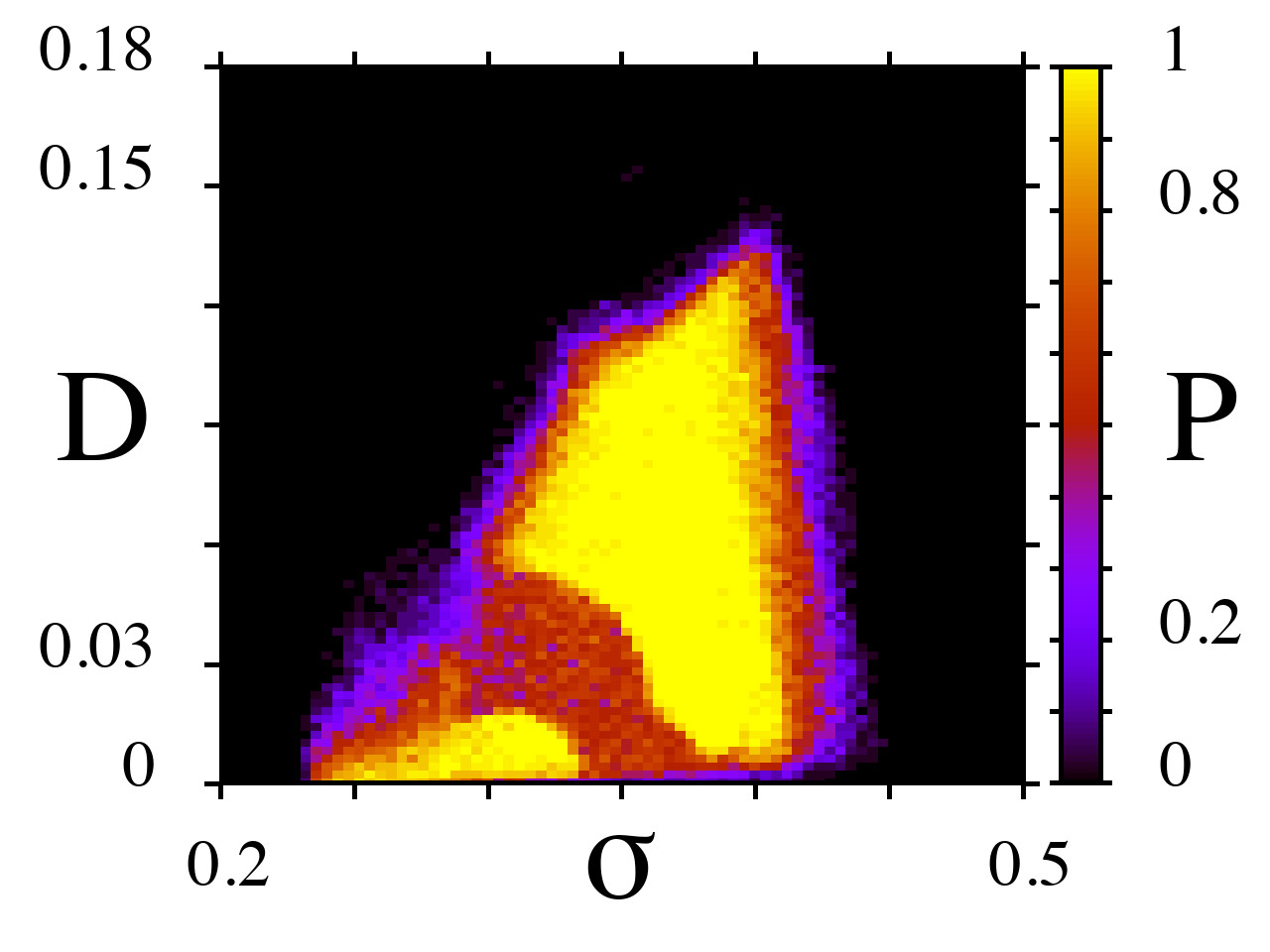}  \\
\hspace{8pt} (a) & \hspace{8pt} (b)\\
\end{tabular}
\begin{tabular}{cc}
\includegraphics[width=.45\columnwidth]{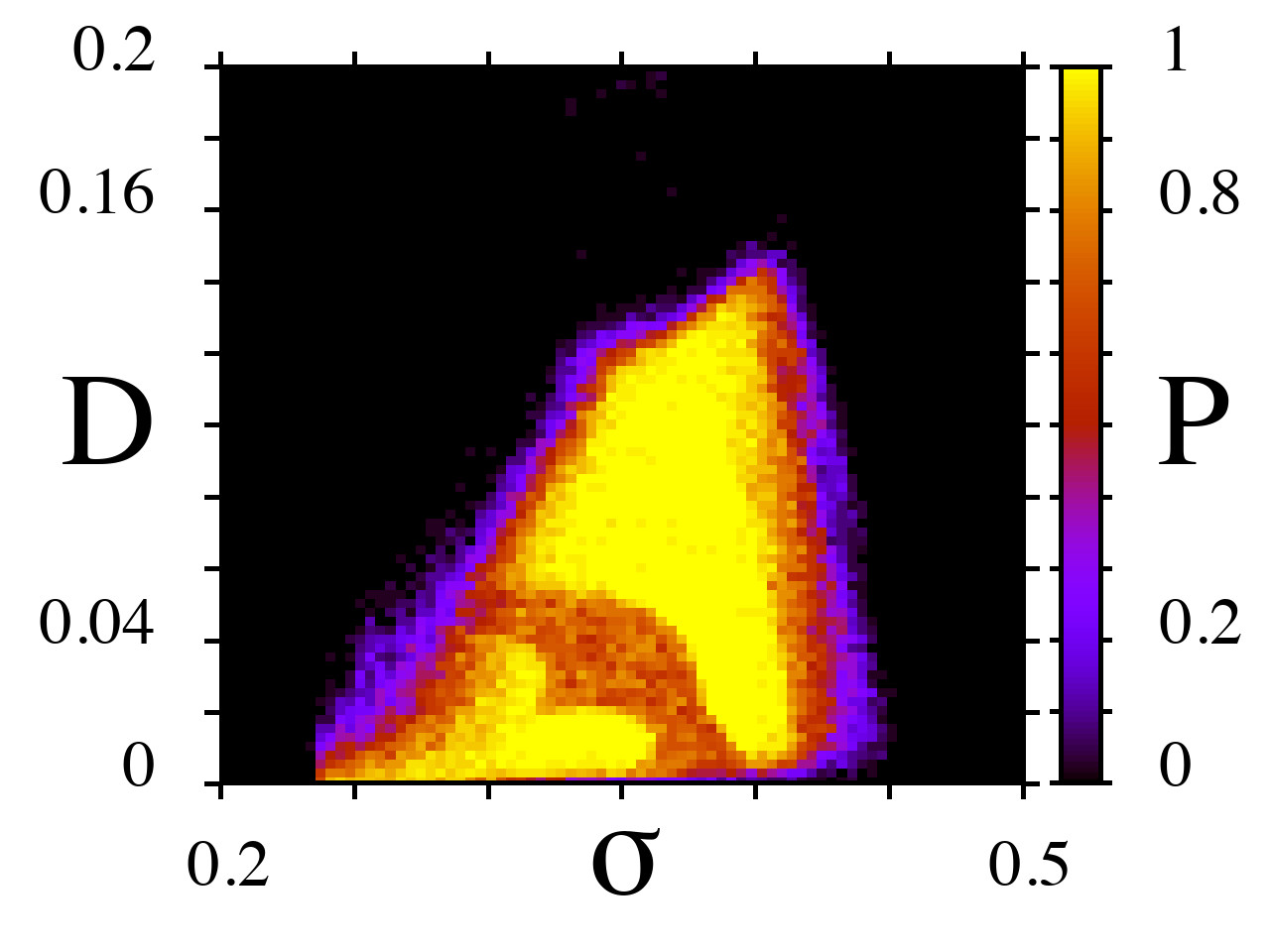} &
\includegraphics[width=.45\columnwidth]{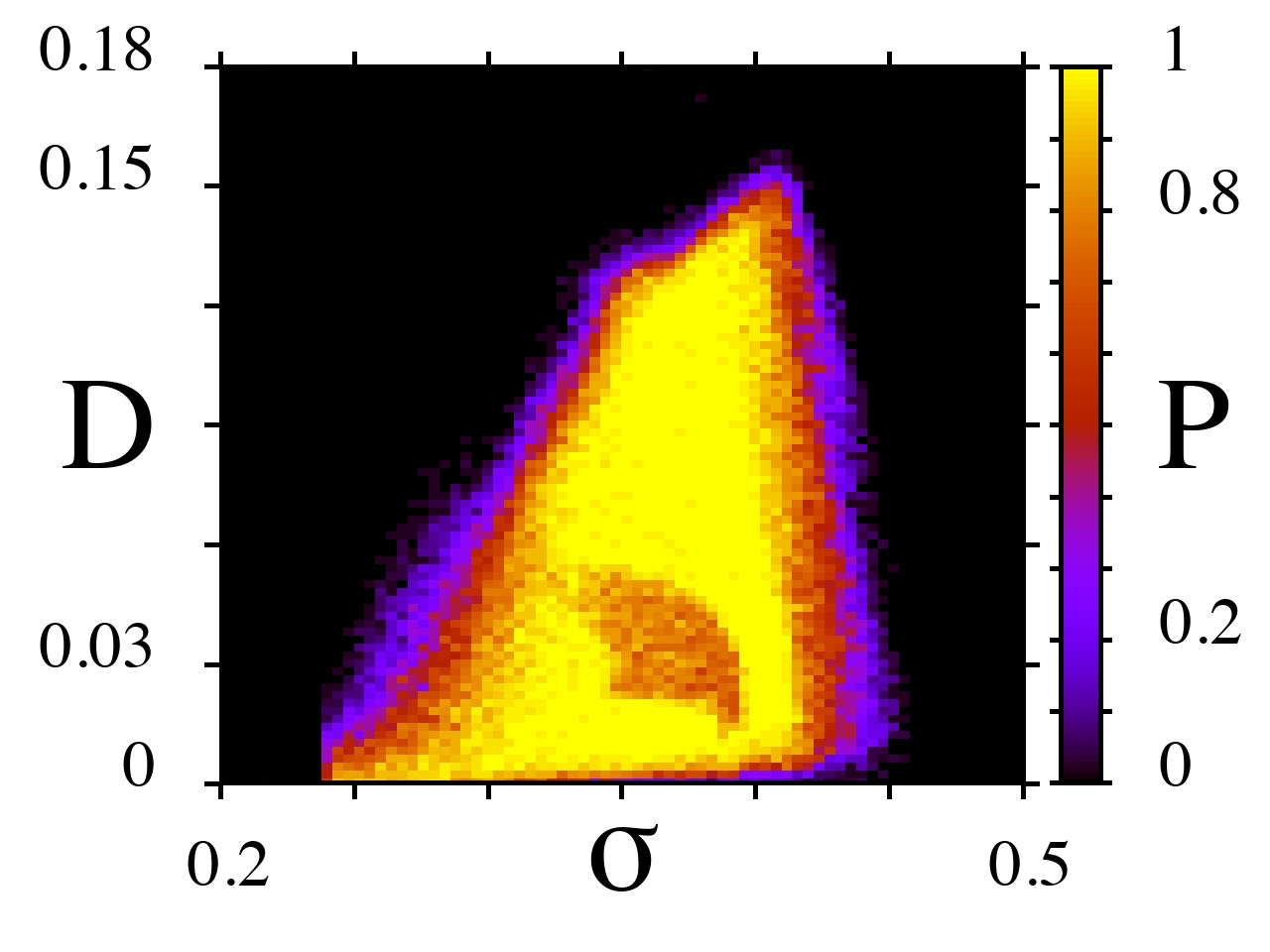}  \\
\hspace{8pt} (c) & \hspace{8pt} (d)\\
\end{tabular}
\begin{tabular}{cc}
\includegraphics[width=.45\columnwidth]{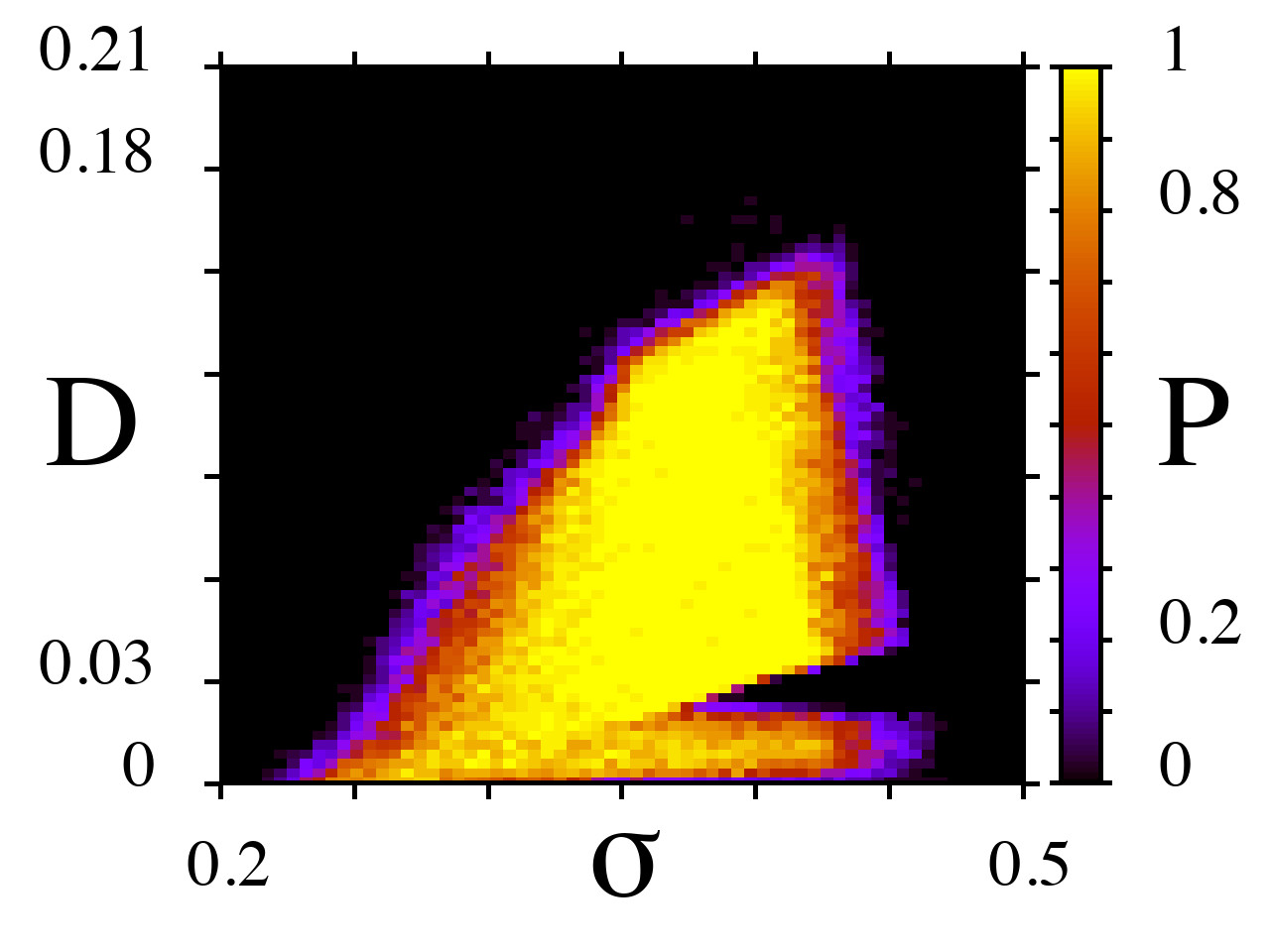} &
\includegraphics[width=.45\columnwidth]{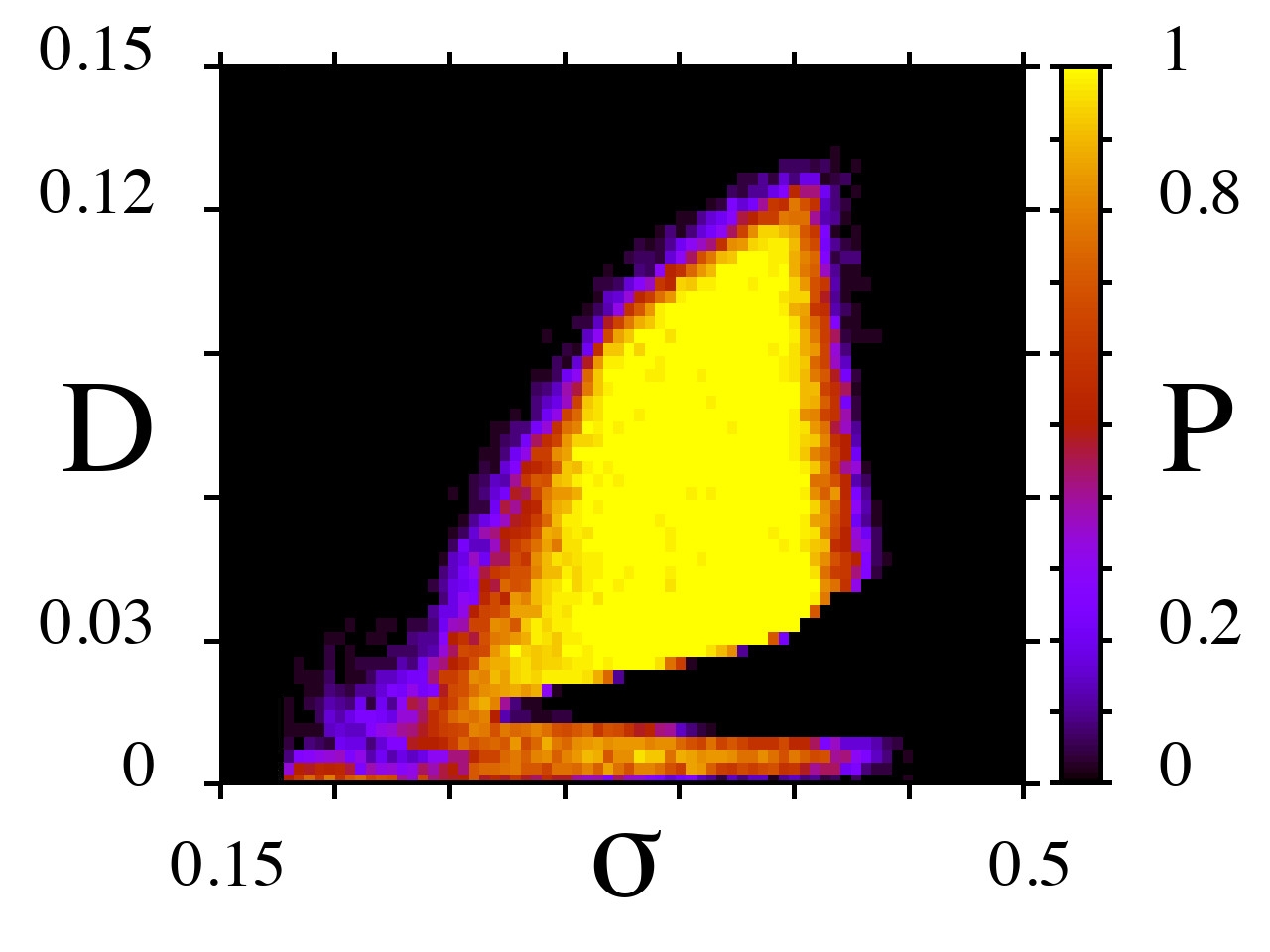}  \\
\hspace{8pt} (e) & \hspace{8pt} (f)\\
\end{tabular}
\caption{Distribution diagrams for the probability $P$ of observing chimera states  in the ($\sigma,D$) parameter plane for the modified Ricker map network  for different values of the local dynamics parameter $\alpha^R$: (a) $\alpha^{R}=17.5$, (b) $\alpha^{R}=20.0$, (c) $\alpha^{R}=21.0$, (d) $\alpha^{R}=22.0$, (e) $\alpha^{R}=30.0$, and (f) $\alpha^{R}=44.0$. The diagrams are plotted using $M=50$ different pairs of realizations of random initial conditions  and noise realizations. Other parameters: $R=320$, $N=1000$.}
	\label{fig_12}
\end{figure}

We now turn to analyze the impact of additive noise on the probability of observing chimera states in the Ricker map network. Numerical results are summarized in the 2D distribution diagrams  for the probability $P$  in the ($\sigma,D$) parameter plane (Fig.~\ref{fig_12}). It has been noted earlier that at $\alpha^{R}<17.2$ there are no chimera states in the network (Fig.~\ref{fig_10}(a),(b)). Introducing 
additive noise of even low intensity induces the appearance of chimera states (with a high probability) in the range of strong coupling. As can be seen from the distribution in Fig.~\ref{fig_12}(a) for $\alpha^R=17.5$, there is a single and rather wide region with respect to both the coupling strength $\sigma$ and the noise intensity $D$ within which the probability $P$ of observing chimeras is essentially equal to 1. Note that at $D=0$, the probability of chimera observation vanishes for all values of $\sigma$ (black and dark-violet color in Fig.~\ref{fig_12}(a)), and only external noise even with an extremely low intensity can increase significantly the probability $P$.

For larger values of $\alpha^R\geq18.5$, the region with a high-probability of chimera observation is split into two parts separated by a channel within which  $P\approx0.5$ (Fig.~\ref{fig_12}(b),(c)). Let us recall that within this channel the probability  vanishes in the case of the logistic map network  (Fig.~\ref{fig_4}(a),(b)) and is $P\approx0.2$ for the Henon map network  (Fig.~\ref{fig_9}(b),(c)). As $\alpha^R$ increases, the channel narrows  and eventually degenerates into a small "island" with a lower probability value  (Fig.~\ref{fig_12}(d)).  

The presented probability distributions for the Ricker map network also verify the manifestation of chimera resonance with respect to both the $\sigma$-interval and the $D$-range corresponding to a high probability of chimera observation. Note that unlike the logistic map and Henon map networks, in this case chimera states exist even for sufficiently strong noise up to $D\approx 0.15$.  

Our numerical studies show that the existence of solitary states within a narrow region in the ($\alpha^R,\sigma$) parameter plane for strong coupling (yellow region around $\sigma \approx 0.4$ in Fig.~\ref{fig_10}(a),(b)) and for large values of $\alpha^R$ can effect the distributions of the probability of observing chimera states already at a low noise intensity. To illustrate this peculiarity two distribution diagrams are presented in Fig.~\ref{fig_12}(e),(f) for large values of $\alpha^R$. 
As can be seen, at low noise levels $D\approx0.025$, $D\approx0.019$ and for sufficiently strong coupling $\sigma \in [0.39,0.45]$, $\sigma \in [0.27,0.44]$ (Fig.~\ref{fig_12}(e),(f), respectively) the region with a high probability of chimera existence is cut by a triangular region in which incoherent dynamics takes place. The appearance of this region with $P=0$ is associated with the presence of solitary states in the Ricker map network at a strong coupling strength. This triangular region is expanded as the local dynamics parameter $\alpha^R$  increases within the range where the solitary state exists in the network.  It has been shown earlier in Ref.~\cite{Rybalova:2022wa} that a sufficiently low noise level is needed to suppress solitary states and to induce a transition to incoherent dynamics. 
Thus, in the considered Ricker map network, the additive noise with low and high intensities can induce chimera states inside the region in which only snapshots with profile discontinuities exist. An intermediate noise intensity shifts the  control parameters to a narrow region of the existence of solitary states and destroys the network dynamics, leading to incoherence. These features cause a gap in the distribution diagrams for the probability of chimera observation for large values of $\alpha^R$ (Fig.~\ref{fig_12}(e),(f)).

\begin{figure}[ht]
	\centering
\begin{tabular}{cccc}
\includegraphics[width=.23\columnwidth]{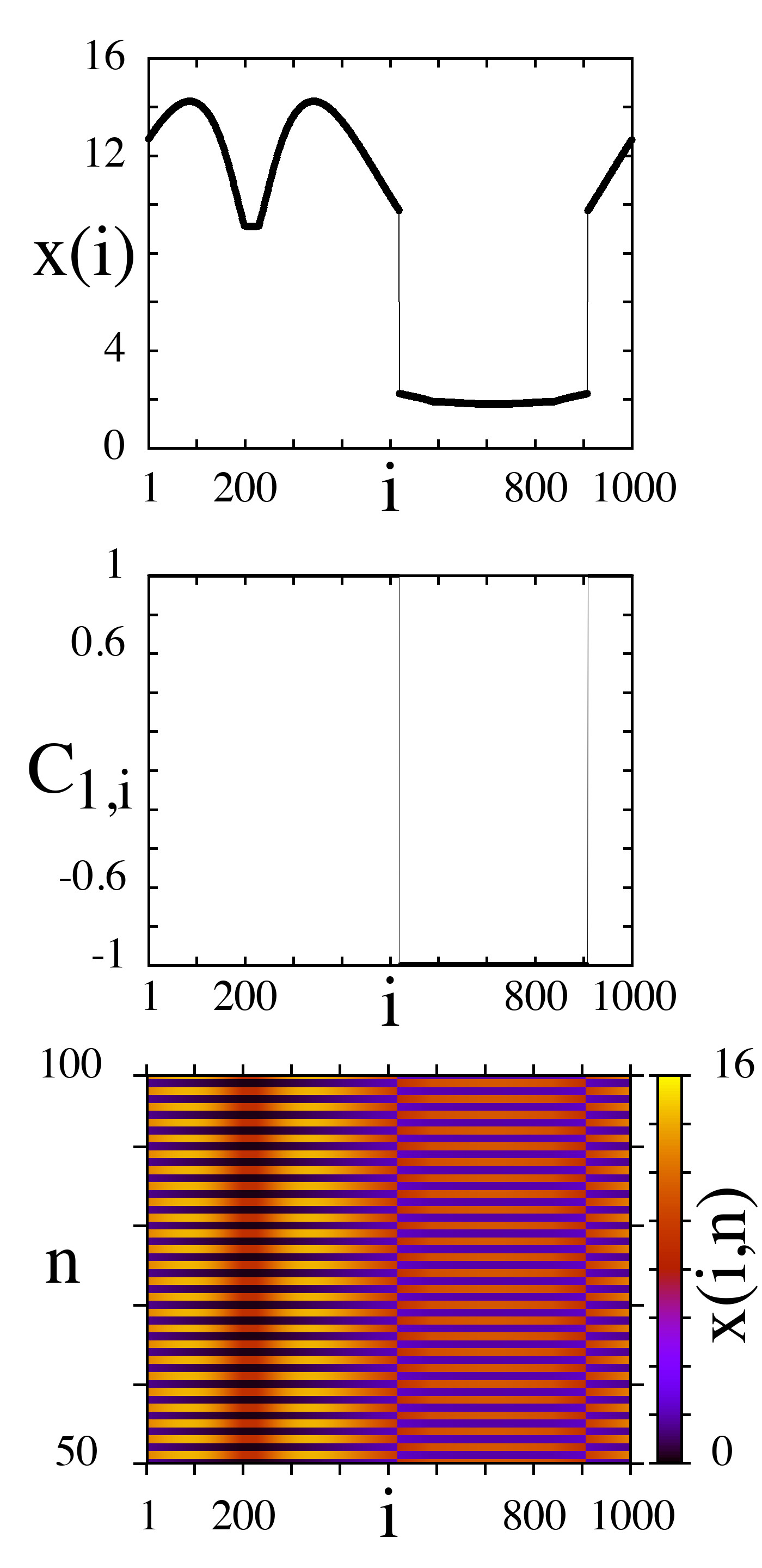} &
\includegraphics[width=.23\columnwidth]{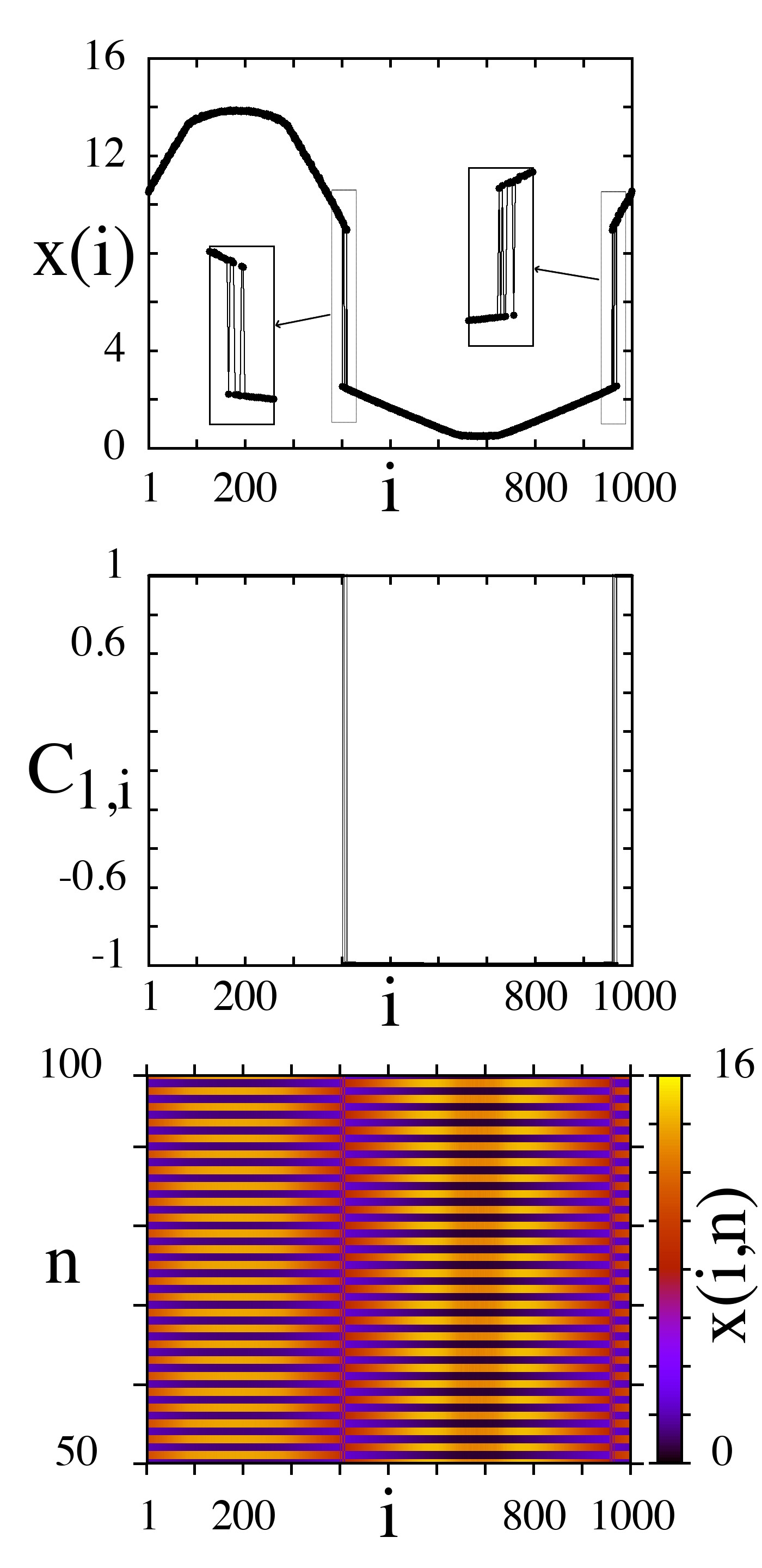}  &
\includegraphics[width=.23\columnwidth]{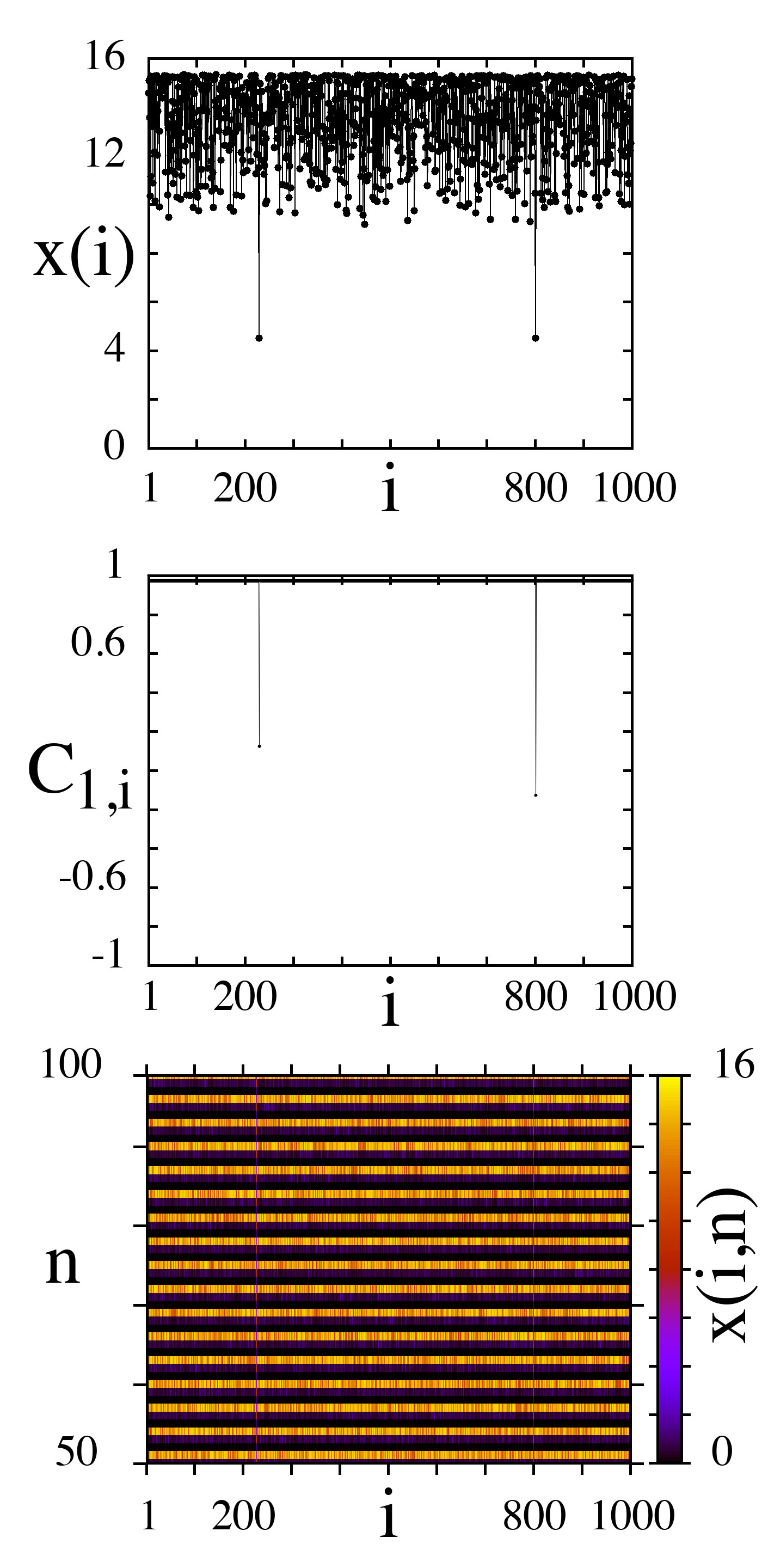} &
\includegraphics[width=.23\columnwidth]{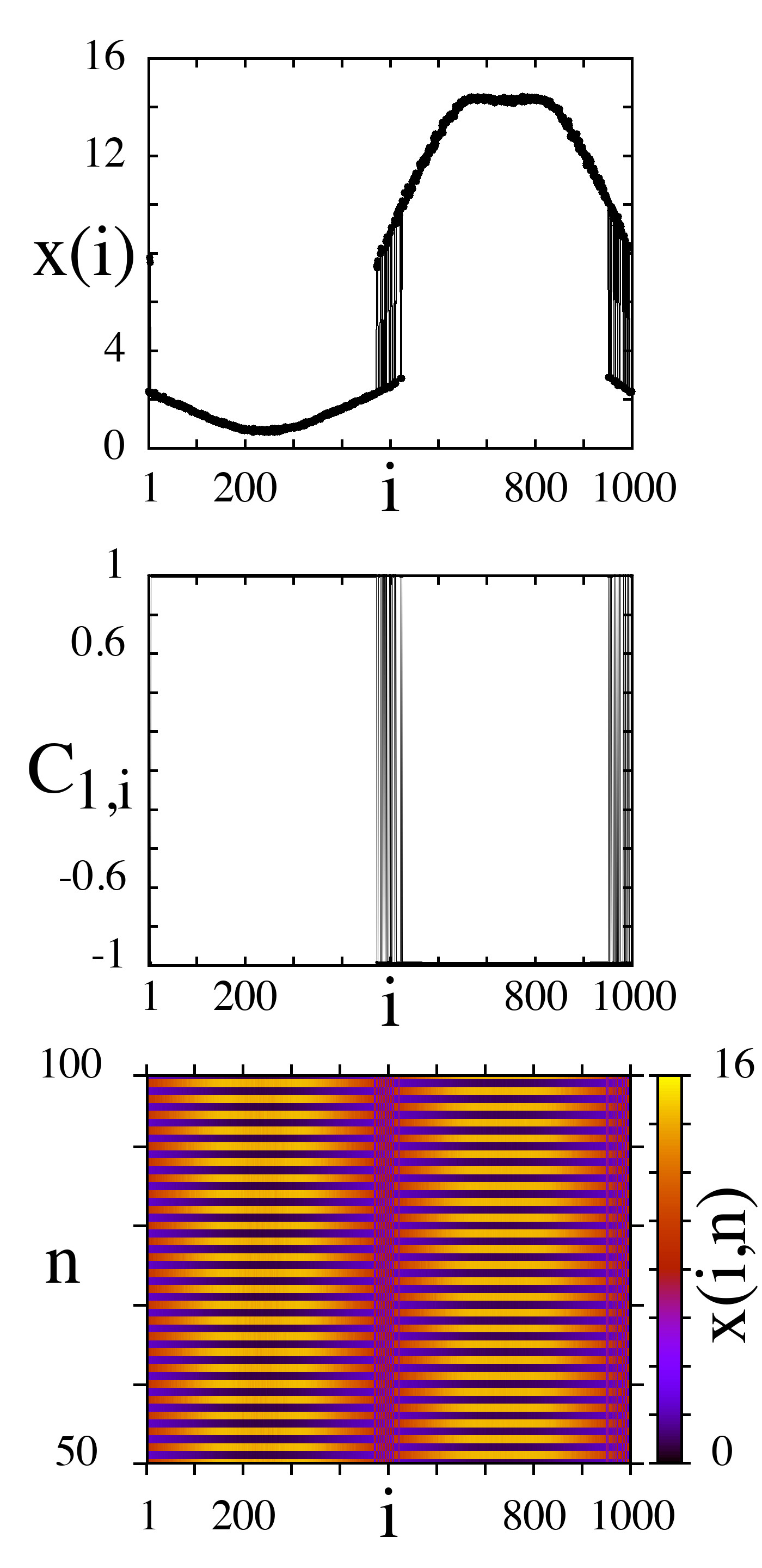} \\
\hspace{8pt} (a) & \hspace{8pt} (b) & \hspace{8pt} (c) & \hspace{8pt} (d)\\
\end{tabular}
\caption{Snapshots of the $x(i)$ variables (upper row), spatial distributions of the cross-correlation coefficient (middle row), and space-time diagrams $x(i,n)$ (lower row) for themodified  Ricker map network at $\sigma=0.333$ corresponding to a snapshot with profile discontinuities at $D=0$ (a) and for different noise intensities: (b) $D=0.00825$, (c) $D=0.0165$, and (d) $D=0.033$.  Other parameters: $\alpha^{R}=44.0$, $R=320$, $N=1000$. The insets in (b) top row show blow-ups.}
	\label{fig_13}
\end{figure}

\begin{figure}[ht]
	\centering
\begin{tabular}{cccc}
\includegraphics[width=.23\columnwidth]{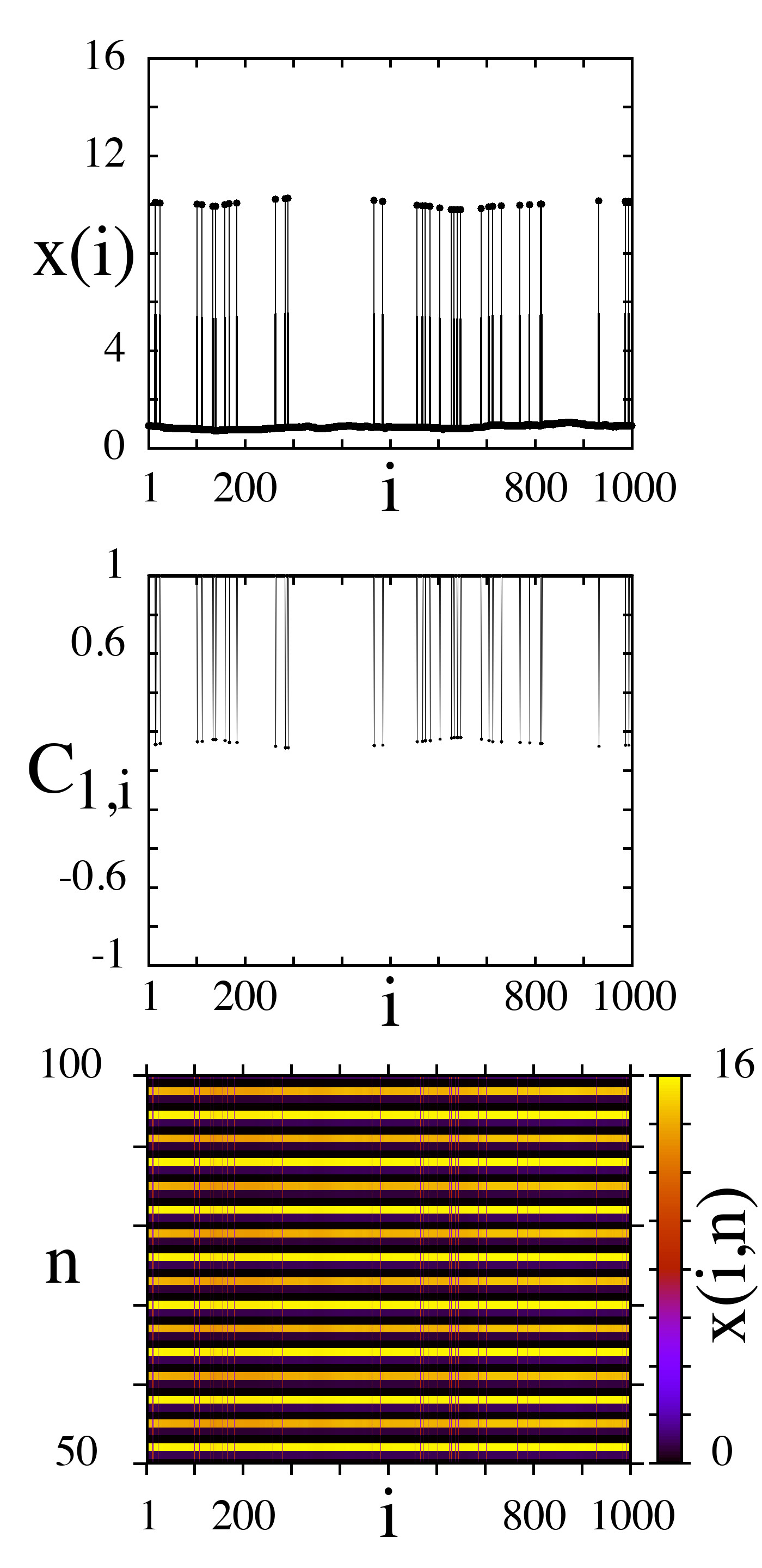} &
\includegraphics[width=.23\columnwidth]{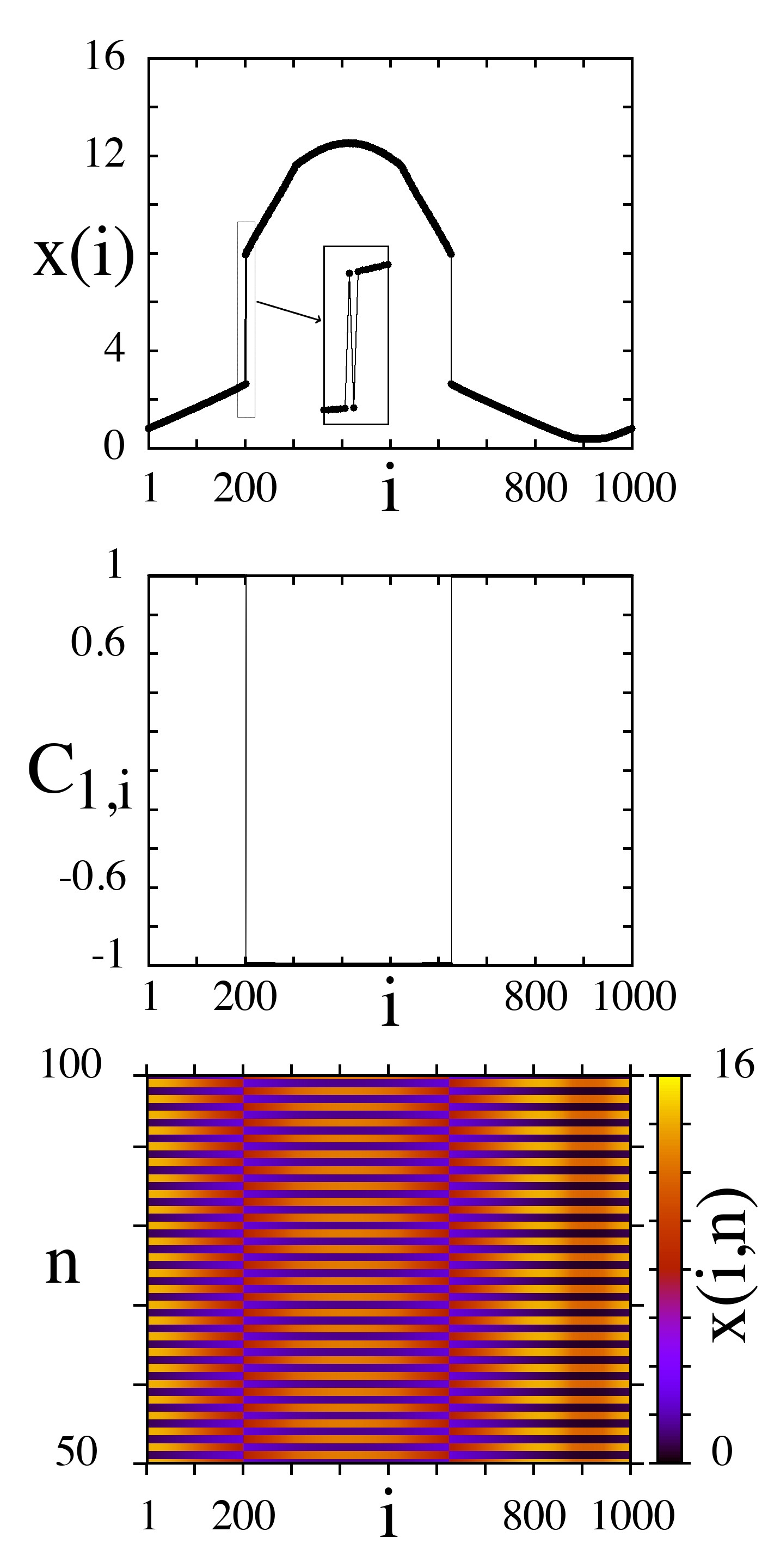}  &
\includegraphics[width=.23\columnwidth]{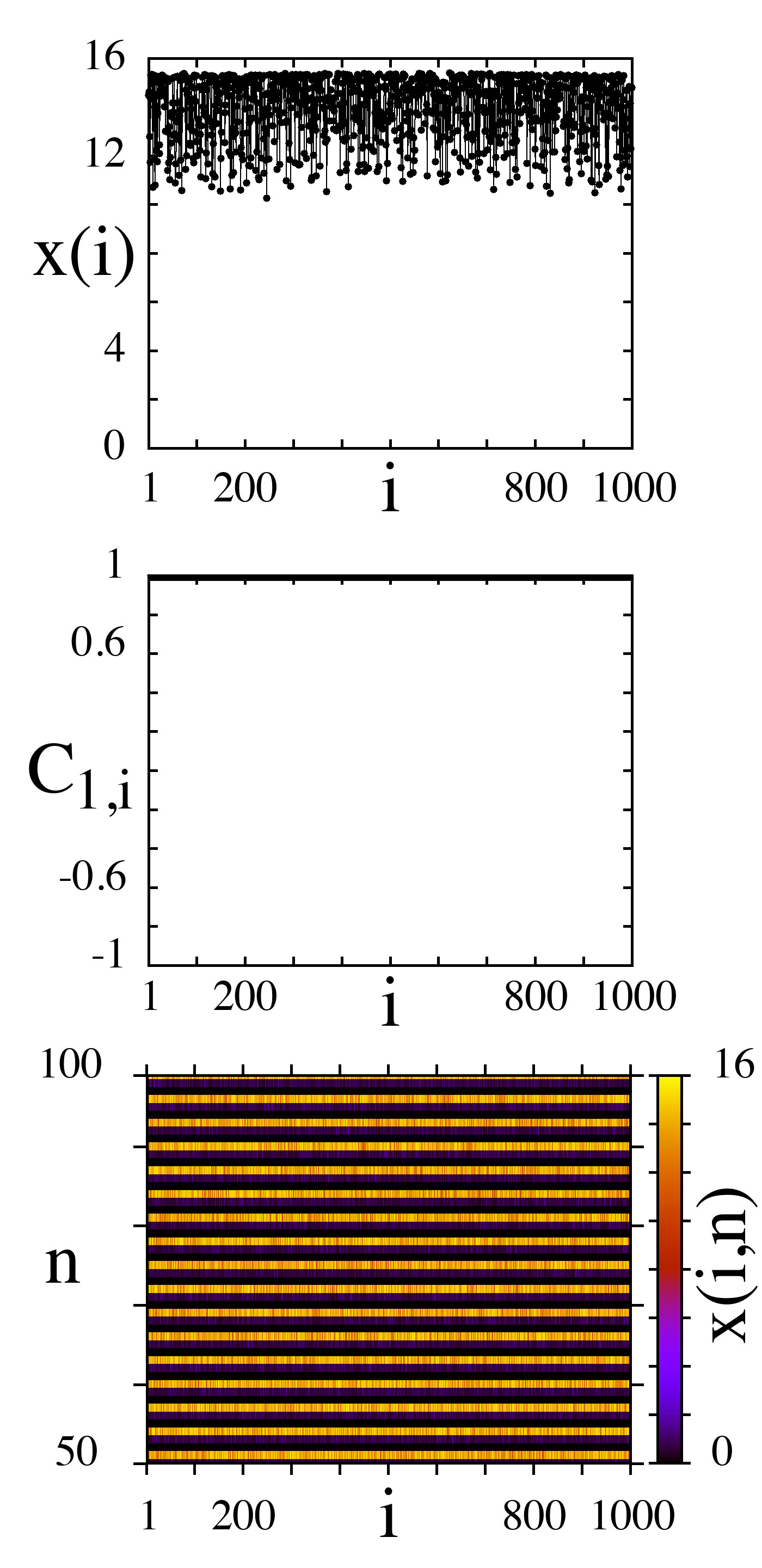} &
\includegraphics[width=.23\columnwidth]{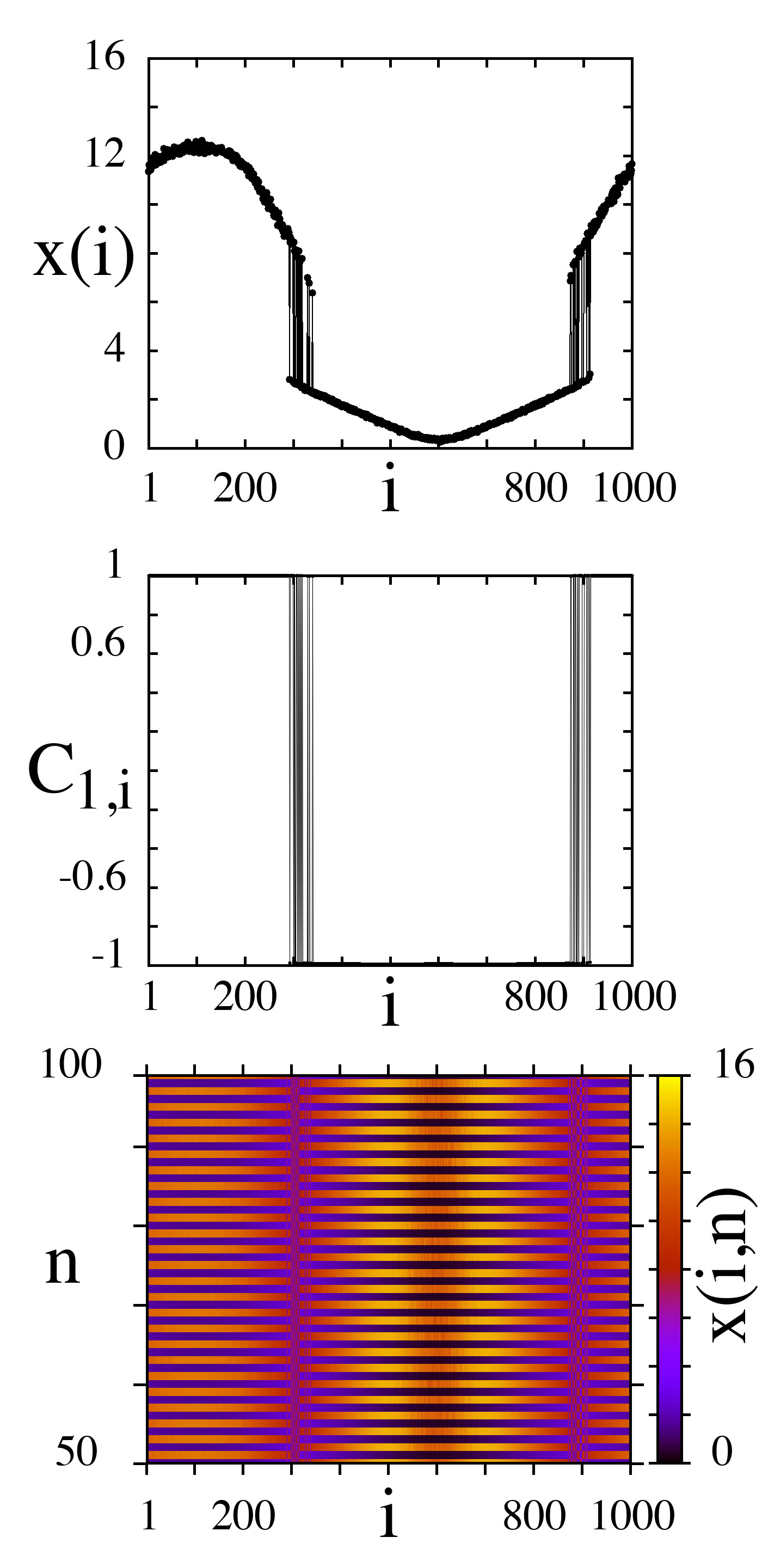} \\
\hspace{0pt} (a) & \hspace{0pt} (b) & \hspace{0pt} (c) & \hspace{0pt} (d)\\
\end{tabular}
\caption{Snapshots of the $x(i)$ variables (upper row), spatial distributions of the cross-correlation coefficient (middle row), and space-time diagrams $x(i,n)$ (lower row) for the Ricker map network at $\sigma=0.378$ corresponding to solitary states at $D=0$ (a) and for different noise intensities: (b) $D=0.00275$, (c) $D=0.01925$, and (d) $D=0.033$. Other parameters: $\alpha^{R}=44.0$, $R=320$, $N=1000$. The inset in (b) top row show blow-ups.}
	\label{fig_14}
\end{figure}

Figures~\ref{fig_13} and \ref{fig_14} illustrate the impact of additive noise of different intensity on the dynamics of the modified Ricker map network for two selected values of the coupling strength, one corresponding to a snapshot with profile discontinuities (Fig.~\ref{fig_13}(a)) and the other corresponding to solitary states (Fig.~\ref{fig_14}(a)) in the noisy-free network of  Ricker maps. In the first case (Fig.~\ref{fig_13}), a low noise intensity induces the appearance of phase chimeras with narrow incoherent clusters (Fig.~\ref{fig_13}(b), $401<i<412$, $958<i<969$). When $D$ increases ($0.0123<D<0.023$, Fig.~\ref{fig_12}(f)), we enter the zero-probability region and the network demonstrates the incoherent regime (Fig.~\ref{fig_13}(c)). And finally, at $D>0.023$, we return again to the region with the high probability of chimera observation and the phase chimera exists in the network (Fig.~\ref{fig_13}(d)). 

A similar evolution of the network dynamics occurs in the presence of noise when solitary states exist in the noise-free case (Fig.~\ref{fig_14}(a)). A low noise intensity $D\lessapprox0.001$ has almost no effect on the regime in the network, the number of solitary nodes  can vary slightly. As the noise level slightly increases, solitary nodes disappear and the network dynamics is characterized by a snapshot with profile discontinuities  or by phase chimeras with narrow incoherent clusters
(Fig.~\ref{fig_14}(b), incoherent cluster $199<i<204$). A further increase in the noise intensity, as in the case of $\sigma=0.333$ (Fig.~\ref{fig_13}),
first leads to entering into the region with zero probability of observing chimeras 
($0.0095<D<0.026$, Fig.~\ref{fig_12}(f)), inside which incoherent dynamics occurs in the network (Fig.~\ref{fig_14},(c)), and then again to the existence of chimera states (Fig.~\ref{fig_14}(d)). Thus, in the solitary state regime which is observed in the noise-free network, a gradual increase in the noise intensity ($0.0095<D<0.026$) suppresses solitary nodes and leads to the stable observation of chimera states at a sufficiently high noise level.

\section{Conclusion}

In this paper we have presented numerical results on the influence of additive Gaussian noise on the spatio-temporal dynamics and especially on the probability of observing chimera states in three different ring networks of nonlocally coupled chaotic discrete-time systems.  The individual nodes are described by the logistic map, the modified Ricker map, and the Henon map.  For each noise-free network, we have constructed two-dimensional diagrams of spatio-temporal regimes in the "local dynamics parameter versus coupling strength" parameter plane and analyzed the peculiarities of the transition from incoherence to complete synchronization as the coupling strength increases. We have found that for all three networks, there is a coherent window  inside the region with profile discontinuities in the diagrams of dynamical regimes, and this feature has led to a significant effect on the probability of observing chimera states in the presence of noise.

To analyze the role of additive noise we have plotted 2D  distribution diagrams for the 
probability of chimera existence in terms of the coupling strength $\sigma$ and the noise intensity $D$. 
Our numerical simulation has shown that in the presence of noise of certain intensities  chimera states can be induced in the networks studied and moreover, the probability of their observation can be significantly increased up to its maximum level (equal to 1) within a rather large interval of the coupling strength $\sigma$. The region in the ($\sigma,D$) parameter plane that corresponds to a high or even maximum probability of observing chimeras can have a different shape depending on the local dynamics parameter of individual nodes in each considered network. In particular, this region is split into two subregions with respect to both $\sigma$ and $D$ if the values of the local dynamics parameter relate to the coherent window in the network dynamics. Within the channel separating the subregions, the probability of chimera observation is either zero (for the logistic map network) or rather low (about 0.2 or 0.5 for the Henon map network and the Ricker map network, respectively). However, after exiting the coherent window with changing local dynamics parameter, the two subregions merge into a single one that has been observed for all the three networks under consideration. 

It has  been established that there is an optimum non-vanishing noise level at which the $\sigma$-interval corresponding to a high or even maximum probability of chimera observation is the largest. The observed phenomenon gives evidence of a beneficial and constructive role of noise in analogy with  stochastic and coherence resonance. In this context, the revealed effect has been called chimera resonance. The value of the coupling strength  $\sigma$ at which chimera states are observed with the maximum probability non-monotonically increases as the noise intensity $D$ grows within the range of chimera existence and decreases as the nonlocal coupling range $R$ decreases. 
We have also found that there is a finite range of the noise intensity $D$ within which chimera states are observed with a high or even maximum probability. This $D$-range is the widest at a certain "resonant" value of the coupling strength $\sigma$. 

Besides the presence of the coherent window in the diagrams of dynamical regimes of the networks studied, the shape of the probability distribution of observing chimeras can also be essentially affected by the existence of solitary states in a network. In our case this has been observed for the modified Ricker map network. It has been shown that at low noise levels and for sufficiently strong coupling $\sigma$ the region with a high probability of chimera existence is cut by a triangular region within which incoherent dynamics takes place in the network. 

 Our results once again show the counterintuitive constructive role of noise in the dynamics of complex networks and the possibility of using external noise as an effective tool for controlling the formation and stability of the observed spatio-temporal structures.

\section*{Acknowledgments}
E.R., V.N., and G.S. acknowledge financial support from the Russian Science Foundation (project No. 20-12-00119).

\section*{Author declarations}
\subsection*{Conflict of Interest}
The authors have no conflicts to disclose.

\subsection*{Author Contributions}
{\bf E. Rybalova:} Conceptualization, Methodology, Software, Validation, Formal analysis, Investigation, Visualization, Writing - Original Draft. 
{\bf V. Nechaev:} Validation, Investigation, Visualization. 
{\bf E. Sch\"{o}ll:} Conceptualization, Writing - Review \& Editing, Supervision. 
{\bf G. Strelkova:} Conceptualization, Methodology, Writing - Original Draft, Writing - Review \& Editing, Supervision.

\section*{References}
\bibliography{thebibliography}
\end{document}